\documentclass[traditabstract]{aa} % for the abstract without structuration
\usepackage{graphicx}
\usepackage{natbib}

\usepackage{comment}
\usepackage{ifthen}
\usepackage{subfig}

\newcommand{\loopand}{\ifnum\value{planetcounter}=2 and \else\fi}
\newcommand{\loopcomma}{\ifnum\value{planetcounter}<2 ,\else. \fi}
\newcommand{\loopcommanoperiod}{\ifnum\value{planetcounter}<2 ,\else \space\fi}
\newcommand{\loopcommanospace}{\ifnum\value{planetcounter}<2 ,\else \fi}
\newcommand{\arcdeg}{^{\circ}}
\newcommand{\kms}{km\,s$^{-1}$}
\newcommand{\ms}{m\,s$^{-1}$}
\newcommand{\arstar}{a/R_{\star}}
\newcommand{\zrstar}{\zeta/R_{\star}}
\newcommand{\rpl}{R_{p}}
\newcommand{\rstar}{R_{\star}}
\newcommand{\mpl}{M_{p}}
\newcommand{\mstar}{M_{\star}}
\newcommand{\mjup}{M_{\mathrm{J}}}
\newcommand{\rjup}{R_{\mathrm{J}}}
\newcommand{\rhopl}{\rho_{p}}
\newcommand{\gcmc}{g\,cm$^{-3}$}
\newcommand{\reftabl}{Table\,\ref}
\newcommand{\refsecl}{Sect.\,\ref}
\newcommand{\teffstar}{\ensuremath{T_{\rm eff\star}}}
\newcommand{\rhostar}{\ensuremath{\rho_\star}}
\newcommand{\lstar}{\ensuremath{L_\star}}

\newcommand{\loggstar}{\ensuremath{\log{g_{\star}}}}
\newcommand{\lc}{light curve}
\newcommand{\masy}{\ensuremath{\rm mas\,yr^{-1}}}
\newcommand{\feh}{\ensuremath{\rm [Fe/H]}}
\newcommand{\vsini}{\ensuremath{v \sin{i_{\star}}}}

\newcommand{\rsun}{\ensuremath{R_{\sun}}}
\newcommand{\msun}{\ensuremath{M_{\sun}}}
\newcommand{\lsun}{\ensuremath{L_{\sun}}}

\newcommand{\reffigl}{Fig.\,\ref}

\newcommand{\hatcurhtrxxxxxA}{HATS582-012}                             % Original HTR name of target
\newcommand{\hatcurfieldxxxxxA}{582}                                   % Original HTR field
\newcommand{\hatcurCCraxxxxxA}{\ensuremath{21^{\mathrm h}07^{\mathrm m}50.88{\mathrm s}}}                            % Right Ascension
\newcommand{\hatcurCCdecxxxxxA}{\ensuremath{-26{\arcdeg}05{\arcmin}48.0{\arcsec}}}                           % Declination
\newcommand{\hatcurCCmagxxxxxA}{13.887}                                % apparent V-band magnitude
\newcommand{\hatcurCCtwomassxxxxxA}{2MASS~21075075-2605479}            % 2MASS identifier
\newcommand{\hatcurCCgscxxxxxA}{GSC~6928-00497}                        % GSC(1.2) identifier
\newcommand{\hatcurCCtassmvxxxxxA}{\ensuremath{13.887\pm0.010}}        % APASS V-band magnitude
\newcommand{\hatcurCCtassmvshortxxxxxA}{\ensuremath{13.9}}             % APASS V-band magnitude
\newcommand{\hatcurCCtassmBxxxxxA}{\ensuremath{14.686\pm0.020}}        % APASS B-band magnitude
\newcommand{\hatcurCCtassmBshortxxxxxA}{\ensuremath{14.7}}             % APASS B-band magnitude
\newcommand{\hatcurCCtassmIxxxxxA}{\ensuremath{100\pm1000}}            % TASS I-band magnitude
\newcommand{\hatcurCCtassmIshortxxxxxA}{\ensuremath{100.0}}            % TASS I-band magnitude
\newcommand{\hatcurCCtassmgxxxxxA}{\ensuremath{14.217\pm0.010}}        % APASS g-band magnitude
\newcommand{\hatcurCCtassmgshortxxxxxA}{\ensuremath{14.2}}             % APASS g-band magnitude
\newcommand{\hatcurCCtassmrxxxxxA}{\ensuremath{13.612\pm0.010}}        % APASS r-band magnitude
\newcommand{\hatcurCCtassmrshortxxxxxA}{\ensuremath{13.6}}             % APASS r-band magnitude
\newcommand{\hatcurCCtassmixxxxxA}{\ensuremath{13.392\pm0.010}}        % APASS i-band magnitude
\newcommand{\hatcurCCtassmishortxxxxxA}{\ensuremath{13.4}}             % APASS i-band magnitude
\newcommand{\hatcurCCtwomassJmagxxxxxA}{\ensuremath{12.439\pm0.021}}   % 2MASS ORIG MAG
\newcommand{\hatcurCCtwomassHmagxxxxxA}{\ensuremath{12.052\pm0.025}}   % 2MASS ORIG MAG
\newcommand{\hatcurCCtwomassKmagxxxxxA}{\ensuremath{11.983\pm0.028}}   % 2MASS ORIG MAG
\newcommand{\hatcurCCcitJmagxxxxxA}{\ensuremath{12.451\pm0.022}}       % 2MASS CIT MAG
\newcommand{\hatcurCCcitHmagxxxxxA}{\ensuremath{12.047\pm0.025}}       % 2MASS CIT MAG
\newcommand{\hatcurCCcitKmagxxxxxA}{\ensuremath{12.007\pm0.028}}       % 2MASS CIT MAG
\newcommand{\hatcurCCbbJmagxxxxxA}{\ensuremath{12.508\pm0.023}}        % 2MASS BB MAG
\newcommand{\hatcurCCbbHmagxxxxxA}{\ensuremath{12.068\pm0.026}}        % 2MASS BB MAG
\newcommand{\hatcurCCbbKmagxxxxxA}{\ensuremath{12.027\pm0.028}}        % 2MASS BB MAG
\newcommand{\hatcurCCesoJmagxxxxxA}{\ensuremath{12.511\pm0.025}}       % 2MASS ESO MAG
\newcommand{\hatcurCCesoHmagxxxxxA}{\ensuremath{12.062\pm0.029}}       % 2MASS ESO MAG
\newcommand{\hatcurCCesoKmagxxxxxA}{\ensuremath{12.025\pm0.029}}       % 2MASS ESO MAG
\newcommand{\hatcurCCesoJHmagxxxxxA}{\ensuremath{0.448\pm0.036}}       % 2MASS ESO JH COLOR
\newcommand{\hatcurCCesoJKmagxxxxxA}{\ensuremath{0.486\pm0.038}}       % 2MASS ESO JK COLOR
\newcommand{\hatcurCCesoHKmagxxxxxA}{\ensuremath{0.038\pm0.042}}       % 2MASS ESO HK COLOR
\newcommand{\hatcurLCdipxxxxxA}{\ensuremath{24.2}}                     % BLS detected dip (mmag)
\newcommand{\hatcurLCrprstarxxxxxA}{\ensuremath{0.1402\pm0.0016}}      % Rp/R*
\newcommand{\hatcurLCbsqxxxxxA}{\ensuremath{0.062_{-0.030}^{+0.035}}}  % impact parameter square
\newcommand{\hatcurLCimpxxxxxA}{\ensuremath{0.249_{-0.071}^{+0.062}}}  % impact parameter
\newcommand{\hatcurLCzetaxxxxxA}{\ensuremath{20.939\pm0.090}}          % zeta/R*
\newcommand{\hatcurLCdurxxxxxA}{\ensuremath{0.10978\pm0.00084}}        % transit duration (days)
\newcommand{\hatcurLCdurshortxxxxxA}{\ensuremath{0.1098}}              % transit duration (days)
\newcommand{\hatcurLCdurhrxxxxxA}{\ensuremath{2.635\pm0.020}}          % transit duration (hours)
\newcommand{\hatcurLCdurhrshortxxxxxA}{\ensuremath{2.635}}             % transit duration (hours)
\newcommand{\hatcurLCqxxxxxA}{\ensuremath{0.03610\pm0.00028}}          % fractional transit duration (days)
\newcommand{\hatcurLCqshortxxxxxA}{\ensuremath{0.036}}                 % fractional transit duration (days)
\newcommand{\hatcurLCingdurxxxxxA}{\ensuremath{0.01430\pm0.00063}}     % ingress/egress duration (days)
\newcommand{\hatcurLCPxxxxxA}{\ensuremath{3.0440499\pm0.0000027}}      % period (days)
\newcommand{\hatcurLCPprecxxxxxA}{\ensuremath{3.0440499}}              % period (days)
\newcommand{\hatcurLCPshortxxxxxA}{\ensuremath{3.0440}}                % period (days)
\newcommand{\hatcurLCTxxxxxA}{\ensuremath{2456340.31705\pm0.00026}}    % epoch (BJD)
\newcommand{\hatcurLCTAxxxxxA}{\ensuremath{2455092.25658\pm0.00099}}   % TA (BJD)
\newcommand{\hatcurLCTBxxxxxA}{\ensuremath{2456474.25524\pm0.00035}}   % TB (BJD)
\newcommand{\hatcurLChatnetmxxxxxA}{\ensuremath{13.66916\pm0.00011}}   % HATNet OOT level
\newcommand{\hatcurLCiblendxxxxxA}{\ensuremath{0.947\pm0.034}}         % HATNet iblend factor
\newcommand{\hatcurSMEiteffxxxxxA}{\ensuremath{5540\pm130}}            % Ini SME, stellar effective temperature
\newcommand{\hatcurSMEizfehxxxxxA}{\ensuremath{0.100\pm0.090}}         % Ini SME, stellar metallicity
\newcommand{\hatcurSMEizfehshortxxxxxA}{\ensuremath{0.10}}             % Ini SME, stellar metallicity
\newcommand{\hatcurSMEiloggxxxxxA}{\ensuremath{4.85\pm0.23}}           % Ini SME, stellar surface gravity
\newcommand{\hatcurSMEivsinxxxxxA}{\ensuremath{3.12\pm0.30}}           % Ini SME, stellar rotational velocity
\newcommand{\hatcurSMEivmacxxxxxA}{\ensuremath{0.0}}                   % Ini SME, stellar macroturbulence
\newcommand{\hatcurSMEivmicxxxxxA}{\ensuremath{0.0}}                   % Ini SME, stellar microturbulence
\newcommand{\hatcurSMEiiteffxxxxxA}{\ensuremath{5523\pm69}}            % Final SME, stellar effective temperature
\newcommand{\hatcurSMEiizfehxxxxxA}{\ensuremath{0.050\pm0.060}}        % Final SME, stellar metallicity
\newcommand{\hatcurSMEiizfehshortxxxxxA}{\ensuremath{0.05}}            % Final SME, stellar metallicity
\newcommand{\hatcurSMEiiloggxxxxxA}{\ensuremath{4.518\pm0.020}}        % Final SME, stellar surface gravity
\newcommand{\hatcurSMEiivsinxxxxxA}{\ensuremath{2.82\pm0.30}}          % Final SME, stellar rotational velocity
\newcommand{\hatcurANUWIFESteffxxxxxA}{\ensuremath{5540\pm300}}           % ANUWIFES stellar effective temperature
\newcommand{\hatcurANUWIFESzfehxxxxxA}{\ensuremath{0.00\pm0.50}}          % ANUWIFES stellar metallicity
\newcommand{\hatcurANUWIFESloggxxxxxA}{\ensuremath{4.50\pm0.30}}          % ANUWIFES stellar surface gravity
\newcommand{\hatcurANUWIFESvsinixxxxxA}{\ensuremath{nff\pmnff}}           % ANUWIFES stellar rotational velocity
\newcommand{\hatcurANUWIFESgammaxxxxxA}{\ensuremath{25.09\pm0.26}}        % ANUWIFES absolute gamma velocity
\newcommand{\hatcurANUWIFESnumspecxxxxxA}{\ensuremath{2}}                 % ANUWIFES number of spectra
\newcommand{\hatcurANUWIFESspanxxxxxA}{\ensuremath{NULL}}                 % ANUWIFES timespan of observations
\newcommand{\hatcurANUWIFESrvrmsxxxxxA}{\ensuremath{0.67}}                % ANUWIFES rms of RV values [km/s]
\newcommand{\hatcurLBizxxxxxA}{\ensuremath{0.2436}}                    % Limb darkening parameters, Gamma1, z-band
\newcommand{\hatcurLBiizxxxxxA}{\ensuremath{0.3108}}                   % Limb darkening parameters, Gamma2, z-band
\newcommand{\hatcurLBiixxxxxA}{\ensuremath{0.3116}}                    % Limb darkening parameters, Gamma1, i-band
\newcommand{\hatcurLBiiixxxxxA}{\ensuremath{0.3063}}                   % Limb darkening parameters, Gamma2, i-band
\newcommand{\hatcurLBiIxxxxxA}{\ensuremath{0.2890}}                    % Limb darkening parameters, Gamma1, I-band
\newcommand{\hatcurLBiiIxxxxxA}{\ensuremath{0.3083}}                   % Limb darkening parameters, Gamma2, I-band
\newcommand{\hatcurLBigxxxxxA}{\ensuremath{0.6213}}                    % Limb darkening parameters, Gamma1, g-band
\newcommand{\hatcurLBiigxxxxxA}{\ensuremath{0.1844}}                   % Limb darkening parameters, Gamma2, g-band
\newcommand{\hatcurLBirxxxxxA}{\ensuremath{0.4107}}                    % Limb darkening parameters, Gamma1, r-band
\newcommand{\hatcurLBiirxxxxxA}{\ensuremath{0.2928}}                   % Limb darkening parameters, Gamma2, r-band
\newcommand{\hatcurLBiRxxxxxA}{\ensuremath{0.3833}}                    % Limb darkening parameters, Gamma1, R-band
\newcommand{\hatcurLBiiRxxxxxA}{\ensuremath{0.2974}}                   % Limb darkening parameters, Gamma2, R-band
\newcommand{\hatcurISOmxxxxxA}{\ensuremath{0.962\pm0.029}}             % stellar mass
\newcommand{\hatcurISOmshortxxxxxA}{\ensuremath{0.96}}                 % stellar mass
\newcommand{\hatcurISOmlongxxxxxA}{\ensuremath{0.962\pm0.029}}         % stellar mass
\newcommand{\hatcurISOrxxxxxA}{\ensuremath{0.887\pm0.019}}             % stellar radius
\newcommand{\hatcurISOrshortxxxxxA}{\ensuremath{0.89}}                 % stellar radius
\newcommand{\hatcurISOrlongxxxxxA}{\ensuremath{0.887\pm0.019}}         % stellar radius
\newcommand{\hatcurISOrhoxxxxxA}{\ensuremath{1.93\pm0.11}}             % stellar density (cgs)
\newcommand{\hatcurISOrholongxxxxxA}{\ensuremath{1.93\pm0.11}}         % stellar density (cgs)
\newcommand{\hatcurISOloggxxxxxA}{\ensuremath{4.524\pm0.017}}          % stellar surface gravity from isochrones
\newcommand{\hatcurISOlumxxxxxA}{\ensuremath{0.650\pm0.051}}           % stellar luminosity
\newcommand{\hatcurISOlumshortxxxxxA}{\ensuremath{0.65}}               % stellar luminosity
\newcommand{\hatcurISOmvxxxxxA}{\ensuremath{5.349\pm0.096}}            % stellar absolute magnitude
\newcommand{\hatcurISOvixxxxxA}{\ensuremath{0.778\pm0.019}}            % stellar V-I index
\newcommand{\hatcurISOagexxxxxA}{\ensuremath{2.5\pm1.7}}               % stellar age
\newcommand{\hatcurISOsigmaxxxxxA}{\ensuremath{0.00090\pm0.00013}}     % system mass-correction sigma parameter
\newcommand{\hatcurISOMJxxxxxA}{\ensuremath{4.070\pm0.066}}            % stellar absolute J magnitude
\newcommand{\hatcurISOMHxxxxxA}{\ensuremath{3.674\pm0.055}}            % stellar absolute H magnitude
\newcommand{\hatcurISOMKxxxxxA}{\ensuremath{3.610\pm0.054}}            % stellar absolute K magnitude
\newcommand{\hatcurISOJKxxxxxA}{\ensuremath{0.460\pm0.020}}            % J-K color index from isochrones.
\newcommand{\hatcurISOspecxxxxxA}{G}                                   % stellar spectral type
\newcommand{\hatcurRVKxxxxxA}{\ensuremath{78\pm10}}                    % RV semi-amplitude [m/s]
\newcommand{\hatcurRVrkxxxxxA}{\ensuremath{0\pm0}}                     % sqrt(e)*cos(omega)
\newcommand{\hatcurRVrhxxxxxA}{\ensuremath{0\pm0}}                     % sqrt(e)*sin(omega)
\newcommand{\hatcurRVkxxxxxA}{\ensuremath{0\pm0}}                      % e*cos(omega)
\newcommand{\hatcurRVhxxxxxA}{\ensuremath{0\pm0}}                      % e*sin(omega)
\newcommand{\hatcurRVtronexxxxxA}{\ensuremath{0\pm0}}                  % RV linear trend tr1 factor
\newcommand{\hatcurRVtrtwoxxxxxA}{\ensuremath{0\pm0}}                  % RV linear trend tr2 factor
\newcommand{\hatcurRVgammaAxxxxxA}{\ensuremath{-6.6\pm6.2}}            % RV gamma velocity, relative scale
\newcommand{\hatcurRVjitterAxxxxxA}{\ensuremath{0.0\pm4.0}}            % RV jitter (m/s)
\newcommand{\hatcurRVfitrmsAxxxxxA}{\ensuremath{21.0}}                 % RVfitrms
\newcommand{\hatcurRVgammaBxxxxxA}{\ensuremath{25807\pm15}}            % RV gamma velocity, relative scale
\newcommand{\hatcurRVjitterBxxxxxA}{\ensuremath{62\pm14}}              % RV jitter (m/s)
\newcommand{\hatcurRVfitrmsBxxxxxA}{\ensuremath{67.5}}                 % RVfitrms
\newcommand{\hatcurRVgammaCxxxxxA}{\ensuremath{25771\pm47}}            % RV gamma velocity, relative scale
\newcommand{\hatcurRVjitterCxxxxxA}{\ensuremath{108\pm41}}             % RV jitter (m/s)
\newcommand{\hatcurRVfitrmsCxxxxxA}{\ensuremath{99.8}}                 % RVfitrms
\newcommand{\hatcurRVgammaDxxxxxA}{\ensuremath{25780\pm140}}           % RV gamma velocity, relative scale
\newcommand{\hatcurRVjitterDxxxxxA}{\ensuremath{220\pm130}}            % RV jitter (m/s)
\newcommand{\hatcurRVfitrmsDxxxxxA}{\ensuremath{188.5}}                % RVfitrms
\newcommand{\hatcurRVeccenxxxxxA}{\ensuremath{0\pm0}}                  % eccentricity
\newcommand{\hatcurRVeccentwosiglimxxxxxA}{\ensuremath{<0.000}}        % eccentricity
\newcommand{\hatcurRVomegaxxxxxA}{\ensuremath{0\pm0}}                  % argument of pericenter
\newcommand{\hatcurPPixxxxxA}{\ensuremath{88.55\pm0.43}}               % orbital inclination
\newcommand{\hatcurPPgxxxxxA}{\ensuremath{9.1\pm1.3}}                  % planetary surface gravity (m/s^2)
\newcommand{\hatcurPPloggxxxxxA}{\ensuremath{2.961\pm0.063}}           % planetary surface gravity (log cgs)
\newcommand{\hatcurPParxxxxxA}{\ensuremath{9.82\pm0.18}}               % relative orbital radius (a/R*)
\newcommand{\hatcurPParelxxxxxA}{\ensuremath{0.04057\pm0.00041}}       % semimajor axis (AU)
\newcommand{\hatcurPPrhoxxxxxA}{\ensuremath{0.377\pm0.058}}            % planetary density (cgs)
\newcommand{\hatcurPPmxxxxxA}{\ensuremath{0.543\pm0.072}}              % planetary mass (M_jup)
\newcommand{\hatcurPPmshortxxxxxA}{\ensuremath{0.54}}                  % planetary mass (M_jup)
\newcommand{\hatcurPPmlongxxxxxA}{\ensuremath{0.543\pm0.072}}          % planetary mass (M_jup)
\newcommand{\hatcurPPmexxxxxA}{\ensuremath{173\pm23}}                  % planetary mass (M_earth)
\newcommand{\hatcurPPmeshortxxxxxA}{\ensuremath{172.5}}                % planetary mass (M_earth)
\newcommand{\hatcurPPmelongxxxxxA}{\ensuremath{173\pm23}}              % planetary mass (M_earth)
\newcommand{\hatcurPPrxxxxxA}{\ensuremath{1.212\pm0.035}}              % planetary radius (R_jup)
\newcommand{\hatcurPPrshortxxxxxA}{\ensuremath{1.21}}                  % planetary radius (R_jup)
\newcommand{\hatcurPPrlongxxxxxA}{\ensuremath{1.212\pm0.035}}          % planetary radius (R_jup)
\newcommand{\hatcurPPrexxxxxA}{\ensuremath{13.58\pm0.39}}              % planetary radius (R_earth)
\newcommand{\hatcurPPreshortxxxxxA}{\ensuremath{13.6}}                 % planetary radius (R_earth)
\newcommand{\hatcurPPrelongxxxxxA}{\ensuremath{13.58\pm0.39}}          % planetary radius (R_earth)
\newcommand{\hatcurPPmrcorrxxxxxA}{\ensuremath{0.07}}                  % mass/radius correlation
\newcommand{\hatcurPPteffxxxxxA}{\ensuremath{1244\pm20}}               % planetary temperature (K)
\newcommand{\hatcurPPthetaxxxxxA}{\ensuremath{0.0377\pm0.0050}}        % Safranov number
\newcommand{\hatcurPPfluxperixxxxxA}{\ensuremath{5.40\pm0.35}}         % flux @ periastron (CGS)
\newcommand{\hatcurPPfluxperidimxxxxxA}{\ensuremath{8}}                % flux @ periastron (CGS) units.
\newcommand{\hatcurPPfluxapxxxxxA}{\ensuremath{5.40\pm0.35}}           % flux @ apastron (CGS)
\newcommand{\hatcurPPfluxapdimxxxxxA}{\ensuremath{8}}                  % flux @ apastron (CGS) units.
\newcommand{\hatcurPPfluxavgxxxxxA}{\ensuremath{5.40\pm0.35}}          % flux on average (CGS)
\newcommand{\hatcurPPfluxavgdimxxxxxA}{\ensuremath{8}}                 % flux average (CGS) units.
\newcommand{\hatcurPPfluxavglogxxxxxA}{\ensuremath{8.732\pm0.028}}     % log10 flux on average (CGS)
\newcommand{\hatcurXsecphasexxxxxA}{\ensuremath{0\pm0}}                % Phase of secondary eclipse
\newcommand{\hatcurXsecondaryxxxxxA}{\ensuremath{2456341.83908\pm0.00026}} % Secondary eclipse epoch
\newcommand{\hatcurXsecdurxxxxxA}{\ensuremath{0.10978\pm0.00084}}      % sec eclipse duration (days)
\newcommand{\hatcurXsecingdurxxxxxA}{\ensuremath{0.01430\pm0.00063}}   % sec I/E duration (days)
\newcommand{\hatcurPPphiconjxxxxxA}{\ensuremath{0\pm0}}                % phase diff between conjunction and periastron
\newcommand{\hatcurPPperixxxxxA}{\ensuremath{2456339.55604\pm0.00026}} % time of periastron passage.
\newcommand{\hatcurPPaequivxxxxxA}{\ensuremath{0.0503\pm0.0016}}       % equivalent semi-major axis
\newcommand{\hatcurPPtcircxxxxxA}{\ensuremath{76\pm14}}                % circularization timescale
\newcommand{\hatcurPPtinfallxxxxxA}{\ensuremath{4690_{-650}^{+860}}}   % infall timescale
\newcommand{\hatcurXdistxxxxxA}{\ensuremath{482\pm14}}                 % distance (pc), no reddenning correction
\newcommand{\hatcurXAvxxxxxA}{\ensuremath{0.152\pm0.062}}              % Av (mag)
\newcommand{\hatcurXdistredxxxxxA}{\ensuremath{476\pm12}}              % distance with Av correction (pc)
\newcommand{\hatcurXEBVxxxxxA}{\ensuremath{0.049\pm0.020}}             % E(B-V) (mag)
\newcommand{\hatcurXmvisoredxxxxxA}{\ensuremath{13.887\pm0.010}}       % Expected m_v with reddening (mag)
\newcommand{\hatcurXmiisoredxxxxxA}{\ensuremath{13.031\pm0.016}}       % Expected m_i with reddening (mag)
\newcommand{\hatcurXmjisoredxxxxxA}{\ensuremath{12.499\pm0.015}}       % Expected m_j with reddening (mag)
\newcommand{\hatcurXmhisoredxxxxxA}{\ensuremath{12.089\pm0.017}}       % Expected m_h with reddening (mag)
\newcommand{\hatcurXmkisoredxxxxxA}{\ensuremath{12.013\pm0.018}}       % Expected m_k with reddening (mag)
\newcommand{\hatcurXviisoredxxxxxA}{\ensuremath{0.856\pm0.017}}        % Expected V-I with reddening (mag)
\newcommand{\hatcurXvkisoredxxxxxA}{\ensuremath{1.874\pm0.021}}        % Expected V-K with reddening (mag)
\newcommand{\hatcurXjhisoredxxxxxA}{\ensuremath{0.4100\pm0.0090}}      % Expected J-H with reddening (mag)
\newcommand{\hatcurXjkisoredxxxxxA}{\ensuremath{0.4850\pm0.0084}}      % Expected J-K with reddening (mag)
\newcommand{\hatcurCCpmraxxxxxA}{\ensuremath{-2\pm14}}                 % proper motion, in RA
\newcommand{\hatcurCCpmdecxxxxxA}{\ensuremath{-9.1\pm1.6}}             % proper motion, in DEC
\newcommand{\hatcurCCpmxxxxxA}{\ensuremath{9\pm14}}                    % proper motion

\newcommand{\hatcurhtrxxxxxB}{HATS582-006}                             % Original HTR name of target
\newcommand{\hatcurfieldxxxxxB}{\ensuremath{string}}                   % HTR field
\newcommand{\hatcurCCraxxxxxB}{\ensuremath{20^{\mathrm h}52^{\mathrm m}51.60{\mathrm s}}}                            % Right Ascension
\newcommand{\hatcurCCdecxxxxxB}{\ensuremath{-25{\arcdeg}41{\arcmin}14.4{\arcsec}}}                           % Declination
\newcommand{\hatcurCCmagxxxxxB}{13.790}                                % apparent V-band magnitude
\newcommand{\hatcurCCtwomassxxxxxB}{2MASS~20525171-2541144}            % 2MASS identifier
\newcommand{\hatcurCCgscxxxxxB}{GSC~6926-00259}                        % GSC(1.2) identifier
\newcommand{\hatcurCCtassmvxxxxxB}{\ensuremath{13.79\pm0.10}}          % APASS V-band magnitude
\newcommand{\hatcurCCtassmvshortxxxxxB}{\ensuremath{13.8}}             % APASS V-band magnitude
\newcommand{\hatcurCCtassmBxxxxxB}{\ensuremath{14.62\pm0.10}}          % APASS B-band magnitude
\newcommand{\hatcurCCtassmBshortxxxxxB}{\ensuremath{14.6}}             % APASS B-band magnitude
\newcommand{\hatcurCCtassmIxxxxxB}{\ensuremath{nff\pmnff}}             % TASS I-band magnitude
\newcommand{\hatcurCCtassmIshortxxxxxB}{\ensuremath{0.0}}              % TASS I-band magnitude
\newcommand{\hatcurCCtassmgxxxxxB}{\ensuremath{nff\pmnff}}             % APASS g-band magnitude
\newcommand{\hatcurCCtassmgshortxxxxxB}{\ensuremath{0.0}}              % APASS g-band magnitude
\newcommand{\hatcurCCtassmrxxxxxB}{\ensuremath{nff\pmnff}}             % APASS r-band magnitude
\newcommand{\hatcurCCtassmrshortxxxxxB}{\ensuremath{0.0}}              % APASS r-band magnitude
\newcommand{\hatcurCCtassmixxxxxB}{\ensuremath{nff\pmnff}}             % APASS i-band magnitude
\newcommand{\hatcurCCtassmishortxxxxxB}{\ensuremath{0.0}}              % APASS i-band magnitude
\newcommand{\hatcurCCtwomassJmagxxxxxB}{\ensuremath{12.518\pm0.026}}   % 2MASS ORIG MAG
\newcommand{\hatcurCCtwomassHmagxxxxxB}{\ensuremath{12.129\pm0.023}}   % 2MASS ORIG MAG
\newcommand{\hatcurCCtwomassKmagxxxxxB}{\ensuremath{12.037\pm0.019}}   % 2MASS ORIG MAG
\newcommand{\hatcurCCcitJmagxxxxxB}{\ensuremath{12.529\pm0.026}}       % 2MASS CIT MAG
\newcommand{\hatcurCCcitHmagxxxxxB}{\ensuremath{12.123\pm0.024}}       % 2MASS CIT MAG
\newcommand{\hatcurCCcitKmagxxxxxB}{\ensuremath{12.061\pm0.019}}       % 2MASS CIT MAG
\newcommand{\hatcurCCbbJmagxxxxxB}{\ensuremath{12.587\pm0.028}}        % 2MASS BB MAG
\newcommand{\hatcurCCbbHmagxxxxxB}{\ensuremath{12.146\pm0.024}}        % 2MASS BB MAG
\newcommand{\hatcurCCbbKmagxxxxxB}{\ensuremath{12.081\pm0.019}}        % 2MASS BB MAG
\newcommand{\hatcurCCesoJmagxxxxxB}{\ensuremath{12.591\pm0.030}}       % 2MASS ESO MAG
\newcommand{\hatcurCCesoHmagxxxxxB}{\ensuremath{12.141\pm0.028}}       % 2MASS ESO MAG
\newcommand{\hatcurCCesoKmagxxxxxB}{\ensuremath{12.079\pm0.020}}       % 2MASS ESO MAG
\newcommand{\hatcurCCesoJHmagxxxxxB}{\ensuremath{0.449\pm0.039}}       % 2MASS ESO JH COLOR
\newcommand{\hatcurCCesoJKmagxxxxxB}{\ensuremath{0.512\pm0.035}}       % 2MASS ESO JK COLOR
\newcommand{\hatcurCCesoHKmagxxxxxB}{\ensuremath{0.061\pm0.034}}       % 2MASS ESO HK COLOR
\newcommand{\hatcurLCdipxxxxxB}{\ensuremath{16.2}}                     % BLS detected dip (mmag)
\newcommand{\hatcurLCrprstarxxxxxB}{\ensuremath{0.1145\pm0.0012}}      % Rp/R*
\newcommand{\hatcurLCbsqxxxxxB}{\ensuremath{0.032_{-0.026}^{+0.058}}}  % impact parameter square
\newcommand{\hatcurLCimpxxxxxB}{\ensuremath{0.18_{-0.10}^{+0.12}}}     % impact parameter
\newcommand{\hatcurLCzetaxxxxxB}{\ensuremath{20.35\pm0.12}}            % zeta/R*
\newcommand{\hatcurLCdurxxxxxB}{\ensuremath{0.11009\pm0.00078}}        % transit duration (days)
\newcommand{\hatcurLCdurshortxxxxxB}{\ensuremath{0.1101}}              % transit duration (days)
\newcommand{\hatcurLCdurhrxxxxxB}{\ensuremath{2.642\pm0.019}}          % transit duration (hours)
\newcommand{\hatcurLCdurhrshortxxxxxB}{\ensuremath{2.642}}             % transit duration (hours)
\newcommand{\hatcurLCqxxxxxB}{\ensuremath{0.03980\pm0.00028}}          % fractional transit duration (days)
\newcommand{\hatcurLCqshortxxxxxB}{\ensuremath{0.040}}                 % fractional transit duration (days)
\newcommand{\hatcurLCingdurxxxxxB}{\ensuremath{0.01168\pm0.00061}}     % ingress/egress duration (days)
\newcommand{\hatcurLCPxxxxxB}{\ensuremath{2.7667641\pm0.0000027}}      % period (days)
\newcommand{\hatcurLCPprecxxxxxB}{\ensuremath{2.7667641}}              % period (days)
\newcommand{\hatcurLCPshortxxxxxB}{\ensuremath{2.7668}}                % period (days)
\newcommand{\hatcurLCTxxxxxB}{\ensuremath{2456408.76462\pm0.00021}}    % epoch (BJD)
\newcommand{\hatcurLCTAxxxxxB}{\ensuremath{2455091.7849\pm0.0012}}     % TA (BJD)
\newcommand{\hatcurLCTBxxxxxB}{\ensuremath{2456455.79960\pm0.00023}}   % TB (BJD)
\newcommand{\hatcurLChatnetmxxxxxB}{\ensuremath{13.776330\pm0.000088}} % HATNet OOT level
\newcommand{\hatcurLCiblendxxxxxB}{\ensuremath{0.929\pm0.035}}         % HATNet iblend factor
\newcommand{\hatcurSMEiteffxxxxxB}{\ensuremath{5361\pm70}}             % Ini SME, stellar effective temperature
\newcommand{\hatcurSMEizfehxxxxxB}{\ensuremath{0.340\pm0.070}}         % Ini SME, stellar metallicity
\newcommand{\hatcurSMEizfehshortxxxxxB}{\ensuremath{0.34}}             % Ini SME, stellar metallicity
\newcommand{\hatcurSMEiloggxxxxxB}{\ensuremath{4.57\pm0.13}}           % Ini SME, stellar surface gravity
\newcommand{\hatcurSMEivsinxxxxxB}{\ensuremath{3.73\pm0.50}}           % Ini SME, stellar rotational velocity
\newcommand{\hatcurSMEivmacxxxxxB}{\ensuremath{0.0}}                   % Ini SME, stellar macroturbulence
\newcommand{\hatcurSMEivmicxxxxxB}{\ensuremath{0.0}}                   % Ini SME, stellar microturbulence
\newcommand{\hatcurSMEiiteffxxxxxB}{\ensuremath{5346\pm60}}            % Final SME, stellar effective temperature
\newcommand{\hatcurSMEiizfehxxxxxB}{\ensuremath{0.330\pm0.060}}        % Final SME, stellar metallicity
\newcommand{\hatcurSMEiizfehshortxxxxxB}{\ensuremath{0.33}}            % Final SME, stellar metallicity
\newcommand{\hatcurSMEiiloggxxxxxB}{\ensuremath{4.490\pm0.030}}        % Final SME, stellar surface gravity
\newcommand{\hatcurSMEiivsinxxxxxB}{\ensuremath{3.8\pm1.2}}            % Final SME, stellar rotational velocity
\newcommand{\hatcurLBizxxxxxB}{\ensuremath{0.2739}}                    % Limb darkening parameters, Gamma1, z-band
\newcommand{\hatcurLBiizxxxxxB}{\ensuremath{0.3035}}                   % Limb darkening parameters, Gamma2, z-band
\newcommand{\hatcurLBiixxxxxB}{\ensuremath{0.3562}}                    % Limb darkening parameters, Gamma1, i-band
\newcommand{\hatcurLBiiixxxxxB}{\ensuremath{0.2871}}                   % Limb darkening parameters, Gamma2, i-band
\newcommand{\hatcurLBiIxxxxxB}{\ensuremath{0.3293}}                    % Limb darkening parameters, Gamma1, I-band
\newcommand{\hatcurLBiiIxxxxxB}{\ensuremath{0.2925}}                   % Limb darkening parameters, Gamma2, I-band
\newcommand{\hatcurLBigxxxxxB}{\ensuremath{0.7052}}                    % Limb darkening parameters, Gamma1, g-band
\newcommand{\hatcurLBiigxxxxxB}{\ensuremath{0.1193}}                   % Limb darkening parameters, Gamma2, g-band
\newcommand{\hatcurLBirxxxxxB}{\ensuremath{0.4725}}                    % Limb darkening parameters, Gamma1, r-band
\newcommand{\hatcurLBiirxxxxxB}{\ensuremath{0.2569}}                   % Limb darkening parameters, Gamma2, r-band
\newcommand{\hatcurLBiRxxxxxB}{\ensuremath{0.4404}}                    % Limb darkening parameters, Gamma1, R-band
\newcommand{\hatcurLBiiRxxxxxB}{\ensuremath{0.2661}}                   % Limb darkening parameters, Gamma2, R-band
\newcommand{\hatcurLBikepxxxxxB}{\ensuremath{0.1000}}                  % Limb darkening parameters, Gamma1, Kep-band
\newcommand{\hatcurLBiikepxxxxxB}{\ensuremath{0.1000}}                 % Limb darkening parameters, Gamma2, Kep-band
\newcommand{\hatcurISOmxxxxxB}{\ensuremath{0.967\pm0.024}}             % stellar mass
\newcommand{\hatcurISOmshortxxxxxB}{\ensuremath{0.97}}                 % stellar mass
\newcommand{\hatcurISOmlongxxxxxB}{\ensuremath{0.967\pm0.024}}         % stellar mass
\newcommand{\hatcurISOrxxxxxB}{\ensuremath{0.933_{-0.015}^{+0.023}}}   % stellar radius
\newcommand{\hatcurISOrshortxxxxxB}{\ensuremath{0.93}}                 % stellar radius
\newcommand{\hatcurISOrlongxxxxxB}{\ensuremath{0.933_{-0.015}^{+0.023}}} % stellar radius
\newcommand{\hatcurISOrhoxxxxxB}{\ensuremath{1.682_{-0.126}^{+0.071}}} % stellar density (cgs)
\newcommand{\hatcurISOrholongxxxxxB}{\ensuremath{1.682_{-0.126}^{+0.071}}} % stellar density (cgs)
\newcommand{\hatcurISOloggxxxxxB}{\ensuremath{4.484\pm0.020}}          % stellar surface gravity from isochrones
\newcommand{\hatcurISOlumxxxxxB}{\ensuremath{0.640\pm0.047}}           % stellar luminosity
\newcommand{\hatcurISOlumshortxxxxxB}{\ensuremath{0.64}}               % stellar luminosity
\newcommand{\hatcurISOmvxxxxxB}{\ensuremath{5.399\pm0.091}}            % stellar absolute magnitude
\newcommand{\hatcurISOvixxxxxB}{\ensuremath{0.834\pm0.016}}            % stellar V-I index
\newcommand{\hatcurISOagexxxxxB}{\ensuremath{4.9\pm1.7}}               % stellar age
\newcommand{\hatcurISOsigmaxxxxxB}{\ensuremath{0.00170\pm0.00014}}     % system mass-correction sigma parameter
\newcommand{\hatcurISOMJxxxxxB}{\ensuremath{4.013\pm0.063}}            % stellar absolute J magnitude
\newcommand{\hatcurISOMHxxxxxB}{\ensuremath{3.594\pm0.054}}            % stellar absolute H magnitude
\newcommand{\hatcurISOMKxxxxxB}{\ensuremath{3.525\pm0.053}}            % stellar absolute K magnitude
\newcommand{\hatcurISOJKxxxxxB}{\ensuremath{0.490\pm0.020}}            % J-K color index from isochrones.
\newcommand{\hatcurISOspecxxxxxB}{G}                                   % stellar spectral type
\newcommand{\hatcurRVKxxxxxB}{\ensuremath{158\pm10}}                   % RV semi-amplitude [m/s]
\newcommand{\hatcurRVrkxxxxxB}{\ensuremath{0\pm0}}                     % sqrt(e)*cos(omega)
\newcommand{\hatcurRVrhxxxxxB}{\ensuremath{0\pm0}}                     % sqrt(e)*sin(omega)
\newcommand{\hatcurRVkxxxxxB}{\ensuremath{0\pm0}}                      % e*cos(omega)
\newcommand{\hatcurRVhxxxxxB}{\ensuremath{0\pm0}}                      % e*sin(omega)
\newcommand{\hatcurRVtronexxxxxB}{\ensuremath{0\pm0}}                  % RV linear trend tr1 factor
\newcommand{\hatcurRVtrtwoxxxxxB}{\ensuremath{0\pm0}}                  % RV linear trend tr2 factor
\newcommand{\hatcurRVgammaAxxxxxB}{\ensuremath{30196\pm18}}            % RV gamma velocity, relative scale
\newcommand{\hatcurRVjitterAxxxxxB}{\ensuremath{0.0\pm1.7}}            % RV jitter (m/s)
\newcommand{\hatcurRVfitrmsAxxxxxB}{\ensuremath{0.0}}                  % RVfitrms
\newcommand{\hatcurRVgammaBxxxxxB}{\ensuremath{30188.0\pm9.2}}         % RV gamma velocity, relative scale
\newcommand{\hatcurRVjitterBxxxxxB}{\ensuremath{0.00\pm0.61}}          % RV jitter (m/s)
\newcommand{\hatcurRVfitrmsBxxxxxB}{\ensuremath{0.0}}                  % RVfitrms
\newcommand{\hatcurRVeccenxxxxxB}{\ensuremath{0\pm0}}                  % eccentricity
\newcommand{\hatcurRVeccentwosiglimxxxxxB}{\ensuremath{<0.000}}        % eccentricity
\newcommand{\hatcurRVomegaxxxxxB}{\ensuremath{0\pm0}}                  % argument of pericenter
\newcommand{\hatcurPPixxxxxB}{\ensuremath{88.83\pm0.66}}               % orbital inclination
\newcommand{\hatcurPPgxxxxxB}{\ensuremath{24.8_{-2.5}^{+1.6}}}         % planetary surface gravity (m/s^2)
\newcommand{\hatcurPPloggxxxxxB}{\ensuremath{3.394_{-0.046}^{+0.026}}} % planetary surface gravity (log cgs)
\newcommand{\hatcurPParxxxxxB}{\ensuremath{8.80_{-0.22}^{+0.12}}}      % relative orbital radius (a/R*)
\newcommand{\hatcurPParelxxxxxB}{\ensuremath{0.03815\pm0.00032}}       % semimajor axis (AU)
\newcommand{\hatcurPPrhoxxxxxB}{\ensuremath{1.191_{-0.140}^{+0.098}}}  % planetary density (cgs)
\newcommand{\hatcurPPmxxxxxB}{\ensuremath{1.071\pm0.070}}              % planetary mass (M_jup)
\newcommand{\hatcurPPmshortxxxxxB}{\ensuremath{1.07}}                  % planetary mass (M_jup)
\newcommand{\hatcurPPmlongxxxxxB}{\ensuremath{1.071\pm0.070}}          % planetary mass (M_jup)
\newcommand{\hatcurPPmexxxxxB}{\ensuremath{340\pm22}}                  % planetary mass (M_earth)
\newcommand{\hatcurPPmeshortxxxxxB}{\ensuremath{340.3}}                % planetary mass (M_earth)
\newcommand{\hatcurPPmelongxxxxxB}{\ensuremath{340\pm22}}              % planetary mass (M_earth)
\newcommand{\hatcurPPrxxxxxB}{\ensuremath{1.039_{-0.022}^{+0.032}}}    % planetary radius (R_jup)
\newcommand{\hatcurPPrshortxxxxxB}{\ensuremath{1.04}}                  % planetary radius (R_jup)
\newcommand{\hatcurPPrlongxxxxxB}{\ensuremath{1.039_{-0.022}^{+0.032}}} % planetary radius (R_jup)
\newcommand{\hatcurPPrexxxxxB}{\ensuremath{11.65_{-0.25}^{+0.36}}}     % planetary radius (R_earth)
\newcommand{\hatcurPPreshortxxxxxB}{\ensuremath{11.6}}                 % planetary radius (R_earth)
\newcommand{\hatcurPPrelongxxxxxB}{\ensuremath{11.65_{-0.25}^{+0.36}}} % planetary radius (R_earth)
\newcommand{\hatcurPPmrcorrxxxxxB}{\ensuremath{-0.08}}                 % mass/radius correlation
\newcommand{\hatcurPPteffxxxxxB}{\ensuremath{1276\pm20}}               % planetary temperature (K)
\newcommand{\hatcurPPthetaxxxxxB}{\ensuremath{0.0814\pm0.0058}}        % Safranov number
\newcommand{\hatcurPPfluxperixxxxxB}{\ensuremath{6.00\pm0.38}}         % flux @ periastron (CGS)
\newcommand{\hatcurPPfluxperidimxxxxxB}{\ensuremath{8}}                % flux @ periastron (CGS) units.
\newcommand{\hatcurPPfluxapxxxxxB}{\ensuremath{6.00\pm0.38}}           % flux @ apastron (CGS)
\newcommand{\hatcurPPfluxapdimxxxxxB}{\ensuremath{8}}                  % flux @ apastron (CGS) units.
\newcommand{\hatcurPPfluxavgxxxxxB}{\ensuremath{6.00\pm0.38}}          % flux on average (CGS)
\newcommand{\hatcurPPfluxavgdimxxxxxB}{\ensuremath{8}}                 % flux average (CGS) units.
\newcommand{\hatcurPPfluxavglogxxxxxB}{\ensuremath{8.778\pm0.027}}     % log10 flux on average (CGS)
\newcommand{\hatcurXsecphasexxxxxB}{\ensuremath{0\pm0}}                % Phase of secondary eclipse
\newcommand{\hatcurXsecondaryxxxxxB}{\ensuremath{2456410.14800\pm0.00021}} % Secondary eclipse epoch
\newcommand{\hatcurXsecdurxxxxxB}{\ensuremath{0.11009\pm0.00078}}      % sec eclipse duration (days)
\newcommand{\hatcurXsecingdurxxxxxB}{\ensuremath{0.01168\pm0.00061}}   % sec I/E duration (days)
\newcommand{\hatcurPPphiconjxxxxxB}{\ensuremath{0\pm0}}                % phase diff between conjunction and periastron
\newcommand{\hatcurPPperixxxxxB}{\ensuremath{2456408.07292\pm0.00021}} % time of periastron passage.
\newcommand{\hatcurPPaequivxxxxxB}{\ensuremath{0.0477\pm0.0015}}       % equivalent semi-major axis
\newcommand{\hatcurPPtcircxxxxxB}{\ensuremath{219_{-36}^{+27}}}        % circularization timescale
\newcommand{\hatcurPPtinfallxxxxxB}{\ensuremath{1240_{-150}^{+110}}}   % infall timescale
\newcommand{\hatcurXdistxxxxxB}{\ensuremath{514\pm14}}                 % distance (pc), no reddenning correction
\newcommand{\hatcurXAvxxxxxB}{\ensuremath{0.000\pm0.033}}              % Av (mag)
\newcommand{\hatcurXdistredxxxxxB}{\ensuremath{513\pm14}}              % distance with Av correction (pc)
\newcommand{\hatcurXEBVxxxxxB}{\ensuremath{0.000\pm0.011}}             % E(B-V) (mag)
\newcommand{\hatcurXmvisoredxxxxxB}{\ensuremath{13.964\pm0.047}}       % Expected m_v with reddening (mag)
\newcommand{\hatcurXmiisoredxxxxxB}{\ensuremath{13.126\pm0.032}}       % Expected m_i with reddening (mag)
\newcommand{\hatcurXmjisoredxxxxxB}{\ensuremath{12.570\pm0.017}}       % Expected m_j with reddening (mag)
\newcommand{\hatcurXmhisoredxxxxxB}{\ensuremath{12.148\pm0.014}}       % Expected m_h with reddening (mag)
\newcommand{\hatcurXmkisoredxxxxxB}{\ensuremath{12.078\pm0.015}}       % Expected m_k with reddening (mag)
\newcommand{\hatcurXviisoredxxxxxB}{\ensuremath{0.837\pm0.019}}        % Expected V-I with reddening (mag)
\newcommand{\hatcurXvkisoredxxxxxB}{\ensuremath{1.887\pm0.049}}        % Expected V-K with reddening (mag)
\newcommand{\hatcurXjhisoredxxxxxB}{\ensuremath{0.422\pm0.012}}        % Expected J-H with reddening (mag)
\newcommand{\hatcurXjkisoredxxxxxB}{\ensuremath{0.492\pm0.014}}        % Expected J-K with reddening (mag)
\newcommand{\hatcurCCpmraxxxxxB}{\ensuremath{0.4\pm1.5}}               % proper motion, in RA
\newcommand{\hatcurCCpmdecxxxxxB}{\ensuremath{-8.8\pm1.3}}             % proper motion, in DEC
\newcommand{\hatcurCCpmxxxxxB}{\ensuremath{8.8\pm2.0}}                 % proper motion

\newcommand{\hatcurANUWIFESgamma}[1]{\ifnum#1=13 %
\hatcurANUWIFESgammaxxxxxA
\else
??????\fi
}
\newcommand{\hatcurANUWIFESlogg}[1]{\ifnum#1=13 %
\hatcurANUWIFESloggxxxxxA
\else
??????\fi
}
\newcommand{\hatcurANUWIFESnumspec}[1]{\ifnum#1=13 %
\hatcurANUWIFESnumspecxxxxxA
\else
??????\fi
}
\newcommand{\hatcurANUWIFESrvrms}[1]{\ifnum#1=13 %
\hatcurANUWIFESrvrmsxxxxxA
\else
??????\fi
}
\newcommand{\hatcurANUWIFESspan}[1]{\ifnum#1=13 %
\hatcurANUWIFESspanxxxxxA
\else
??????\fi
}
\newcommand{\hatcurANUWIFESteff}[1]{\ifnum#1=13 %
\hatcurANUWIFESteffxxxxxA
\else
??????\fi
}
\newcommand{\hatcurANUWIFESvsini}[1]{\ifnum#1=13 %
\hatcurANUWIFESvsinixxxxxA
\else
??????\fi
}
\newcommand{\hatcurANUWIFESzfeh}[1]{\ifnum#1=13 %
\hatcurANUWIFESzfehxxxxxA
\else
??????\fi
}
\newcommand{\hatcurCCbbHmag}[1]{\ifnum#1=13 %
\hatcurCCbbHmagxxxxxA
\else
\ifnum#1=14 %
\hatcurCCbbHmagxxxxxB
\else
??????\fi
\fi
}
\newcommand{\hatcurCCbbJmag}[1]{\ifnum#1=13 %
\hatcurCCbbJmagxxxxxA
\else
\ifnum#1=14 %
\hatcurCCbbJmagxxxxxB
\else
??????\fi
\fi
}
\newcommand{\hatcurCCbbKmag}[1]{\ifnum#1=13 %
\hatcurCCbbKmagxxxxxA
\else
\ifnum#1=14 %
\hatcurCCbbKmagxxxxxB
\else
??????\fi
\fi
}
\newcommand{\hatcurCCcitHmag}[1]{\ifnum#1=13 %
\hatcurCCcitHmagxxxxxA
\else
\ifnum#1=14 %
\hatcurCCcitHmagxxxxxB
\else
??????\fi
\fi
}
\newcommand{\hatcurCCcitJmag}[1]{\ifnum#1=13 %
\hatcurCCcitJmagxxxxxA
\else
\ifnum#1=14 %
\hatcurCCcitJmagxxxxxB
\else
??????\fi
\fi
}
\newcommand{\hatcurCCcitKmag}[1]{\ifnum#1=13 %
\hatcurCCcitKmagxxxxxA
\else
\ifnum#1=14 %
\hatcurCCcitKmagxxxxxB
\else
??????\fi
\fi
}
\newcommand{\hatcurCCdec}[1]{\ifnum#1=13 %
\hatcurCCdecxxxxxA
\else
\ifnum#1=14 %
\hatcurCCdecxxxxxB
\else
??????\fi
\fi
}
\newcommand{\hatcurCCesoHKmag}[1]{\ifnum#1=13 %
\hatcurCCesoHKmagxxxxxA
\else
\ifnum#1=14 %
\hatcurCCesoHKmagxxxxxB
\else
??????\fi
\fi
}
\newcommand{\hatcurCCesoHmag}[1]{\ifnum#1=13 %
\hatcurCCesoHmagxxxxxA
\else
\ifnum#1=14 %
\hatcurCCesoHmagxxxxxB
\else
??????\fi
\fi
}
\newcommand{\hatcurCCesoJHmag}[1]{\ifnum#1=13 %
\hatcurCCesoJHmagxxxxxA
\else
\ifnum#1=14 %
\hatcurCCesoJHmagxxxxxB
\else
??????\fi
\fi
}
\newcommand{\hatcurCCesoJKmag}[1]{\ifnum#1=13 %
\hatcurCCesoJKmagxxxxxA
\else
\ifnum#1=14 %
\hatcurCCesoJKmagxxxxxB
\else
??????\fi
\fi
}
\newcommand{\hatcurCCesoJmag}[1]{\ifnum#1=13 %
\hatcurCCesoJmagxxxxxA
\else
\ifnum#1=14 %
\hatcurCCesoJmagxxxxxB
\else
??????\fi
\fi
}
\newcommand{\hatcurCCesoKmag}[1]{\ifnum#1=13 %
\hatcurCCesoKmagxxxxxA
\else
\ifnum#1=14 %
\hatcurCCesoKmagxxxxxB
\else
??????\fi
\fi
}
\newcommand{\hatcurCCgsc}[1]{\ifnum#1=13 %
\hatcurCCgscxxxxxA
\else
\ifnum#1=14 %
\hatcurCCgscxxxxxB
\else
??????\fi
\fi
}
\newcommand{\hatcurCCmag}[1]{\ifnum#1=13 %
\hatcurCCmagxxxxxA
\else
\ifnum#1=14 %
\hatcurCCmagxxxxxB
\else
??????\fi
\fi
}
\newcommand{\hatcurCCpm}[1]{\ifnum#1=13 %
\hatcurCCpmxxxxxA
\else
\ifnum#1=14 %
\hatcurCCpmxxxxxB
\else
??????\fi
\fi
}
\newcommand{\hatcurCCpmdec}[1]{\ifnum#1=13 %
\hatcurCCpmdecxxxxxA
\else
\ifnum#1=14 %
\hatcurCCpmdecxxxxxB
\else
??????\fi
\fi
}
\newcommand{\hatcurCCpmra}[1]{\ifnum#1=13 %
\hatcurCCpmraxxxxxA
\else
\ifnum#1=14 %
\hatcurCCpmraxxxxxB
\else
??????\fi
\fi
}
\newcommand{\hatcurCCra}[1]{\ifnum#1=13 %
\hatcurCCraxxxxxA
\else
\ifnum#1=14 %
\hatcurCCraxxxxxB
\else
??????\fi
\fi
}
\newcommand{\hatcurCCtassmB}[1]{\ifnum#1=13 %
\hatcurCCtassmBxxxxxA
\else
\ifnum#1=14 %
\hatcurCCtassmBxxxxxB
\else
??????\fi
\fi
}
\newcommand{\hatcurCCtassmBshort}[1]{\ifnum#1=13 %
\hatcurCCtassmBshortxxxxxA
\else
\ifnum#1=14 %
\hatcurCCtassmBshortxxxxxB
\else
??????\fi
\fi
}
\newcommand{\hatcurCCtassmg}[1]{\ifnum#1=13 %
\hatcurCCtassmgxxxxxA
\else
\ifnum#1=14 %
\hatcurCCtassmgxxxxxB
\else
??????\fi
\fi
}
\newcommand{\hatcurCCtassmgshort}[1]{\ifnum#1=13 %
\hatcurCCtassmgshortxxxxxA
\else
\ifnum#1=14 %
\hatcurCCtassmgshortxxxxxB
\else
??????\fi
\fi
}
\newcommand{\hatcurCCtassmi}[1]{\ifnum#1=13 %
\hatcurCCtassmixxxxxA
\else
\ifnum#1=14 %
\hatcurCCtassmixxxxxB
\else
??????\fi
\fi
}
\newcommand{\hatcurCCtassmI}[1]{\ifnum#1=13 %
\hatcurCCtassmIxxxxxA
\else
\ifnum#1=14 %
\hatcurCCtassmIxxxxxB
\else
??????\fi
\fi
}
\newcommand{\hatcurCCtassmishort}[1]{\ifnum#1=13 %
\hatcurCCtassmishortxxxxxA
\else
\ifnum#1=14 %
\hatcurCCtassmishortxxxxxB
\else
??????\fi
\fi
}
\newcommand{\hatcurCCtassmIshort}[1]{\ifnum#1=13 %
\hatcurCCtassmIshortxxxxxA
\else
\ifnum#1=14 %
\hatcurCCtassmIshortxxxxxB
\else
??????\fi
\fi
}
\newcommand{\hatcurCCtassmr}[1]{\ifnum#1=13 %
\hatcurCCtassmrxxxxxA
\else
\ifnum#1=14 %
\hatcurCCtassmrxxxxxB
\else
??????\fi
\fi
}
\newcommand{\hatcurCCtassmrshort}[1]{\ifnum#1=13 %
\hatcurCCtassmrshortxxxxxA
\else
\ifnum#1=14 %
\hatcurCCtassmrshortxxxxxB
\else
??????\fi
\fi
}
\newcommand{\hatcurCCtassmv}[1]{\ifnum#1=13 %
\hatcurCCtassmvxxxxxA
\else
\ifnum#1=14 %
\hatcurCCtassmvxxxxxB
\else
??????\fi
\fi
}
\newcommand{\hatcurCCtassmvshort}[1]{\ifnum#1=13 %
\hatcurCCtassmvshortxxxxxA
\else
\ifnum#1=14 %
\hatcurCCtassmvshortxxxxxB
\else
??????\fi
\fi
}
\newcommand{\hatcurCCtwomass}[1]{\ifnum#1=13 %
\hatcurCCtwomassxxxxxA
\else
\ifnum#1=14 %
\hatcurCCtwomassxxxxxB
\else
??????\fi
\fi
}
\newcommand{\hatcurCCtwomassHmag}[1]{\ifnum#1=13 %
\hatcurCCtwomassHmagxxxxxA
\else
\ifnum#1=14 %
\hatcurCCtwomassHmagxxxxxB
\else
??????\fi
\fi
}
\newcommand{\hatcurCCtwomassJmag}[1]{\ifnum#1=13 %
\hatcurCCtwomassJmagxxxxxA
\else
\ifnum#1=14 %
\hatcurCCtwomassJmagxxxxxB
\else
??????\fi
\fi
}
\newcommand{\hatcurCCtwomassKmag}[1]{\ifnum#1=13 %
\hatcurCCtwomassKmagxxxxxA
\else
\ifnum#1=14 %
\hatcurCCtwomassKmagxxxxxB
\else
??????\fi
\fi
}
\newcommand{\hatcurfield}[1]{\ifnum#1=13 %
\hatcurfieldxxxxxA
\else
\ifnum#1=14 %
\hatcurfieldxxxxxB
\else
??????\fi
\fi
}
\newcommand{\hatcurhtr}[1]{\ifnum#1=13 %
\hatcurhtrxxxxxA
\else
\ifnum#1=14 %
\hatcurhtrxxxxxB
\else
??????\fi
\fi
}
\newcommand{\hatcurISOage}[1]{\ifnum#1=13 %
\hatcurISOagexxxxxA
\else
\ifnum#1=14 %
\hatcurISOagexxxxxB
\else
??????\fi
\fi
}
\newcommand{\hatcurISOJK}[1]{\ifnum#1=13 %
\hatcurISOJKxxxxxA
\else
\ifnum#1=14 %
\hatcurISOJKxxxxxB
\else
??????\fi
\fi
}
\newcommand{\hatcurISOlogg}[1]{\ifnum#1=13 %
\hatcurISOloggxxxxxA
\else
\ifnum#1=14 %
\hatcurISOloggxxxxxB
\else
??????\fi
\fi
}
\newcommand{\hatcurISOlum}[1]{\ifnum#1=13 %
\hatcurISOlumxxxxxA
\else
\ifnum#1=14 %
\hatcurISOlumxxxxxB
\else
??????\fi
\fi
}
\newcommand{\hatcurISOlumshort}[1]{\ifnum#1=13 %
\hatcurISOlumshortxxxxxA
\else
\ifnum#1=14 %
\hatcurISOlumshortxxxxxB
\else
??????\fi
\fi
}
\newcommand{\hatcurISOm}[1]{\ifnum#1=13 %
\hatcurISOmxxxxxA
\else
\ifnum#1=14 %
\hatcurISOmxxxxxB
\else
??????\fi
\fi
}
\newcommand{\hatcurISOMH}[1]{\ifnum#1=13 %
\hatcurISOMHxxxxxA
\else
\ifnum#1=14 %
\hatcurISOMHxxxxxB
\else
??????\fi
\fi
}
\newcommand{\hatcurISOMJ}[1]{\ifnum#1=13 %
\hatcurISOMJxxxxxA
\else
\ifnum#1=14 %
\hatcurISOMJxxxxxB
\else
??????\fi
\fi
}
\newcommand{\hatcurISOMK}[1]{\ifnum#1=13 %
\hatcurISOMKxxxxxA
\else
\ifnum#1=14 %
\hatcurISOMKxxxxxB
\else
??????\fi
\fi
}
\newcommand{\hatcurISOmlong}[1]{\ifnum#1=13 %
\hatcurISOmlongxxxxxA
\else
\ifnum#1=14 %
\hatcurISOmlongxxxxxB
\else
??????\fi
\fi
}
\newcommand{\hatcurISOmshort}[1]{\ifnum#1=13 %
\hatcurISOmshortxxxxxA
\else
\ifnum#1=14 %
\hatcurISOmshortxxxxxB
\else
??????\fi
\fi
}
\newcommand{\hatcurISOmv}[1]{\ifnum#1=13 %
\hatcurISOmvxxxxxA
\else
\ifnum#1=14 %
\hatcurISOmvxxxxxB
\else
??????\fi
\fi
}
\newcommand{\hatcurISOr}[1]{\ifnum#1=13 %
\hatcurISOrxxxxxA
\else
\ifnum#1=14 %
\hatcurISOrxxxxxB
\else
??????\fi
\fi
}
\newcommand{\hatcurISOrho}[1]{\ifnum#1=13 %
\hatcurISOrhoxxxxxA
\else
\ifnum#1=14 %
\hatcurISOrhoxxxxxB
\else
??????\fi
\fi
}
\newcommand{\hatcurISOrholong}[1]{\ifnum#1=13 %
\hatcurISOrholongxxxxxA
\else
\ifnum#1=14 %
\hatcurISOrholongxxxxxB
\else
??????\fi
\fi
}
\newcommand{\hatcurISOrlong}[1]{\ifnum#1=13 %
\hatcurISOrlongxxxxxA
\else
\ifnum#1=14 %
\hatcurISOrlongxxxxxB
\else
??????\fi
\fi
}
\newcommand{\hatcurISOrshort}[1]{\ifnum#1=13 %
\hatcurISOrshortxxxxxA
\else
\ifnum#1=14 %
\hatcurISOrshortxxxxxB
\else
??????\fi
\fi
}
\newcommand{\hatcurISOsigma}[1]{\ifnum#1=13 %
\hatcurISOsigmaxxxxxA
\else
\ifnum#1=14 %
\hatcurISOsigmaxxxxxB
\else
??????\fi
\fi
}
\newcommand{\hatcurISOspec}[1]{\ifnum#1=13 %
\hatcurISOspecxxxxxA
\else
\ifnum#1=14 %
\hatcurISOspecxxxxxB
\else
??????\fi
\fi
}
\newcommand{\hatcurISOvi}[1]{\ifnum#1=13 %
\hatcurISOvixxxxxA
\else
\ifnum#1=14 %
\hatcurISOvixxxxxB
\else
??????\fi
\fi
}
\newcommand{\hatcurLBig}[1]{\ifnum#1=13 %
\hatcurLBigxxxxxA
\else
\ifnum#1=14 %
\hatcurLBigxxxxxB
\else
??????\fi
\fi
}
\newcommand{\hatcurLBii}[1]{\ifnum#1=13 %
\hatcurLBiixxxxxA
\else
\ifnum#1=14 %
\hatcurLBiixxxxxB
\else
??????\fi
\fi
}
\newcommand{\hatcurLBiI}[1]{\ifnum#1=13 %
\hatcurLBiIxxxxxA
\else
\ifnum#1=14 %
\hatcurLBiIxxxxxB
\else
??????\fi
\fi
}
\newcommand{\hatcurLBiig}[1]{\ifnum#1=13 %
\hatcurLBiigxxxxxA
\else
\ifnum#1=14 %
\hatcurLBiigxxxxxB
\else
??????\fi
\fi
}
\newcommand{\hatcurLBiii}[1]{\ifnum#1=13 %
\hatcurLBiiixxxxxA
\else
\ifnum#1=14 %
\hatcurLBiiixxxxxB
\else
??????\fi
\fi
}
\newcommand{\hatcurLBiiI}[1]{\ifnum#1=13 %
\hatcurLBiiIxxxxxA
\else
\ifnum#1=14 %
\hatcurLBiiIxxxxxB
\else
??????\fi
\fi
}
\newcommand{\hatcurLBiikep}[1]{\ifnum#1=14 %
\hatcurLBiikepxxxxxB
\else
??????\fi
}
\newcommand{\hatcurLBiir}[1]{\ifnum#1=13 %
\hatcurLBiirxxxxxA
\else
\ifnum#1=14 %
\hatcurLBiirxxxxxB
\else
??????\fi
\fi
}
\newcommand{\hatcurLBiiR}[1]{\ifnum#1=13 %
\hatcurLBiiRxxxxxA
\else
\ifnum#1=14 %
\hatcurLBiiRxxxxxB
\else
??????\fi
\fi
}
\newcommand{\hatcurLBiiz}[1]{\ifnum#1=13 %
\hatcurLBiizxxxxxA
\else
\ifnum#1=14 %
\hatcurLBiizxxxxxB
\else
??????\fi
\fi
}
\newcommand{\hatcurLBikep}[1]{\ifnum#1=14 %
\hatcurLBikepxxxxxB
\else
??????\fi
}
\newcommand{\hatcurLBir}[1]{\ifnum#1=13 %
\hatcurLBirxxxxxA
\else
\ifnum#1=14 %
\hatcurLBirxxxxxB
\else
??????\fi
\fi
}
\newcommand{\hatcurLBiR}[1]{\ifnum#1=13 %
\hatcurLBiRxxxxxA
\else
\ifnum#1=14 %
\hatcurLBiRxxxxxB
\else
??????\fi
\fi
}
\newcommand{\hatcurLBiz}[1]{\ifnum#1=13 %
\hatcurLBizxxxxxA
\else
\ifnum#1=14 %
\hatcurLBizxxxxxB
\else
??????\fi
\fi
}
\newcommand{\hatcurLCbsq}[1]{\ifnum#1=13 %
\hatcurLCbsqxxxxxA
\else
\ifnum#1=14 %
\hatcurLCbsqxxxxxB
\else
??????\fi
\fi
}
\newcommand{\hatcurLCdip}[1]{\ifnum#1=13 %
\hatcurLCdipxxxxxA
\else
\ifnum#1=14 %
\hatcurLCdipxxxxxB
\else
??????\fi
\fi
}
\newcommand{\hatcurLCdur}[1]{\ifnum#1=13 %
\hatcurLCdurxxxxxA
\else
\ifnum#1=14 %
\hatcurLCdurxxxxxB
\else
??????\fi
\fi
}
\newcommand{\hatcurLCdurhr}[1]{\ifnum#1=13 %
\hatcurLCdurhrxxxxxA
\else
\ifnum#1=14 %
\hatcurLCdurhrxxxxxB
\else
??????\fi
\fi
}
\newcommand{\hatcurLCdurhrshort}[1]{\ifnum#1=13 %
\hatcurLCdurhrshortxxxxxA
\else
\ifnum#1=14 %
\hatcurLCdurhrshortxxxxxB
\else
??????\fi
\fi
}
\newcommand{\hatcurLCdurshort}[1]{\ifnum#1=13 %
\hatcurLCdurshortxxxxxA
\else
\ifnum#1=14 %
\hatcurLCdurshortxxxxxB
\else
??????\fi
\fi
}
\newcommand{\hatcurLChatnetm}[1]{\ifnum#1=13 %
\hatcurLChatnetmxxxxxA
\else
\ifnum#1=14 %
\hatcurLChatnetmxxxxxB
\else
??????\fi
\fi
}
\newcommand{\hatcurLCiblend}[1]{\ifnum#1=13 %
\hatcurLCiblendxxxxxA
\else
\ifnum#1=14 %
\hatcurLCiblendxxxxxB
\else
??????\fi
\fi
}
\newcommand{\hatcurLCimp}[1]{\ifnum#1=13 %
\hatcurLCimpxxxxxA
\else
\ifnum#1=14 %
\hatcurLCimpxxxxxB
\else
??????\fi
\fi
}
\newcommand{\hatcurLCingdur}[1]{\ifnum#1=13 %
\hatcurLCingdurxxxxxA
\else
\ifnum#1=14 %
\hatcurLCingdurxxxxxB
\else
??????\fi
\fi
}
\newcommand{\hatcurLCP}[1]{\ifnum#1=13 %
\hatcurLCPxxxxxA
\else
\ifnum#1=14 %
\hatcurLCPxxxxxB
\else
??????\fi
\fi
}
\newcommand{\hatcurLCPprec}[1]{\ifnum#1=13 %
\hatcurLCPprecxxxxxA
\else
\ifnum#1=14 %
\hatcurLCPprecxxxxxB
\else
??????\fi
\fi
}
\newcommand{\hatcurLCPshort}[1]{\ifnum#1=13 %
\hatcurLCPshortxxxxxA
\else
\ifnum#1=14 %
\hatcurLCPshortxxxxxB
\else
??????\fi
\fi
}
\newcommand{\hatcurLCq}[1]{\ifnum#1=13 %
\hatcurLCqxxxxxA
\else
\ifnum#1=14 %
\hatcurLCqxxxxxB
\else
??????\fi
\fi
}
\newcommand{\hatcurLCqshort}[1]{\ifnum#1=13 %
\hatcurLCqshortxxxxxA
\else
\ifnum#1=14 %
\hatcurLCqshortxxxxxB
\else
??????\fi
\fi
}
\newcommand{\hatcurLCrprstar}[1]{\ifnum#1=13 %
\hatcurLCrprstarxxxxxA
\else
\ifnum#1=14 %
\hatcurLCrprstarxxxxxB
\else
??????\fi
\fi
}
\newcommand{\hatcurLCT}[1]{\ifnum#1=13 %
\hatcurLCTxxxxxA
\else
\ifnum#1=14 %
\hatcurLCTxxxxxB
\else
??????\fi
\fi
}
\newcommand{\hatcurLCTA}[1]{\ifnum#1=13 %
\hatcurLCTAxxxxxA
\else
\ifnum#1=14 %
\hatcurLCTAxxxxxB
\else
??????\fi
\fi
}
\newcommand{\hatcurLCTB}[1]{\ifnum#1=13 %
\hatcurLCTBxxxxxA
\else
\ifnum#1=14 %
\hatcurLCTBxxxxxB
\else
??????\fi
\fi
}
\newcommand{\hatcurLCzeta}[1]{\ifnum#1=13 %
\hatcurLCzetaxxxxxA
\else
\ifnum#1=14 %
\hatcurLCzetaxxxxxB
\else
??????\fi
\fi
}
\newcommand{\hatcurPPaequiv}[1]{\ifnum#1=13 %
\hatcurPPaequivxxxxxA
\else
\ifnum#1=14 %
\hatcurPPaequivxxxxxB
\else
??????\fi
\fi
}
\newcommand{\hatcurPPar}[1]{\ifnum#1=13 %
\hatcurPParxxxxxA
\else
\ifnum#1=14 %
\hatcurPParxxxxxB
\else
??????\fi
\fi
}
\newcommand{\hatcurPParel}[1]{\ifnum#1=13 %
\hatcurPParelxxxxxA
\else
\ifnum#1=14 %
\hatcurPParelxxxxxB
\else
??????\fi
\fi
}
\newcommand{\hatcurPPfluxap}[1]{\ifnum#1=13 %
\hatcurPPfluxapxxxxxA
\else
\ifnum#1=14 %
\hatcurPPfluxapxxxxxB
\else
??????\fi
\fi
}
\newcommand{\hatcurPPfluxapdim}[1]{\ifnum#1=13 %
\hatcurPPfluxapdimxxxxxA
\else
\ifnum#1=14 %
\hatcurPPfluxapdimxxxxxB
\else
??????\fi
\fi
}
\newcommand{\hatcurPPfluxavg}[1]{\ifnum#1=13 %
\hatcurPPfluxavgxxxxxA
\else
\ifnum#1=14 %
\hatcurPPfluxavgxxxxxB
\else
??????\fi
\fi
}
\newcommand{\hatcurPPfluxavgdim}[1]{\ifnum#1=13 %
\hatcurPPfluxavgdimxxxxxA
\else
\ifnum#1=14 %
\hatcurPPfluxavgdimxxxxxB
\else
??????\fi
\fi
}
\newcommand{\hatcurPPfluxavglog}[1]{\ifnum#1=13 %
\hatcurPPfluxavglogxxxxxA
\else
\ifnum#1=14 %
\hatcurPPfluxavglogxxxxxB
\else
??????\fi
\fi
}
\newcommand{\hatcurPPfluxperi}[1]{\ifnum#1=13 %
\hatcurPPfluxperixxxxxA
\else
\ifnum#1=14 %
\hatcurPPfluxperixxxxxB
\else
??????\fi
\fi
}
\newcommand{\hatcurPPfluxperidim}[1]{\ifnum#1=13 %
\hatcurPPfluxperidimxxxxxA
\else
\ifnum#1=14 %
\hatcurPPfluxperidimxxxxxB
\else
??????\fi
\fi
}
\newcommand{\hatcurPPg}[1]{\ifnum#1=13 %
\hatcurPPgxxxxxA
\else
\ifnum#1=14 %
\hatcurPPgxxxxxB
\else
??????\fi
\fi
}
\newcommand{\hatcurPPi}[1]{\ifnum#1=13 %
\hatcurPPixxxxxA
\else
\ifnum#1=14 %
\hatcurPPixxxxxB
\else
??????\fi
\fi
}
\newcommand{\hatcurPPlogg}[1]{\ifnum#1=13 %
\hatcurPPloggxxxxxA
\else
\ifnum#1=14 %
\hatcurPPloggxxxxxB
\else
??????\fi
\fi
}
\newcommand{\hatcurPPm}[1]{\ifnum#1=13 %
\hatcurPPmxxxxxA
\else
\ifnum#1=14 %
\hatcurPPmxxxxxB
\else
??????\fi
\fi
}
\newcommand{\hatcurPPme}[1]{\ifnum#1=13 %
\hatcurPPmexxxxxA
\else
\ifnum#1=14 %
\hatcurPPmexxxxxB
\else
??????\fi
\fi
}
\newcommand{\hatcurPPmelong}[1]{\ifnum#1=13 %
\hatcurPPmelongxxxxxA
\else
\ifnum#1=14 %
\hatcurPPmelongxxxxxB
\else
??????\fi
\fi
}
\newcommand{\hatcurPPmeshort}[1]{\ifnum#1=13 %
\hatcurPPmeshortxxxxxA
\else
\ifnum#1=14 %
\hatcurPPmeshortxxxxxB
\else
??????\fi
\fi
}
\newcommand{\hatcurPPmlong}[1]{\ifnum#1=13 %
\hatcurPPmlongxxxxxA
\else
\ifnum#1=14 %
\hatcurPPmlongxxxxxB
\else
??????\fi
\fi
}
\newcommand{\hatcurPPmrcorr}[1]{\ifnum#1=13 %
\hatcurPPmrcorrxxxxxA
\else
\ifnum#1=14 %
\hatcurPPmrcorrxxxxxB
\else
??????\fi
\fi
}
\newcommand{\hatcurPPmshort}[1]{\ifnum#1=13 %
\hatcurPPmshortxxxxxA
\else
\ifnum#1=14 %
\hatcurPPmshortxxxxxB
\else
??????\fi
\fi
}
\newcommand{\hatcurPPperi}[1]{\ifnum#1=13 %
\hatcurPPperixxxxxA
\else
\ifnum#1=14 %
\hatcurPPperixxxxxB
\else
??????\fi
\fi
}
\newcommand{\hatcurPPphiconj}[1]{\ifnum#1=13 %
\hatcurPPphiconjxxxxxA
\else
\ifnum#1=14 %
\hatcurPPphiconjxxxxxB
\else
??????\fi
\fi
}
\newcommand{\hatcurPPr}[1]{\ifnum#1=13 %
\hatcurPPrxxxxxA
\else
\ifnum#1=14 %
\hatcurPPrxxxxxB
\else
??????\fi
\fi
}
\newcommand{\hatcurPPre}[1]{\ifnum#1=13 %
\hatcurPPrexxxxxA
\else
\ifnum#1=14 %
\hatcurPPrexxxxxB
\else
??????\fi
\fi
}
\newcommand{\hatcurPPrelong}[1]{\ifnum#1=13 %
\hatcurPPrelongxxxxxA
\else
\ifnum#1=14 %
\hatcurPPrelongxxxxxB
\else
??????\fi
\fi
}
\newcommand{\hatcurPPreshort}[1]{\ifnum#1=13 %
\hatcurPPreshortxxxxxA
\else
\ifnum#1=14 %
\hatcurPPreshortxxxxxB
\else
??????\fi
\fi
}
\newcommand{\hatcurPPrho}[1]{\ifnum#1=13 %
\hatcurPPrhoxxxxxA
\else
\ifnum#1=14 %
\hatcurPPrhoxxxxxB
\else
??????\fi
\fi
}
\newcommand{\hatcurPPrlong}[1]{\ifnum#1=13 %
\hatcurPPrlongxxxxxA
\else
\ifnum#1=14 %
\hatcurPPrlongxxxxxB
\else
??????\fi
\fi
}
\newcommand{\hatcurPPrshort}[1]{\ifnum#1=13 %
\hatcurPPrshortxxxxxA
\else
\ifnum#1=14 %
\hatcurPPrshortxxxxxB
\else
??????\fi
\fi
}
\newcommand{\hatcurPPtcirc}[1]{\ifnum#1=13 %
\hatcurPPtcircxxxxxA
\else
\ifnum#1=14 %
\hatcurPPtcircxxxxxB
\else
??????\fi
\fi
}
\newcommand{\hatcurPPteff}[1]{\ifnum#1=13 %
\hatcurPPteffxxxxxA
\else
\ifnum#1=14 %
\hatcurPPteffxxxxxB
\else
??????\fi
\fi
}
\newcommand{\hatcurPPtheta}[1]{\ifnum#1=13 %
\hatcurPPthetaxxxxxA
\else
\ifnum#1=14 %
\hatcurPPthetaxxxxxB
\else
??????\fi
\fi
}
\newcommand{\hatcurPPtinfall}[1]{\ifnum#1=13 %
\hatcurPPtinfallxxxxxA
\else
\ifnum#1=14 %
\hatcurPPtinfallxxxxxB
\else
??????\fi
\fi
}
\newcommand{\hatcurRVeccen}[1]{\ifnum#1=13 %
\hatcurRVeccenxxxxxA
\else
\ifnum#1=14 %
\hatcurRVeccenxxxxxB
\else
??????\fi
\fi
}
\newcommand{\hatcurRVeccentwosiglim}[1]{\ifnum#1=13 %
\hatcurRVeccentwosiglimxxxxxA
\else
\ifnum#1=14 %
\hatcurRVeccentwosiglimxxxxxB
\else
??????\fi
\fi
}
\newcommand{\hatcurRVfitrmsA}[1]{\ifnum#1=13 %
\hatcurRVfitrmsAxxxxxA
\else
\ifnum#1=14 %
\hatcurRVfitrmsAxxxxxB
\else
??????\fi
\fi
}
\newcommand{\hatcurRVfitrmsB}[1]{\ifnum#1=13 %
\hatcurRVfitrmsBxxxxxA
\else
\ifnum#1=14 %
\hatcurRVfitrmsBxxxxxB
\else
??????\fi
\fi
}
\newcommand{\hatcurRVfitrmsC}[1]{\ifnum#1=13 %
\hatcurRVfitrmsCxxxxxA
\else
??????\fi
}
\newcommand{\hatcurRVfitrmsD}[1]{\ifnum#1=13 %
\hatcurRVfitrmsDxxxxxA
\else
??????\fi
}
\newcommand{\hatcurRVgammaA}[1]{\ifnum#1=13 %
\hatcurRVgammaAxxxxxA
\else
\ifnum#1=14 %
\hatcurRVgammaAxxxxxB
\else
??????\fi
\fi
}
\newcommand{\hatcurRVgammaB}[1]{\ifnum#1=13 %
\hatcurRVgammaBxxxxxA
\else
\ifnum#1=14 %
\hatcurRVgammaBxxxxxB
\else
??????\fi
\fi
}
\newcommand{\hatcurRVgammaC}[1]{\ifnum#1=13 %
\hatcurRVgammaCxxxxxA
\else
??????\fi
}
\newcommand{\hatcurRVgammaD}[1]{\ifnum#1=13 %
\hatcurRVgammaDxxxxxA
\else
??????\fi
}
\newcommand{\hatcurRVh}[1]{\ifnum#1=13 %
\hatcurRVhxxxxxA
\else
\ifnum#1=14 %
\hatcurRVhxxxxxB
\else
??????\fi
\fi
}
\newcommand{\hatcurRVjitterA}[1]{\ifnum#1=13 %
\hatcurRVjitterAxxxxxA
\else
\ifnum#1=14 %
\hatcurRVjitterAxxxxxB
\else
??????\fi
\fi
}
\newcommand{\hatcurRVjitterB}[1]{\ifnum#1=13 %
\hatcurRVjitterBxxxxxA
\else
\ifnum#1=14 %
\hatcurRVjitterBxxxxxB
\else
??????\fi
\fi
}
\newcommand{\hatcurRVjitterC}[1]{\ifnum#1=13 %
\hatcurRVjitterCxxxxxA
\else
??????\fi
}
\newcommand{\hatcurRVjitterD}[1]{\ifnum#1=13 %
\hatcurRVjitterDxxxxxA
\else
??????\fi
}
\newcommand{\hatcurRVk}[1]{\ifnum#1=13 %
\hatcurRVkxxxxxA
\else
\ifnum#1=14 %
\hatcurRVkxxxxxB
\else
??????\fi
\fi
}
\newcommand{\hatcurRVK}[1]{\ifnum#1=13 %
\hatcurRVKxxxxxA
\else
\ifnum#1=14 %
\hatcurRVKxxxxxB
\else
??????\fi
\fi
}
\newcommand{\hatcurRVomega}[1]{\ifnum#1=13 %
\hatcurRVomegaxxxxxA
\else
\ifnum#1=14 %
\hatcurRVomegaxxxxxB
\else
??????\fi
\fi
}
\newcommand{\hatcurRVrh}[1]{\ifnum#1=13 %
\hatcurRVrhxxxxxA
\else
\ifnum#1=14 %
\hatcurRVrhxxxxxB
\else
??????\fi
\fi
}
\newcommand{\hatcurRVrk}[1]{\ifnum#1=13 %
\hatcurRVrkxxxxxA
\else
\ifnum#1=14 %
\hatcurRVrkxxxxxB
\else
??????\fi
\fi
}
\newcommand{\hatcurRVtrone}[1]{\ifnum#1=13 %
\hatcurRVtronexxxxxA
\else
\ifnum#1=14 %
\hatcurRVtronexxxxxB
\else
??????\fi
\fi
}
\newcommand{\hatcurRVtrtwo}[1]{\ifnum#1=13 %
\hatcurRVtrtwoxxxxxA
\else
\ifnum#1=14 %
\hatcurRVtrtwoxxxxxB
\else
??????\fi
\fi
}
\newcommand{\hatcurSMEiilogg}[1]{\ifnum#1=13 %
\hatcurSMEiiloggxxxxxA
\else
\ifnum#1=14 %
\hatcurSMEiiloggxxxxxB
\else
??????\fi
\fi
}
\newcommand{\hatcurSMEiiteff}[1]{\ifnum#1=13 %
\hatcurSMEiiteffxxxxxA
\else
\ifnum#1=14 %
\hatcurSMEiiteffxxxxxB
\else
??????\fi
\fi
}
\newcommand{\hatcurSMEiivsin}[1]{\ifnum#1=13 %
\hatcurSMEiivsinxxxxxA
\else
\ifnum#1=14 %
\hatcurSMEiivsinxxxxxB
\else
??????\fi
\fi
}
\newcommand{\hatcurSMEiizfeh}[1]{\ifnum#1=13 %
\hatcurSMEiizfehxxxxxA
\else
\ifnum#1=14 %
\hatcurSMEiizfehxxxxxB
\else
??????\fi
\fi
}
\newcommand{\hatcurSMEiizfehshort}[1]{\ifnum#1=13 %
\hatcurSMEiizfehshortxxxxxA
\else
\ifnum#1=14 %
\hatcurSMEiizfehshortxxxxxB
\else
??????\fi
\fi
}
\newcommand{\hatcurSMEilogg}[1]{\ifnum#1=13 %
\hatcurSMEiloggxxxxxA
\else
\ifnum#1=14 %
\hatcurSMEiloggxxxxxB
\else
??????\fi
\fi
}
\newcommand{\hatcurSMEiteff}[1]{\ifnum#1=13 %
\hatcurSMEiteffxxxxxA
\else
\ifnum#1=14 %
\hatcurSMEiteffxxxxxB
\else
??????\fi
\fi
}
\newcommand{\hatcurSMEivmac}[1]{\ifnum#1=13 %
\hatcurSMEivmacxxxxxA
\else
\ifnum#1=14 %
\hatcurSMEivmacxxxxxB
\else
??????\fi
\fi
}
\newcommand{\hatcurSMEivmic}[1]{\ifnum#1=13 %
\hatcurSMEivmicxxxxxA
\else
\ifnum#1=14 %
\hatcurSMEivmicxxxxxB
\else
??????\fi
\fi
}
\newcommand{\hatcurSMEivsin}[1]{\ifnum#1=13 %
\hatcurSMEivsinxxxxxA
\else
\ifnum#1=14 %
\hatcurSMEivsinxxxxxB
\else
??????\fi
\fi
}
\newcommand{\hatcurSMEizfeh}[1]{\ifnum#1=13 %
\hatcurSMEizfehxxxxxA
\else
\ifnum#1=14 %
\hatcurSMEizfehxxxxxB
\else
??????\fi
\fi
}
\newcommand{\hatcurSMEizfehshort}[1]{\ifnum#1=13 %
\hatcurSMEizfehshortxxxxxA
\else
\ifnum#1=14 %
\hatcurSMEizfehshortxxxxxB
\else
??????\fi
\fi
}
\newcommand{\hatcurXAv}[1]{\ifnum#1=13 %
\hatcurXAvxxxxxA
\else
\ifnum#1=14 %
\hatcurXAvxxxxxB
\else
??????\fi
\fi
}
\newcommand{\hatcurXdist}[1]{\ifnum#1=13 %
\hatcurXdistxxxxxA
\else
\ifnum#1=14 %
\hatcurXdistxxxxxB
\else
??????\fi
\fi
}
\newcommand{\hatcurXdistred}[1]{\ifnum#1=13 %
\hatcurXdistredxxxxxA
\else
\ifnum#1=14 %
\hatcurXdistredxxxxxB
\else
??????\fi
\fi
}
\newcommand{\hatcurXEBV}[1]{\ifnum#1=13 %
\hatcurXEBVxxxxxA
\else
\ifnum#1=14 %
\hatcurXEBVxxxxxB
\else
??????\fi
\fi
}
\newcommand{\hatcurXjhisored}[1]{\ifnum#1=13 %
\hatcurXjhisoredxxxxxA
\else
\ifnum#1=14 %
\hatcurXjhisoredxxxxxB
\else
??????\fi
\fi
}
\newcommand{\hatcurXjkisored}[1]{\ifnum#1=13 %
\hatcurXjkisoredxxxxxA
\else
\ifnum#1=14 %
\hatcurXjkisoredxxxxxB
\else
??????\fi
\fi
}
\newcommand{\hatcurXmhisored}[1]{\ifnum#1=13 %
\hatcurXmhisoredxxxxxA
\else
\ifnum#1=14 %
\hatcurXmhisoredxxxxxB
\else
??????\fi
\fi
}
\newcommand{\hatcurXmiisored}[1]{\ifnum#1=13 %
\hatcurXmiisoredxxxxxA
\else
\ifnum#1=14 %
\hatcurXmiisoredxxxxxB
\else
??????\fi
\fi
}
\newcommand{\hatcurXmjisored}[1]{\ifnum#1=13 %
\hatcurXmjisoredxxxxxA
\else
\ifnum#1=14 %
\hatcurXmjisoredxxxxxB
\else
??????\fi
\fi
}
\newcommand{\hatcurXmkisored}[1]{\ifnum#1=13 %
\hatcurXmkisoredxxxxxA
\else
\ifnum#1=14 %
\hatcurXmkisoredxxxxxB
\else
??????\fi
\fi
}
\newcommand{\hatcurXmvisored}[1]{\ifnum#1=13 %
\hatcurXmvisoredxxxxxA
\else
\ifnum#1=14 %
\hatcurXmvisoredxxxxxB
\else
??????\fi
\fi
}
\newcommand{\hatcurXsecdur}[1]{\ifnum#1=13 %
\hatcurXsecdurxxxxxA
\else
\ifnum#1=14 %
\hatcurXsecdurxxxxxB
\else
??????\fi
\fi
}
\newcommand{\hatcurXsecingdur}[1]{\ifnum#1=13 %
\hatcurXsecingdurxxxxxA
\else
\ifnum#1=14 %
\hatcurXsecingdurxxxxxB
\else
??????\fi
\fi
}
\newcommand{\hatcurXsecondary}[1]{\ifnum#1=13 %
\hatcurXsecondaryxxxxxA
\else
\ifnum#1=14 %
\hatcurXsecondaryxxxxxB
\else
??????\fi
\fi
}
\newcommand{\hatcurXsecphase}[1]{\ifnum#1=13 %
\hatcurXsecphasexxxxxA
\else
\ifnum#1=14 %
\hatcurXsecphasexxxxxB
\else
??????\fi
\fi
}
\newcommand{\hatcurXviisored}[1]{\ifnum#1=13 %
\hatcurXviisoredxxxxxA
\else
\ifnum#1=14 %
\hatcurXviisoredxxxxxB
\else
??????\fi
\fi
}
\newcommand{\hatcurXvkisored}[1]{\ifnum#1=13 %
\hatcurXvkisoredxxxxxA
\else
\ifnum#1=14 %
\hatcurXvkisoredxxxxxB
\else
??????\fi
\fi
}

\newcommand{\hatcurhtreccenxxxxxA}{HATS582-012}                        % Original HTR name of target
\newcommand{\hatcurfieldeccenxxxxxA}{582}                              % Original HTR field
\newcommand{\hatcurCCraeccenxxxxxA}{\ensuremath{21^{\mathrm h}07^{\mathrm m}50.88{\mathrm s}}}                       % Right Ascension
\newcommand{\hatcurCCdececcenxxxxxA}{\ensuremath{-26{\arcdeg}05{\arcmin}48.0{\arcsec}}}                      % Declination
\newcommand{\hatcurCCmageccenxxxxxA}{13.887}                           % apparent V-band magnitude
\newcommand{\hatcurCCtwomasseccenxxxxxA}{2MASS~21075075-2605479}       % 2MASS identifier
\newcommand{\hatcurCCgsceccenxxxxxA}{GSC~6928-00497}                   % GSC(1.2) identifier
\newcommand{\hatcurCCtassmveccenxxxxxA}{\ensuremath{13.887\pm0.010}}   % APASS V-band magnitude
\newcommand{\hatcurCCtassmvshorteccenxxxxxA}{\ensuremath{13.9}}        % APASS V-band magnitude
\newcommand{\hatcurCCtassmBeccenxxxxxA}{\ensuremath{14.686\pm0.020}}   % APASS B-band magnitude
\newcommand{\hatcurCCtassmBshorteccenxxxxxA}{\ensuremath{14.7}}        % APASS B-band magnitude
\newcommand{\hatcurCCtassmIeccenxxxxxA}{\ensuremath{100\pm1000}}       % TASS I-band magnitude
\newcommand{\hatcurCCtassmIshorteccenxxxxxA}{\ensuremath{100.0}}       % TASS I-band magnitude
\newcommand{\hatcurCCtassmgeccenxxxxxA}{\ensuremath{14.217\pm0.010}}   % APASS g-band magnitude
\newcommand{\hatcurCCtassmgshorteccenxxxxxA}{\ensuremath{14.2}}        % APASS g-band magnitude
\newcommand{\hatcurCCtassmreccenxxxxxA}{\ensuremath{13.612\pm0.010}}   % APASS r-band magnitude
\newcommand{\hatcurCCtassmrshorteccenxxxxxA}{\ensuremath{13.6}}        % APASS r-band magnitude
\newcommand{\hatcurCCtassmieccenxxxxxA}{\ensuremath{13.392\pm0.010}}   % APASS i-band magnitude
\newcommand{\hatcurCCtassmishorteccenxxxxxA}{\ensuremath{13.4}}        % APASS i-band magnitude
\newcommand{\hatcurCCtwomassJmageccenxxxxxA}{\ensuremath{12.439\pm0.021}} % 2MASS ORIG MAG
\newcommand{\hatcurCCtwomassHmageccenxxxxxA}{\ensuremath{12.052\pm0.025}} % 2MASS ORIG MAG
\newcommand{\hatcurCCtwomassKmageccenxxxxxA}{\ensuremath{11.983\pm0.028}} % 2MASS ORIG MAG
\newcommand{\hatcurCCcitJmageccenxxxxxA}{\ensuremath{12.451\pm0.022}}  % 2MASS CIT MAG
\newcommand{\hatcurCCcitHmageccenxxxxxA}{\ensuremath{12.047\pm0.025}}  % 2MASS CIT MAG
\newcommand{\hatcurCCcitKmageccenxxxxxA}{\ensuremath{12.007\pm0.028}}  % 2MASS CIT MAG
\newcommand{\hatcurCCbbJmageccenxxxxxA}{\ensuremath{12.508\pm0.023}}   % 2MASS BB MAG
\newcommand{\hatcurCCbbHmageccenxxxxxA}{\ensuremath{12.068\pm0.026}}   % 2MASS BB MAG
\newcommand{\hatcurCCbbKmageccenxxxxxA}{\ensuremath{12.027\pm0.028}}   % 2MASS BB MAG
\newcommand{\hatcurCCesoJmageccenxxxxxA}{\ensuremath{12.511\pm0.025}}  % 2MASS ESO MAG
\newcommand{\hatcurCCesoHmageccenxxxxxA}{\ensuremath{12.062\pm0.029}}  % 2MASS ESO MAG
\newcommand{\hatcurCCesoKmageccenxxxxxA}{\ensuremath{12.025\pm0.029}}  % 2MASS ESO MAG
\newcommand{\hatcurCCesoJHmageccenxxxxxA}{\ensuremath{0.448\pm0.036}}  % 2MASS ESO JH COLOR
\newcommand{\hatcurCCesoJKmageccenxxxxxA}{\ensuremath{0.486\pm0.038}}  % 2MASS ESO JK COLOR
\newcommand{\hatcurCCesoHKmageccenxxxxxA}{\ensuremath{0.038\pm0.042}}  % 2MASS ESO HK COLOR
\newcommand{\hatcurLCdipeccenxxxxxA}{\ensuremath{24.2}}                % BLS detected dip (mmag)
\newcommand{\hatcurLCrprstareccenxxxxxA}{\ensuremath{0.1402\pm0.0019}} % Rp/R*
\newcommand{\hatcurLCbsqeccenxxxxxA}{\ensuremath{0.067_{-0.038}^{+0.038}}} % impact parameter square
\newcommand{\hatcurLCimpeccenxxxxxA}{\ensuremath{0.259_{-0.088}^{+0.065}}} % impact parameter
\newcommand{\hatcurLCzetaeccenxxxxxA}{\ensuremath{20.962\pm0.091}}     % zeta/R*
\newcommand{\hatcurLCdureccenxxxxxA}{\ensuremath{0.1096\pm0.0010}}     % transit duration (days)
\newcommand{\hatcurLCdurshorteccenxxxxxA}{\ensuremath{0.1096}}         % transit duration (days)
\newcommand{\hatcurLCdurhreccenxxxxxA}{\ensuremath{2.631\pm0.024}}     % transit duration (hours)
\newcommand{\hatcurLCdurhrshorteccenxxxxxA}{\ensuremath{2.631}}        % transit duration (hours)
\newcommand{\hatcurLCqeccenxxxxxA}{\ensuremath{0.03600\pm0.00033}}     % fractional transit duration (days)
\newcommand{\hatcurLCqshorteccenxxxxxA}{\ensuremath{0.036}}            % fractional transit duration (days)
\newcommand{\hatcurLCingdureccenxxxxxA}{\ensuremath{0.01432\pm0.00077}} % ingress/egress duration (days)
\newcommand{\hatcurLCPeccenxxxxxA}{\ensuremath{3.0440492\pm0.0000025}} % period (days)
\newcommand{\hatcurLCPprececcenxxxxxA}{\ensuremath{3.0440492}}         % period (days)
\newcommand{\hatcurLCPshorteccenxxxxxA}{\ensuremath{3.0440}}           % period (days)
\newcommand{\hatcurLCTeccenxxxxxA}{\ensuremath{2456297.70033\pm0.00019}} % epoch (BJD)
\newcommand{\hatcurLCTAeccenxxxxxA}{\ensuremath{2455092.25686\pm0.00092}} % TA (BJD)
\newcommand{\hatcurLCTBeccenxxxxxA}{\ensuremath{2456474.25520\pm0.00028}} % TB (BJD)
\newcommand{\hatcurLChatnetmeccenxxxxxA}{\ensuremath{13.66914\pm0.00011}} % HATNet OOT level
\newcommand{\hatcurLCiblendeccenxxxxxA}{\ensuremath{0.955\pm0.040}}    % HATNet iblend factor
\newcommand{\hatcurSMEiteffeccenxxxxxA}{\ensuremath{5540\pm130}}       % Ini SME, stellar effective temperature
\newcommand{\hatcurSMEizfeheccenxxxxxA}{\ensuremath{0.100\pm0.090}}    % Ini SME, stellar metallicity
\newcommand{\hatcurSMEizfehshorteccenxxxxxA}{\ensuremath{0.10}}        % Ini SME, stellar metallicity
\newcommand{\hatcurSMEiloggeccenxxxxxA}{\ensuremath{4.85\pm0.23}}      % Ini SME, stellar surface gravity
\newcommand{\hatcurSMEivsineccenxxxxxA}{\ensuremath{3.12\pm0.30}}      % Ini SME, stellar rotational velocity
\newcommand{\hatcurSMEivmaceccenxxxxxA}{\ensuremath{0.0}}              % Ini SME, stellar macroturbulence
\newcommand{\hatcurSMEivmiceccenxxxxxA}{\ensuremath{0.0}}              % Ini SME, stellar microturbulence
\newcommand{\hatcurSMEiiteffeccenxxxxxA}{\ensuremath{5523\pm69}}       % Final SME, stellar effective temperature
\newcommand{\hatcurSMEiizfeheccenxxxxxA}{\ensuremath{0.050\pm0.060}}   % Final SME, stellar metallicity
\newcommand{\hatcurSMEiizfehshorteccenxxxxxA}{\ensuremath{0.05}}       % Final SME, stellar metallicity
\newcommand{\hatcurSMEiiloggeccenxxxxxA}{\ensuremath{4.518\pm0.020}}   % Final SME, stellar surface gravity
\newcommand{\hatcurSMEiivsineccenxxxxxA}{\ensuremath{2.82\pm0.30}}     % Final SME, stellar rotational velocity
\newcommand{\hatcurANUWIFESteffeccenxxxxxA}{\ensuremath{5540\pm300}}      % ANUWIFES stellar effective temperature
\newcommand{\hatcurANUWIFESzfeheccenxxxxxA}{\ensuremath{0.00\pm0.50}}     % ANUWIFES stellar metallicity
\newcommand{\hatcurANUWIFESloggeccenxxxxxA}{\ensuremath{4.50\pm0.30}}     % ANUWIFES stellar surface gravity
\newcommand{\hatcurANUWIFESvsinieccenxxxxxA}{\ensuremath{nff\pmnff}}      % ANUWIFES stellar rotational velocity
\newcommand{\hatcurANUWIFESgammaeccenxxxxxA}{\ensuremath{25.09\pm0.26}}   % ANUWIFES absolute gamma velocity
\newcommand{\hatcurANUWIFESnumspececcenxxxxxA}{\ensuremath{2}}            % ANUWIFES number of spectra
\newcommand{\hatcurANUWIFESspaneccenxxxxxA}{\ensuremath{NULL}}            % ANUWIFES timespan of observations
\newcommand{\hatcurANUWIFESrvrmseccenxxxxxA}{\ensuremath{0.67}}           % ANUWIFES rms of RV values [km/s]
\newcommand{\hatcurLBizeccenxxxxxA}{\ensuremath{0.2436}}               % Limb darkening parameters, Gamma1, z-band
\newcommand{\hatcurLBiizeccenxxxxxA}{\ensuremath{0.3108}}              % Limb darkening parameters, Gamma2, z-band
\newcommand{\hatcurLBiieccenxxxxxA}{\ensuremath{0.3116}}               % Limb darkening parameters, Gamma1, i-band
\newcommand{\hatcurLBiiieccenxxxxxA}{\ensuremath{0.3063}}              % Limb darkening parameters, Gamma2, i-band
\newcommand{\hatcurLBiIeccenxxxxxA}{\ensuremath{0.2890}}               % Limb darkening parameters, Gamma1, I-band
\newcommand{\hatcurLBiiIeccenxxxxxA}{\ensuremath{0.3083}}              % Limb darkening parameters, Gamma2, I-band
\newcommand{\hatcurLBigeccenxxxxxA}{\ensuremath{0.6213}}               % Limb darkening parameters, Gamma1, g-band
\newcommand{\hatcurLBiigeccenxxxxxA}{\ensuremath{0.1844}}              % Limb darkening parameters, Gamma2, g-band
\newcommand{\hatcurLBireccenxxxxxA}{\ensuremath{0.4107}}               % Limb darkening parameters, Gamma1, r-band
\newcommand{\hatcurLBiireccenxxxxxA}{\ensuremath{0.2928}}              % Limb darkening parameters, Gamma2, r-band
\newcommand{\hatcurLBiReccenxxxxxA}{\ensuremath{0.3833}}               % Limb darkening parameters, Gamma1, R-band
\newcommand{\hatcurLBiiReccenxxxxxA}{\ensuremath{0.2974}}              % Limb darkening parameters, Gamma2, R-band
\newcommand{\hatcurISOmeccenxxxxxA}{\ensuremath{0.954\pm0.031}}        % stellar mass
\newcommand{\hatcurISOmshorteccenxxxxxA}{\ensuremath{0.95}}            % stellar mass
\newcommand{\hatcurISOmlongeccenxxxxxA}{\ensuremath{0.954\pm0.031}}    % stellar mass
\newcommand{\hatcurISOreccenxxxxxA}{\ensuremath{0.900\pm0.049}}        % stellar radius
\newcommand{\hatcurISOrshorteccenxxxxxA}{\ensuremath{0.90}}            % stellar radius
\newcommand{\hatcurISOrlongeccenxxxxxA}{\ensuremath{0.900\pm0.049}}    % stellar radius
\newcommand{\hatcurISOrhoeccenxxxxxA}{\ensuremath{1.86\pm0.26}}        % stellar density (cgs)
\newcommand{\hatcurISOrholongeccenxxxxxA}{\ensuremath{1.86\pm0.26}}    % stellar density (cgs)
\newcommand{\hatcurISOloggeccenxxxxxA}{\ensuremath{4.511\pm0.047}}     % stellar surface gravity from isochrones
\newcommand{\hatcurISOlumeccenxxxxxA}{\ensuremath{0.671\pm0.094}}      % stellar luminosity
\newcommand{\hatcurISOlumshorteccenxxxxxA}{\ensuremath{0.67}}          % stellar luminosity
\newcommand{\hatcurISOmveccenxxxxxA}{\ensuremath{5.31\pm0.15}}         % stellar absolute magnitude
\newcommand{\hatcurISOvieccenxxxxxA}{\ensuremath{0.779\pm0.020}}       % stellar V-I index
\newcommand{\hatcurISOageeccenxxxxxA}{\ensuremath{3.4_{-2.2}^{+3.1}}}  % stellar age
\newcommand{\hatcurISOsigmaeccenxxxxxA}{\ensuremath{0.00120\pm0.00046}} % system mass-correction sigma parameter
\newcommand{\hatcurISOMJeccenxxxxxA}{\ensuremath{4.04\pm0.13}}         % stellar absolute J magnitude
\newcommand{\hatcurISOMHeccenxxxxxA}{\ensuremath{3.64\pm0.12}}         % stellar absolute H magnitude
\newcommand{\hatcurISOMKeccenxxxxxA}{\ensuremath{3.58\pm0.12}}         % stellar absolute K magnitude
\newcommand{\hatcurISOJKeccenxxxxxA}{\ensuremath{0.460\pm0.020}}       % J-K color index from isochrones.
\newcommand{\hatcurISOspececcenxxxxxA}{G}                              % stellar spectral type
\newcommand{\hatcurRVKeccenxxxxxA}{\ensuremath{79\pm11}}               % RV semi-amplitude [m/s]
\newcommand{\hatcurRVrkeccenxxxxxA}{\ensuremath{0.12_{-0.25}^{+0.10}}} % sqrt(e)*cos(omega)
\newcommand{\hatcurRVrheccenxxxxxA}{\ensuremath{0.07\pm0.15}}          % sqrt(e)*sin(omega)
\newcommand{\hatcurRVkeccenxxxxxA}{\ensuremath{0.023_{-0.055}^{+0.032}}} % e*cos(omega)
\newcommand{\hatcurRVheccenxxxxxA}{\ensuremath{0.012_{-0.027}^{+0.050}}} % e*sin(omega)
\newcommand{\hatcurRVtroneeccenxxxxxA}{\ensuremath{0\pm0}}             % RV linear trend tr1 factor
\newcommand{\hatcurRVtrtwoeccenxxxxxA}{\ensuremath{0\pm0}}             % RV linear trend tr2 factor
\newcommand{\hatcurRVgammaAeccenxxxxxA}{\ensuremath{-6.5\pm5.7}}       % RV gamma velocity, relative scale
\newcommand{\hatcurRVjitterAeccenxxxxxA}{\ensuremath{0.1\pm3.8}}       % RV jitter (m/s)
\newcommand{\hatcurRVfitrmsAeccenxxxxxA}{\ensuremath{21.0}}            % RVfitrms
\newcommand{\hatcurRVgammaBeccenxxxxxA}{\ensuremath{25803\pm17}}       % RV gamma velocity, relative scale
\newcommand{\hatcurRVjitterBeccenxxxxxA}{\ensuremath{64\pm14}}         % RV jitter (m/s)
\newcommand{\hatcurRVfitrmsBeccenxxxxxA}{\ensuremath{67.5}}            % RVfitrms
\newcommand{\hatcurRVgammaCeccenxxxxxA}{\ensuremath{25781\pm41}}       % RV gamma velocity, relative scale
\newcommand{\hatcurRVjitterCeccenxxxxxA}{\ensuremath{123\pm44}}        % RV jitter (m/s)
\newcommand{\hatcurRVfitrmsCeccenxxxxxA}{\ensuremath{99.8}}            % RVfitrms
\newcommand{\hatcurRVgammaDeccenxxxxxA}{\ensuremath{25800\pm180}}      % RV gamma velocity, relative scale
\newcommand{\hatcurRVjitterDeccenxxxxxA}{\ensuremath{230\pm160}}       % RV jitter (m/s)
\newcommand{\hatcurRVfitrmsDeccenxxxxxA}{\ensuremath{188.5}}           % RVfitrms
\newcommand{\hatcurRVecceneccenxxxxxA}{\ensuremath{0.045\pm0.053}}     % eccentricity
\newcommand{\hatcurRVeccentwosiglimeccenxxxxxA}{\ensuremath{<0.181}}   % eccentricity
\newcommand{\hatcurRVomegaeccenxxxxxA}{\ensuremath{110\pm120}}         % argument of pericenter
\newcommand{\hatcurPPieccenxxxxxA}{\ensuremath{88.44\pm0.56}}          % orbital inclination
\newcommand{\hatcurPPgeccenxxxxxA}{\ensuremath{9.12_{-2.22}^{+0.98}}}  % planetary surface gravity (m/s^2)
\newcommand{\hatcurPPloggeccenxxxxxA}{\ensuremath{2.960_{-0.122}^{+0.044}}} % planetary surface gravity (log cgs)
\newcommand{\hatcurPPareccenxxxxxA}{\ensuremath{9.69\pm0.49}}          % relative orbital radius (a/R*)
\newcommand{\hatcurPPareleccenxxxxxA}{\ensuremath{0.04047\pm0.00044}}  % semimajor axis (AU)
\newcommand{\hatcurPPrhoeccenxxxxxA}{\ensuremath{0.373_{-0.101}^{+0.045}}} % planetary density (cgs)
\newcommand{\hatcurPPmeccenxxxxxA}{\ensuremath{0.540\pm0.075}}         % planetary mass (M_jup)
\newcommand{\hatcurPPmshorteccenxxxxxA}{\ensuremath{0.54}}             % planetary mass (M_jup)
\newcommand{\hatcurPPmlongeccenxxxxxA}{\ensuremath{0.540\pm0.075}}     % planetary mass (M_jup)
\newcommand{\hatcurPPmeeccenxxxxxA}{\ensuremath{172\pm24}}             % planetary mass (M_earth)
\newcommand{\hatcurPPmeshorteccenxxxxxA}{\ensuremath{171.5}}           % planetary mass (M_earth)
\newcommand{\hatcurPPmelongeccenxxxxxA}{\ensuremath{172\pm24}}         % planetary mass (M_earth)
\newcommand{\hatcurPPreccenxxxxxA}{\ensuremath{1.230\pm0.070}}         % planetary radius (R_jup)
\newcommand{\hatcurPPrshorteccenxxxxxA}{\ensuremath{1.23}}             % planetary radius (R_jup)
\newcommand{\hatcurPPrlongeccenxxxxxA}{\ensuremath{1.230\pm0.070}}     % planetary radius (R_jup)
\newcommand{\hatcurPPreeccenxxxxxA}{\ensuremath{13.78\pm0.78}}         % planetary radius (R_earth)
\newcommand{\hatcurPPreshorteccenxxxxxA}{\ensuremath{13.8}}            % planetary radius (R_earth)
\newcommand{\hatcurPPrelongeccenxxxxxA}{\ensuremath{13.78\pm0.78}}     % planetary radius (R_earth)
\newcommand{\hatcurPPmrcorreccenxxxxxA}{\ensuremath{-0.03}}            % mass/radius correlation
\newcommand{\hatcurPPteffeccenxxxxxA}{\ensuremath{1255\pm41}}          % planetary temperature (K)
\newcommand{\hatcurPPthetaeccenxxxxxA}{\ensuremath{0.0372\pm0.0056}}   % Safranov number
\newcommand{\hatcurPPfluxperieccenxxxxxA}{\ensuremath{6.11_{-0.71}^{+1.51}}} % flux @ periastron (CGS)
\newcommand{\hatcurPPfluxperidimeccenxxxxxA}{\ensuremath{8}}           % flux @ periastron (CGS) units.
\newcommand{\hatcurPPfluxapeccenxxxxxA}{\ensuremath{5.07\pm0.56}}      % flux @ apastron (CGS)
\newcommand{\hatcurPPfluxapdimeccenxxxxxA}{\ensuremath{8}}             % flux @ apastron (CGS) units.
\newcommand{\hatcurPPfluxavgeccenxxxxxA}{\ensuremath{5.60\pm0.79}}     % flux on average (CGS)
\newcommand{\hatcurPPfluxavgdimeccenxxxxxA}{\ensuremath{8}}            % flux average (CGS) units.
\newcommand{\hatcurPPfluxavglogeccenxxxxxA}{\ensuremath{8.748\pm0.055}} % log10 flux on average (CGS)
\newcommand{\hatcurXsecphaseeccenxxxxxA}{\ensuremath{0.514\pm0.038}}   % Phase of secondary eclipse
\newcommand{\hatcurXsecondaryeccenxxxxxA}{\ensuremath{2456299.27\pm0.11}} % Secondary eclipse epoch
\newcommand{\hatcurXsecdureccenxxxxxA}{\ensuremath{0.112\pm0.011}}     % sec eclipse duration (days)
\newcommand{\hatcurXsecingdureccenxxxxxA}{\ensuremath{0.0147\pm0.0020}} % sec I/E duration (days)
\newcommand{\hatcurPPphiconjeccenxxxxxA}{\ensuremath{0.13_{-0.28}^{+0.15}}} % phase diff between conjunction and periastron
\newcommand{\hatcurPPperieccenxxxxxA}{\ensuremath{2456297.31\pm0.73}}  % time of periastron passage.
\newcommand{\hatcurPPaequiveccenxxxxxA}{\ensuremath{0.0494\pm0.0030}}  % equivalent semi-major axis
\newcommand{\hatcurPPtcirceccenxxxxxA}{\ensuremath{72_{-29}^{+15}}}    % circularization timescale
\newcommand{\hatcurPPtinfalleccenxxxxxA}{\ensuremath{4230_{-900}^{+1330}}} % infall timescale
\newcommand{\hatcurXdisteccenxxxxxA}{\ensuremath{489\pm29}}            % distance (pc), no reddenning correction
\newcommand{\hatcurXAveccenxxxxxA}{\ensuremath{0.150\pm0.065}}         % Av (mag)
\newcommand{\hatcurXdistredeccenxxxxxA}{\ensuremath{483\pm27}}         % distance with Av correction (pc)
\newcommand{\hatcurXEBVeccenxxxxxA}{\ensuremath{0.048\pm0.021}}        % E(B-V) (mag)
\newcommand{\hatcurXmvisoredeccenxxxxxA}{\ensuremath{13.887\pm0.010}}  % Expected m_v with reddening (mag)
\newcommand{\hatcurXmiisoredeccenxxxxxA}{\ensuremath{13.030\pm0.016}}  % Expected m_i with reddening (mag)
\newcommand{\hatcurXmjisoredeccenxxxxxA}{\ensuremath{12.498\pm0.015}}  % Expected m_j with reddening (mag)
\newcommand{\hatcurXmhisoredeccenxxxxxA}{\ensuremath{12.089\pm0.016}}  % Expected m_h with reddening (mag)
\newcommand{\hatcurXmkisoredeccenxxxxxA}{\ensuremath{12.013\pm0.018}}  % Expected m_k with reddening (mag)
\newcommand{\hatcurXviisoredeccenxxxxxA}{\ensuremath{0.857\pm0.017}}   % Expected V-I with reddening (mag)
\newcommand{\hatcurXvkisoredeccenxxxxxA}{\ensuremath{1.874\pm0.021}}   % Expected V-K with reddening (mag)
\newcommand{\hatcurXjhisoredeccenxxxxxA}{\ensuremath{0.4100\pm0.0098}} % Expected J-H with reddening (mag)
\newcommand{\hatcurXjkisoredeccenxxxxxA}{\ensuremath{0.4860\pm0.0090}} % Expected J-K with reddening (mag)
\newcommand{\hatcurCCpmraeccenxxxxxA}{\ensuremath{-2\pm14}}            % proper motion, in RA
\newcommand{\hatcurCCpmdececcenxxxxxA}{\ensuremath{-9.1\pm1.6}}        % proper motion, in DEC
\newcommand{\hatcurCCpmeccenxxxxxA}{\ensuremath{9\pm14}}               % proper motion

\newcommand{\hatcurhtreccenxxxxxB}{HATS582-006}                        % Original HTR name of target
\newcommand{\hatcurfieldeccenxxxxxB}{\ensuremath{string}}              % HTR field
\newcommand{\hatcurCCraeccenxxxxxB}{\ensuremath{20^{\mathrm h}52^{\mathrm m}51.60{\mathrm s}}}                       % Right Ascension
\newcommand{\hatcurCCdececcenxxxxxB}{\ensuremath{-25{\arcdeg}41{\arcmin}14.4{\arcsec}}}                      % Declination
\newcommand{\hatcurCCmageccenxxxxxB}{13.790}                           % apparent V-band magnitude
\newcommand{\hatcurCCtwomasseccenxxxxxB}{2MASS~20525171-2541144}       % 2MASS identifier
\newcommand{\hatcurCCgsceccenxxxxxB}{GSC~6926-00259}                   % GSC(1.2) identifier
\newcommand{\hatcurCCtassmveccenxxxxxB}{\ensuremath{13.79\pm0.10}}     % APASS V-band magnitude
\newcommand{\hatcurCCtassmvshorteccenxxxxxB}{\ensuremath{13.8}}        % APASS V-band magnitude
\newcommand{\hatcurCCtassmBeccenxxxxxB}{\ensuremath{14.62\pm0.10}}     % APASS B-band magnitude
\newcommand{\hatcurCCtassmBshorteccenxxxxxB}{\ensuremath{14.6}}        % APASS B-band magnitude
\newcommand{\hatcurCCtassmIeccenxxxxxB}{\ensuremath{nff\pmnff}}        % TASS I-band magnitude
\newcommand{\hatcurCCtassmIshorteccenxxxxxB}{\ensuremath{0.0}}         % TASS I-band magnitude
\newcommand{\hatcurCCtassmgeccenxxxxxB}{\ensuremath{nff\pmnff}}        % APASS g-band magnitude
\newcommand{\hatcurCCtassmgshorteccenxxxxxB}{\ensuremath{0.0}}         % APASS g-band magnitude
\newcommand{\hatcurCCtassmreccenxxxxxB}{\ensuremath{nff\pmnff}}        % APASS r-band magnitude
\newcommand{\hatcurCCtassmrshorteccenxxxxxB}{\ensuremath{0.0}}         % APASS r-band magnitude
\newcommand{\hatcurCCtassmieccenxxxxxB}{\ensuremath{nff\pmnff}}        % APASS i-band magnitude
\newcommand{\hatcurCCtassmishorteccenxxxxxB}{\ensuremath{0.0}}         % APASS i-band magnitude
\newcommand{\hatcurCCtwomassJmageccenxxxxxB}{\ensuremath{12.518\pm0.026}} % 2MASS ORIG MAG
\newcommand{\hatcurCCtwomassHmageccenxxxxxB}{\ensuremath{12.129\pm0.023}} % 2MASS ORIG MAG
\newcommand{\hatcurCCtwomassKmageccenxxxxxB}{\ensuremath{12.037\pm0.019}} % 2MASS ORIG MAG
\newcommand{\hatcurCCcitJmageccenxxxxxB}{\ensuremath{12.529\pm0.026}}  % 2MASS CIT MAG
\newcommand{\hatcurCCcitHmageccenxxxxxB}{\ensuremath{12.123\pm0.024}}  % 2MASS CIT MAG
\newcommand{\hatcurCCcitKmageccenxxxxxB}{\ensuremath{12.061\pm0.019}}  % 2MASS CIT MAG
\newcommand{\hatcurCCbbJmageccenxxxxxB}{\ensuremath{12.587\pm0.028}}   % 2MASS BB MAG
\newcommand{\hatcurCCbbHmageccenxxxxxB}{\ensuremath{12.146\pm0.024}}   % 2MASS BB MAG
\newcommand{\hatcurCCbbKmageccenxxxxxB}{\ensuremath{12.081\pm0.019}}   % 2MASS BB MAG
\newcommand{\hatcurCCesoJmageccenxxxxxB}{\ensuremath{12.591\pm0.030}}  % 2MASS ESO MAG
\newcommand{\hatcurCCesoHmageccenxxxxxB}{\ensuremath{12.141\pm0.028}}  % 2MASS ESO MAG
\newcommand{\hatcurCCesoKmageccenxxxxxB}{\ensuremath{12.079\pm0.020}}  % 2MASS ESO MAG
\newcommand{\hatcurCCesoJHmageccenxxxxxB}{\ensuremath{0.449\pm0.039}}  % 2MASS ESO JH COLOR
\newcommand{\hatcurCCesoJKmageccenxxxxxB}{\ensuremath{0.512\pm0.035}}  % 2MASS ESO JK COLOR
\newcommand{\hatcurCCesoHKmageccenxxxxxB}{\ensuremath{0.061\pm0.034}}  % 2MASS ESO HK COLOR
\newcommand{\hatcurLCdipeccenxxxxxB}{\ensuremath{16.2}}                % BLS detected dip (mmag)
\newcommand{\hatcurLCrprstareccenxxxxxB}{\ensuremath{0.1145\pm0.0011}} % Rp/R*
\newcommand{\hatcurLCbsqeccenxxxxxB}{\ensuremath{0.034_{-0.025}^{+0.047}}} % impact parameter square
\newcommand{\hatcurLCimpeccenxxxxxB}{\ensuremath{0.184_{-0.089}^{+0.101}}} % impact parameter
\newcommand{\hatcurLCzetaeccenxxxxxB}{\ensuremath{20.319_{-0.071}^{+0.133}}} % zeta/R*
\newcommand{\hatcurLCdureccenxxxxxB}{\ensuremath{0.11015\pm0.00066}}   % transit duration (days)
\newcommand{\hatcurLCdurshorteccenxxxxxB}{\ensuremath{0.1102}}         % transit duration (days)
\newcommand{\hatcurLCdurhreccenxxxxxB}{\ensuremath{2.644\pm0.016}}     % transit duration (hours)
\newcommand{\hatcurLCdurhrshorteccenxxxxxB}{\ensuremath{2.644}}        % transit duration (hours)
\newcommand{\hatcurLCqeccenxxxxxB}{\ensuremath{0.03980\pm0.00024}}     % fractional transit duration (days)
\newcommand{\hatcurLCqshorteccenxxxxxB}{\ensuremath{0.040}}            % fractional transit duration (days)
\newcommand{\hatcurLCingdureccenxxxxxB}{\ensuremath{0.01165\pm0.00055}} % ingress/egress duration (days)
\newcommand{\hatcurLCPeccenxxxxxB}{\ensuremath{2.7667635\pm0.0000025}} % period (days)
\newcommand{\hatcurLCPprececcenxxxxxB}{\ensuremath{2.7667635}}         % period (days)
\newcommand{\hatcurLCPshorteccenxxxxxB}{\ensuremath{2.7668}}           % period (days)
\newcommand{\hatcurLCTeccenxxxxxB}{\ensuremath{2456408.76462\pm0.00020}} % epoch (BJD)
\newcommand{\hatcurLCTAeccenxxxxxB}{\ensuremath{2455091.7853\pm0.0012}} % TA (BJD)
\newcommand{\hatcurLCTBeccenxxxxxB}{\ensuremath{2456455.79959\pm0.00020}} % TB (BJD)
\newcommand{\hatcurLChatnetmeccenxxxxxB}{\ensuremath{13.77631\pm0.00010}} % HATNet OOT level
\newcommand{\hatcurLCiblendeccenxxxxxB}{\ensuremath{0.937\pm0.038}}    % HATNet iblend factor
\newcommand{\hatcurSMEiteffeccenxxxxxB}{\ensuremath{5361\pm70}}        % Ini SME, stellar effective temperature
\newcommand{\hatcurSMEizfeheccenxxxxxB}{\ensuremath{0.340\pm0.070}}    % Ini SME, stellar metallicity
\newcommand{\hatcurSMEizfehshorteccenxxxxxB}{\ensuremath{0.34}}        % Ini SME, stellar metallicity
\newcommand{\hatcurSMEiloggeccenxxxxxB}{\ensuremath{4.57\pm0.13}}      % Ini SME, stellar surface gravity
\newcommand{\hatcurSMEivsineccenxxxxxB}{\ensuremath{3.73\pm0.50}}      % Ini SME, stellar rotational velocity
\newcommand{\hatcurSMEivmaceccenxxxxxB}{\ensuremath{0.0}}              % Ini SME, stellar macroturbulence
\newcommand{\hatcurSMEivmiceccenxxxxxB}{\ensuremath{0.0}}              % Ini SME, stellar microturbulence
\newcommand{\hatcurSMEiiteffeccenxxxxxB}{\ensuremath{5346\pm60}}       % Final SME, stellar effective temperature
\newcommand{\hatcurSMEiizfeheccenxxxxxB}{\ensuremath{0.330\pm0.060}}   % Final SME, stellar metallicity
\newcommand{\hatcurSMEiizfehshorteccenxxxxxB}{\ensuremath{0.33}}       % Final SME, stellar metallicity
\newcommand{\hatcurSMEiiloggeccenxxxxxB}{\ensuremath{4.490\pm0.030}}   % Final SME, stellar surface gravity
\newcommand{\hatcurSMEiivsineccenxxxxxB}{\ensuremath{3.8\pm1.2}}       % Final SME, stellar rotational velocity
\newcommand{\hatcurLBizeccenxxxxxB}{\ensuremath{0.2739}}               % Limb darkening parameters, Gamma1, z-band
\newcommand{\hatcurLBiizeccenxxxxxB}{\ensuremath{0.3035}}              % Limb darkening parameters, Gamma2, z-band
\newcommand{\hatcurLBiieccenxxxxxB}{\ensuremath{0.3562}}               % Limb darkening parameters, Gamma1, i-band
\newcommand{\hatcurLBiiieccenxxxxxB}{\ensuremath{0.2871}}              % Limb darkening parameters, Gamma2, i-band
\newcommand{\hatcurLBiIeccenxxxxxB}{\ensuremath{0.3293}}               % Limb darkening parameters, Gamma1, I-band
\newcommand{\hatcurLBiiIeccenxxxxxB}{\ensuremath{0.2925}}              % Limb darkening parameters, Gamma2, I-band
\newcommand{\hatcurLBigeccenxxxxxB}{\ensuremath{0.7052}}               % Limb darkening parameters, Gamma1, g-band
\newcommand{\hatcurLBiigeccenxxxxxB}{\ensuremath{0.1193}}              % Limb darkening parameters, Gamma2, g-band
\newcommand{\hatcurLBireccenxxxxxB}{\ensuremath{0.4725}}               % Limb darkening parameters, Gamma1, r-band
\newcommand{\hatcurLBiireccenxxxxxB}{\ensuremath{0.2569}}              % Limb darkening parameters, Gamma2, r-band
\newcommand{\hatcurLBiReccenxxxxxB}{\ensuremath{0.4404}}               % Limb darkening parameters, Gamma1, R-band
\newcommand{\hatcurLBiiReccenxxxxxB}{\ensuremath{0.2661}}              % Limb darkening parameters, Gamma2, R-band
\newcommand{\hatcurLBikepeccenxxxxxB}{\ensuremath{0.1000}}             % Limb darkening parameters, Gamma1, Kep-band
\newcommand{\hatcurLBiikepeccenxxxxxB}{\ensuremath{0.1000}}            % Limb darkening parameters, Gamma2, Kep-band
\newcommand{\hatcurISOmeccenxxxxxB}{\ensuremath{0.967\pm0.025}}        % stellar mass
\newcommand{\hatcurISOmshorteccenxxxxxB}{\ensuremath{0.97}}            % stellar mass
\newcommand{\hatcurISOmlongeccenxxxxxB}{\ensuremath{0.967\pm0.025}}    % stellar mass
\newcommand{\hatcurISOreccenxxxxxB}{\ensuremath{0.931_{-0.038}^{+0.069}}} % stellar radius
\newcommand{\hatcurISOrshorteccenxxxxxB}{\ensuremath{0.93}}            % stellar radius
\newcommand{\hatcurISOrlongeccenxxxxxB}{\ensuremath{0.931_{-0.038}^{+0.069}}} % stellar radius
\newcommand{\hatcurISOrhoeccenxxxxxB}{\ensuremath{1.70_{-0.36}^{+0.21}}} % stellar density (cgs)
\newcommand{\hatcurISOrholongeccenxxxxxB}{\ensuremath{1.70_{-0.36}^{+0.21}}} % stellar density (cgs)
\newcommand{\hatcurISOloggeccenxxxxxB}{\ensuremath{4.487\pm0.058}}     % stellar surface gravity from isochrones
\newcommand{\hatcurISOlumeccenxxxxxB}{\ensuremath{0.636_{-0.068}^{+0.103}}} % stellar luminosity
\newcommand{\hatcurISOlumshorteccenxxxxxB}{\ensuremath{0.64}}          % stellar luminosity
\newcommand{\hatcurISOmveccenxxxxxB}{\ensuremath{5.41\pm0.16}}         % stellar absolute magnitude
\newcommand{\hatcurISOvieccenxxxxxB}{\ensuremath{0.835\pm0.016}}       % stellar V-I index
\newcommand{\hatcurISOageeccenxxxxxB}{\ensuremath{4.7_{-2.8}^{+3.8}}}  % stellar age
\newcommand{\hatcurISOsigmaeccenxxxxxB}{\ensuremath{0.00190\pm0.00038}} % system mass-correction sigma parameter
\newcommand{\hatcurISOMJeccenxxxxxB}{\ensuremath{4.02\pm0.15}}         % stellar absolute J magnitude
\newcommand{\hatcurISOMHeccenxxxxxB}{\ensuremath{3.60\pm0.14}}         % stellar absolute H magnitude
\newcommand{\hatcurISOMKeccenxxxxxB}{\ensuremath{3.53\pm0.14}}         % stellar absolute K magnitude
\newcommand{\hatcurISOJKeccenxxxxxB}{\ensuremath{0.490\pm0.010}}       % J-K color index from isochrones.
\newcommand{\hatcurISOspececcenxxxxxB}{G}                              % stellar spectral type
\newcommand{\hatcurRVKeccenxxxxxB}{\ensuremath{162\pm12}}              % RV semi-amplitude [m/s]
\newcommand{\hatcurRVrkeccenxxxxxB}{\ensuremath{0.01\pm0.13}}          % sqrt(e)*cos(omega)
\newcommand{\hatcurRVrheccenxxxxxB}{\ensuremath{-0.01\pm0.19}}         % sqrt(e)*sin(omega)
\newcommand{\hatcurRVkeccenxxxxxB}{\ensuremath{0.001\pm0.033}}         % e*cos(omega)
\newcommand{\hatcurRVheccenxxxxxB}{\ensuremath{-0.001_{-0.046}^{+0.066}}} % e*sin(omega)
\newcommand{\hatcurRVtroneeccenxxxxxB}{\ensuremath{0\pm0}}             % RV linear trend tr1 factor
\newcommand{\hatcurRVtrtwoeccenxxxxxB}{\ensuremath{0\pm0}}             % RV linear trend tr2 factor
\newcommand{\hatcurRVgammaAeccenxxxxxB}{\ensuremath{30192\pm21}}       % RV gamma velocity, relative scale
\newcommand{\hatcurRVjitterAeccenxxxxxB}{\ensuremath{0.00\pm0.58}}     % RV jitter (m/s)
\newcommand{\hatcurRVfitrmsAeccenxxxxxB}{\ensuremath{0.0}}             % RVfitrms
\newcommand{\hatcurRVgammaBeccenxxxxxB}{\ensuremath{30190\pm11}}       % RV gamma velocity, relative scale
\newcommand{\hatcurRVjitterBeccenxxxxxB}{\ensuremath{0.0\pm1.2}}       % RV jitter (m/s)
\newcommand{\hatcurRVfitrmsBeccenxxxxxB}{\ensuremath{0.0}}             % RVfitrms
\newcommand{\hatcurRVecceneccenxxxxxB}{\ensuremath{0.044\pm0.045}}     % eccentricity
\newcommand{\hatcurRVeccentwosiglimeccenxxxxxB}{\ensuremath{<0.142}}   % eccentricity
\newcommand{\hatcurRVomegaeccenxxxxxB}{\ensuremath{180\pm100}}         % argument of pericenter
\newcommand{\hatcurPPieccenxxxxxB}{\ensuremath{88.88_{-0.85}^{+0.46}}} % orbital inclination
\newcommand{\hatcurPPgeccenxxxxxB}{\ensuremath{25.1_{-4.2}^{+2.6}}}    % planetary surface gravity (m/s^2)
\newcommand{\hatcurPPloggeccenxxxxxB}{\ensuremath{3.400_{-0.080}^{+0.042}}} % planetary surface gravity (log cgs)
\newcommand{\hatcurPPareccenxxxxxB}{\ensuremath{8.83_{-0.67}^{+0.35}}} % relative orbital radius (a/R*)
\newcommand{\hatcurPPareleccenxxxxxB}{\ensuremath{0.03814\pm0.00033}}  % semimajor axis (AU)
\newcommand{\hatcurPPrhoeccenxxxxxB}{\ensuremath{1.22_{-0.27}^{+0.17}}} % planetary density (cgs)
\newcommand{\hatcurPPmeccenxxxxxB}{\ensuremath{1.087\pm0.083}}         % planetary mass (M_jup)
\newcommand{\hatcurPPmshorteccenxxxxxB}{\ensuremath{1.09}}             % planetary mass (M_jup)
\newcommand{\hatcurPPmlongeccenxxxxxB}{\ensuremath{1.087\pm0.083}}     % planetary mass (M_jup)
\newcommand{\hatcurPPmeeccenxxxxxB}{\ensuremath{345\pm26}}             % planetary mass (M_earth)
\newcommand{\hatcurPPmeshorteccenxxxxxB}{\ensuremath{345.4}}           % planetary mass (M_earth)
\newcommand{\hatcurPPmelongeccenxxxxxB}{\ensuremath{345\pm26}}         % planetary mass (M_earth)
\newcommand{\hatcurPPreccenxxxxxB}{\ensuremath{1.039_{-0.047}^{+0.079}}} % planetary radius (R_jup)
\newcommand{\hatcurPPrshorteccenxxxxxB}{\ensuremath{1.04}}             % planetary radius (R_jup)
\newcommand{\hatcurPPrlongeccenxxxxxB}{\ensuremath{1.039_{-0.047}^{+0.079}}} % planetary radius (R_jup)
\newcommand{\hatcurPPreeccenxxxxxB}{\ensuremath{11.64_{-0.53}^{+0.89}}} % planetary radius (R_earth)
\newcommand{\hatcurPPreshorteccenxxxxxB}{\ensuremath{11.6}}            % planetary radius (R_earth)
\newcommand{\hatcurPPrelongeccenxxxxxB}{\ensuremath{11.64_{-0.53}^{+0.89}}} % planetary radius (R_earth)
\newcommand{\hatcurPPmrcorreccenxxxxxB}{\ensuremath{0.21}}             % mass/radius correlation
\newcommand{\hatcurPPteffeccenxxxxxB}{\ensuremath{1275_{-34}^{+51}}}   % planetary temperature (K)
\newcommand{\hatcurPPthetaeccenxxxxxB}{\ensuremath{0.0820_{-0.0085}^{+0.0065}}} % Safranov number
\newcommand{\hatcurPPfluxperieccenxxxxxB}{\ensuremath{6.33_{-0.49}^{+1.76}}} % flux @ periastron (CGS)
\newcommand{\hatcurPPfluxperidimeccenxxxxxB}{\ensuremath{8}}           % flux @ periastron (CGS) units.
\newcommand{\hatcurPPfluxapeccenxxxxxB}{\ensuremath{5.63_{-0.85}^{+0.53}}} % flux @ apastron (CGS)
\newcommand{\hatcurPPfluxapdimeccenxxxxxB}{\ensuremath{8}}             % flux @ apastron (CGS) units.
\newcommand{\hatcurPPfluxavgeccenxxxxxB}{\ensuremath{5.96_{-0.61}^{+1.01}}} % flux on average (CGS)
\newcommand{\hatcurPPfluxavgdimeccenxxxxxB}{\ensuremath{8}}            % flux average (CGS) units.
\newcommand{\hatcurPPfluxavglogeccenxxxxxB}{\ensuremath{8.775_{-0.047}^{+0.068}}} % log10 flux on average (CGS)
\newcommand{\hatcurXsecphaseeccenxxxxxB}{\ensuremath{0.500\pm0.021}}   % Phase of secondary eclipse
\newcommand{\hatcurXsecondaryeccenxxxxxB}{\ensuremath{2456410.149\pm0.059}} % Secondary eclipse epoch
\newcommand{\hatcurXsecdureccenxxxxxB}{\ensuremath{0.110\pm0.014}}     % sec eclipse duration (days)
\newcommand{\hatcurXsecingdureccenxxxxxB}{\ensuremath{0.0116\pm0.0019}} % sec I/E duration (days)
\newcommand{\hatcurPPphiconjeccenxxxxxB}{\ensuremath{0.02\pm0.30}}     % phase diff between conjunction and periastron
\newcommand{\hatcurPPperieccenxxxxxB}{\ensuremath{2456408.72\pm0.82}}  % time of periastron passage.
\newcommand{\hatcurPPaequiveccenxxxxxB}{\ensuremath{0.0479_{-0.0036}^{+0.0027}}} % equivalent semi-major axis
\newcommand{\hatcurPPtcirceccenxxxxxB}{\ensuremath{222_{-81}^{+49}}}   % circularization timescale
\newcommand{\hatcurPPtinfalleccenxxxxxB}{\ensuremath{1210\pm370}}      % infall timescale
\newcommand{\hatcurXdisteccenxxxxxB}{\ensuremath{512_{-24}^{+38}}}     % distance (pc), no reddenning correction
\newcommand{\hatcurXAveccenxxxxxB}{\ensuremath{0.000\pm0.031}}         % Av (mag)
\newcommand{\hatcurXdistredeccenxxxxxB}{\ensuremath{512_{-24}^{+38}}}  % distance with Av correction (pc)
\newcommand{\hatcurXEBVeccenxxxxxB}{\ensuremath{0.000\pm0.010}}        % E(B-V) (mag)
\newcommand{\hatcurXmvisoredeccenxxxxxB}{\ensuremath{13.968\pm0.046}}  % Expected m_v with reddening (mag)
\newcommand{\hatcurXmiisoredeccenxxxxxB}{\ensuremath{13.128\pm0.031}}  % Expected m_i with reddening (mag)
\newcommand{\hatcurXmjisoredeccenxxxxxB}{\ensuremath{12.570\pm0.017}}  % Expected m_j with reddening (mag)
\newcommand{\hatcurXmhisoredeccenxxxxxB}{\ensuremath{12.148\pm0.014}}  % Expected m_h with reddening (mag)
\newcommand{\hatcurXmkisoredeccenxxxxxB}{\ensuremath{12.077\pm0.015}}  % Expected m_k with reddening (mag)
\newcommand{\hatcurXviisoredeccenxxxxxB}{\ensuremath{0.838\pm0.019}}   % Expected V-I with reddening (mag)
\newcommand{\hatcurXvkisoredeccenxxxxxB}{\ensuremath{1.891\pm0.048}}   % Expected V-K with reddening (mag)
\newcommand{\hatcurXjhisoredeccenxxxxxB}{\ensuremath{0.422\pm0.012}}   % Expected J-H with reddening (mag)
\newcommand{\hatcurXjkisoredeccenxxxxxB}{\ensuremath{0.493\pm0.014}}   % Expected J-K with reddening (mag)
\newcommand{\hatcurCCpmraeccenxxxxxB}{\ensuremath{0.4\pm1.5}}          % proper motion, in RA
\newcommand{\hatcurCCpmdececcenxxxxxB}{\ensuremath{-8.8\pm1.3}}        % proper motion, in DEC
\newcommand{\hatcurCCpmeccenxxxxxB}{\ensuremath{8.8\pm2.0}}            % proper motion

\newcommand{\hatcurANUWIFESgammaeccen}[1]{\ifnum#1=13 %
\hatcurANUWIFESgammaeccenxxxxxA
\else
??????\fi
}
\newcommand{\hatcurANUWIFESloggeccen}[1]{\ifnum#1=13 %
\hatcurANUWIFESloggeccenxxxxxA
\else
??????\fi
}
\newcommand{\hatcurANUWIFESnumspececcen}[1]{\ifnum#1=13 %
\hatcurANUWIFESnumspececcenxxxxxA
\else
??????\fi
}
\newcommand{\hatcurANUWIFESrvrmseccen}[1]{\ifnum#1=13 %
\hatcurANUWIFESrvrmseccenxxxxxA
\else
??????\fi
}
\newcommand{\hatcurANUWIFESspaneccen}[1]{\ifnum#1=13 %
\hatcurANUWIFESspaneccenxxxxxA
\else
??????\fi
}
\newcommand{\hatcurANUWIFESteffeccen}[1]{\ifnum#1=13 %
\hatcurANUWIFESteffeccenxxxxxA
\else
??????\fi
}
\newcommand{\hatcurANUWIFESvsinieccen}[1]{\ifnum#1=13 %
\hatcurANUWIFESvsinieccenxxxxxA
\else
??????\fi
}
\newcommand{\hatcurANUWIFESzfeheccen}[1]{\ifnum#1=13 %
\hatcurANUWIFESzfeheccenxxxxxA
\else
??????\fi
}
\newcommand{\hatcurCCbbHmageccen}[1]{\ifnum#1=13 %
\hatcurCCbbHmageccenxxxxxA
\else
\ifnum#1=14 %
\hatcurCCbbHmageccenxxxxxB
\else
??????\fi
\fi
}
\newcommand{\hatcurCCbbJmageccen}[1]{\ifnum#1=13 %
\hatcurCCbbJmageccenxxxxxA
\else
\ifnum#1=14 %
\hatcurCCbbJmageccenxxxxxB
\else
??????\fi
\fi
}
\newcommand{\hatcurCCbbKmageccen}[1]{\ifnum#1=13 %
\hatcurCCbbKmageccenxxxxxA
\else
\ifnum#1=14 %
\hatcurCCbbKmageccenxxxxxB
\else
??????\fi
\fi
}
\newcommand{\hatcurCCcitHmageccen}[1]{\ifnum#1=13 %
\hatcurCCcitHmageccenxxxxxA
\else
\ifnum#1=14 %
\hatcurCCcitHmageccenxxxxxB
\else
??????\fi
\fi
}
\newcommand{\hatcurCCcitJmageccen}[1]{\ifnum#1=13 %
\hatcurCCcitJmageccenxxxxxA
\else
\ifnum#1=14 %
\hatcurCCcitJmageccenxxxxxB
\else
??????\fi
\fi
}
\newcommand{\hatcurCCcitKmageccen}[1]{\ifnum#1=13 %
\hatcurCCcitKmageccenxxxxxA
\else
\ifnum#1=14 %
\hatcurCCcitKmageccenxxxxxB
\else
??????\fi
\fi
}
\newcommand{\hatcurCCdececcen}[1]{\ifnum#1=13 %
\hatcurCCdececcenxxxxxA
\else
\ifnum#1=14 %
\hatcurCCdececcenxxxxxB
\else
??????\fi
\fi
}
\newcommand{\hatcurCCesoHKmageccen}[1]{\ifnum#1=13 %
\hatcurCCesoHKmageccenxxxxxA
\else
\ifnum#1=14 %
\hatcurCCesoHKmageccenxxxxxB
\else
??????\fi
\fi
}
\newcommand{\hatcurCCesoHmageccen}[1]{\ifnum#1=13 %
\hatcurCCesoHmageccenxxxxxA
\else
\ifnum#1=14 %
\hatcurCCesoHmageccenxxxxxB
\else
??????\fi
\fi
}
\newcommand{\hatcurCCesoJHmageccen}[1]{\ifnum#1=13 %
\hatcurCCesoJHmageccenxxxxxA
\else
\ifnum#1=14 %
\hatcurCCesoJHmageccenxxxxxB
\else
??????\fi
\fi
}
\newcommand{\hatcurCCesoJKmageccen}[1]{\ifnum#1=13 %
\hatcurCCesoJKmageccenxxxxxA
\else
\ifnum#1=14 %
\hatcurCCesoJKmageccenxxxxxB
\else
??????\fi
\fi
}
\newcommand{\hatcurCCesoJmageccen}[1]{\ifnum#1=13 %
\hatcurCCesoJmageccenxxxxxA
\else
\ifnum#1=14 %
\hatcurCCesoJmageccenxxxxxB
\else
??????\fi
\fi
}
\newcommand{\hatcurCCesoKmageccen}[1]{\ifnum#1=13 %
\hatcurCCesoKmageccenxxxxxA
\else
\ifnum#1=14 %
\hatcurCCesoKmageccenxxxxxB
\else
??????\fi
\fi
}
\newcommand{\hatcurCCgsceccen}[1]{\ifnum#1=13 %
\hatcurCCgsceccenxxxxxA
\else
\ifnum#1=14 %
\hatcurCCgsceccenxxxxxB
\else
??????\fi
\fi
}
\newcommand{\hatcurCCmageccen}[1]{\ifnum#1=13 %
\hatcurCCmageccenxxxxxA
\else
\ifnum#1=14 %
\hatcurCCmageccenxxxxxB
\else
??????\fi
\fi
}
\newcommand{\hatcurCCpmeccen}[1]{\ifnum#1=13 %
\hatcurCCpmeccenxxxxxA
\else
\ifnum#1=14 %
\hatcurCCpmeccenxxxxxB
\else
??????\fi
\fi
}
\newcommand{\hatcurCCpmdececcen}[1]{\ifnum#1=13 %
\hatcurCCpmdececcenxxxxxA
\else
\ifnum#1=14 %
\hatcurCCpmdececcenxxxxxB
\else
??????\fi
\fi
}
\newcommand{\hatcurCCpmraeccen}[1]{\ifnum#1=13 %
\hatcurCCpmraeccenxxxxxA
\else
\ifnum#1=14 %
\hatcurCCpmraeccenxxxxxB
\else
??????\fi
\fi
}
\newcommand{\hatcurCCraeccen}[1]{\ifnum#1=13 %
\hatcurCCraeccenxxxxxA
\else
\ifnum#1=14 %
\hatcurCCraeccenxxxxxB
\else
??????\fi
\fi
}
\newcommand{\hatcurCCtassmBeccen}[1]{\ifnum#1=13 %
\hatcurCCtassmBeccenxxxxxA
\else
\ifnum#1=14 %
\hatcurCCtassmBeccenxxxxxB
\else
??????\fi
\fi
}
\newcommand{\hatcurCCtassmBshorteccen}[1]{\ifnum#1=13 %
\hatcurCCtassmBshorteccenxxxxxA
\else
\ifnum#1=14 %
\hatcurCCtassmBshorteccenxxxxxB
\else
??????\fi
\fi
}
\newcommand{\hatcurCCtassmgeccen}[1]{\ifnum#1=13 %
\hatcurCCtassmgeccenxxxxxA
\else
\ifnum#1=14 %
\hatcurCCtassmgeccenxxxxxB
\else
??????\fi
\fi
}
\newcommand{\hatcurCCtassmgshorteccen}[1]{\ifnum#1=13 %
\hatcurCCtassmgshorteccenxxxxxA
\else
\ifnum#1=14 %
\hatcurCCtassmgshorteccenxxxxxB
\else
??????\fi
\fi
}
\newcommand{\hatcurCCtassmieccen}[1]{\ifnum#1=13 %
\hatcurCCtassmieccenxxxxxA
\else
\ifnum#1=14 %
\hatcurCCtassmieccenxxxxxB
\else
??????\fi
\fi
}
\newcommand{\hatcurCCtassmIeccen}[1]{\ifnum#1=13 %
\hatcurCCtassmIeccenxxxxxA
\else
\ifnum#1=14 %
\hatcurCCtassmIeccenxxxxxB
\else
??????\fi
\fi
}
\newcommand{\hatcurCCtassmishorteccen}[1]{\ifnum#1=13 %
\hatcurCCtassmishorteccenxxxxxA
\else
\ifnum#1=14 %
\hatcurCCtassmishorteccenxxxxxB
\else
??????\fi
\fi
}
\newcommand{\hatcurCCtassmIshorteccen}[1]{\ifnum#1=13 %
\hatcurCCtassmIshorteccenxxxxxA
\else
\ifnum#1=14 %
\hatcurCCtassmIshorteccenxxxxxB
\else
??????\fi
\fi
}
\newcommand{\hatcurCCtassmreccen}[1]{\ifnum#1=13 %
\hatcurCCtassmreccenxxxxxA
\else
\ifnum#1=14 %
\hatcurCCtassmreccenxxxxxB
\else
??????\fi
\fi
}
\newcommand{\hatcurCCtassmrshorteccen}[1]{\ifnum#1=13 %
\hatcurCCtassmrshorteccenxxxxxA
\else
\ifnum#1=14 %
\hatcurCCtassmrshorteccenxxxxxB
\else
??????\fi
\fi
}
\newcommand{\hatcurCCtassmveccen}[1]{\ifnum#1=13 %
\hatcurCCtassmveccenxxxxxA
\else
\ifnum#1=14 %
\hatcurCCtassmveccenxxxxxB
\else
??????\fi
\fi
}
\newcommand{\hatcurCCtassmvshorteccen}[1]{\ifnum#1=13 %
\hatcurCCtassmvshorteccenxxxxxA
\else
\ifnum#1=14 %
\hatcurCCtassmvshorteccenxxxxxB
\else
??????\fi
\fi
}
\newcommand{\hatcurCCtwomasseccen}[1]{\ifnum#1=13 %
\hatcurCCtwomasseccenxxxxxA
\else
\ifnum#1=14 %
\hatcurCCtwomasseccenxxxxxB
\else
??????\fi
\fi
}
\newcommand{\hatcurCCtwomassHmageccen}[1]{\ifnum#1=13 %
\hatcurCCtwomassHmageccenxxxxxA
\else
\ifnum#1=14 %
\hatcurCCtwomassHmageccenxxxxxB
\else
??????\fi
\fi
}
\newcommand{\hatcurCCtwomassJmageccen}[1]{\ifnum#1=13 %
\hatcurCCtwomassJmageccenxxxxxA
\else
\ifnum#1=14 %
\hatcurCCtwomassJmageccenxxxxxB
\else
??????\fi
\fi
}
\newcommand{\hatcurCCtwomassKmageccen}[1]{\ifnum#1=13 %
\hatcurCCtwomassKmageccenxxxxxA
\else
\ifnum#1=14 %
\hatcurCCtwomassKmageccenxxxxxB
\else
??????\fi
\fi
}
\newcommand{\hatcurfieldeccen}[1]{\ifnum#1=13 %
\hatcurfieldeccenxxxxxA
\else
\ifnum#1=14 %
\hatcurfieldeccenxxxxxB
\else
??????\fi
\fi
}
\newcommand{\hatcurhtreccen}[1]{\ifnum#1=13 %
\hatcurhtreccenxxxxxA
\else
\ifnum#1=14 %
\hatcurhtreccenxxxxxB
\else
??????\fi
\fi
}
\newcommand{\hatcurISOageeccen}[1]{\ifnum#1=13 %
\hatcurISOageeccenxxxxxA
\else
\ifnum#1=14 %
\hatcurISOageeccenxxxxxB
\else
??????\fi
\fi
}
\newcommand{\hatcurISOJKeccen}[1]{\ifnum#1=13 %
\hatcurISOJKeccenxxxxxA
\else
\ifnum#1=14 %
\hatcurISOJKeccenxxxxxB
\else
??????\fi
\fi
}
\newcommand{\hatcurISOloggeccen}[1]{\ifnum#1=13 %
\hatcurISOloggeccenxxxxxA
\else
\ifnum#1=14 %
\hatcurISOloggeccenxxxxxB
\else
??????\fi
\fi
}
\newcommand{\hatcurISOlumeccen}[1]{\ifnum#1=13 %
\hatcurISOlumeccenxxxxxA
\else
\ifnum#1=14 %
\hatcurISOlumeccenxxxxxB
\else
??????\fi
\fi
}
\newcommand{\hatcurISOlumshorteccen}[1]{\ifnum#1=13 %
\hatcurISOlumshorteccenxxxxxA
\else
\ifnum#1=14 %
\hatcurISOlumshorteccenxxxxxB
\else
??????\fi
\fi
}
\newcommand{\hatcurISOmeccen}[1]{\ifnum#1=13 %
\hatcurISOmeccenxxxxxA
\else
\ifnum#1=14 %
\hatcurISOmeccenxxxxxB
\else
??????\fi
\fi
}
\newcommand{\hatcurISOMHeccen}[1]{\ifnum#1=13 %
\hatcurISOMHeccenxxxxxA
\else
\ifnum#1=14 %
\hatcurISOMHeccenxxxxxB
\else
??????\fi
\fi
}
\newcommand{\hatcurISOMJeccen}[1]{\ifnum#1=13 %
\hatcurISOMJeccenxxxxxA
\else
\ifnum#1=14 %
\hatcurISOMJeccenxxxxxB
\else
??????\fi
\fi
}
\newcommand{\hatcurISOMKeccen}[1]{\ifnum#1=13 %
\hatcurISOMKeccenxxxxxA
\else
\ifnum#1=14 %
\hatcurISOMKeccenxxxxxB
\else
??????\fi
\fi
}
\newcommand{\hatcurISOmlongeccen}[1]{\ifnum#1=13 %
\hatcurISOmlongeccenxxxxxA
\else
\ifnum#1=14 %
\hatcurISOmlongeccenxxxxxB
\else
??????\fi
\fi
}
\newcommand{\hatcurISOmshorteccen}[1]{\ifnum#1=13 %
\hatcurISOmshorteccenxxxxxA
\else
\ifnum#1=14 %
\hatcurISOmshorteccenxxxxxB
\else
??????\fi
\fi
}
\newcommand{\hatcurISOmveccen}[1]{\ifnum#1=13 %
\hatcurISOmveccenxxxxxA
\else
\ifnum#1=14 %
\hatcurISOmveccenxxxxxB
\else
??????\fi
\fi
}
\newcommand{\hatcurISOreccen}[1]{\ifnum#1=13 %
\hatcurISOreccenxxxxxA
\else
\ifnum#1=14 %
\hatcurISOreccenxxxxxB
\else
??????\fi
\fi
}
\newcommand{\hatcurISOrhoeccen}[1]{\ifnum#1=13 %
\hatcurISOrhoeccenxxxxxA
\else
\ifnum#1=14 %
\hatcurISOrhoeccenxxxxxB
\else
??????\fi
\fi
}
\newcommand{\hatcurISOrholongeccen}[1]{\ifnum#1=13 %
\hatcurISOrholongeccenxxxxxA
\else
\ifnum#1=14 %
\hatcurISOrholongeccenxxxxxB
\else
??????\fi
\fi
}
\newcommand{\hatcurISOrlongeccen}[1]{\ifnum#1=13 %
\hatcurISOrlongeccenxxxxxA
\else
\ifnum#1=14 %
\hatcurISOrlongeccenxxxxxB
\else
??????\fi
\fi
}
\newcommand{\hatcurISOrshorteccen}[1]{\ifnum#1=13 %
\hatcurISOrshorteccenxxxxxA
\else
\ifnum#1=14 %
\hatcurISOrshorteccenxxxxxB
\else
??????\fi
\fi
}
\newcommand{\hatcurISOsigmaeccen}[1]{\ifnum#1=13 %
\hatcurISOsigmaeccenxxxxxA
\else
\ifnum#1=14 %
\hatcurISOsigmaeccenxxxxxB
\else
??????\fi
\fi
}
\newcommand{\hatcurISOspececcen}[1]{\ifnum#1=13 %
\hatcurISOspececcenxxxxxA
\else
\ifnum#1=14 %
\hatcurISOspececcenxxxxxB
\else
??????\fi
\fi
}
\newcommand{\hatcurISOvieccen}[1]{\ifnum#1=13 %
\hatcurISOvieccenxxxxxA
\else
\ifnum#1=14 %
\hatcurISOvieccenxxxxxB
\else
??????\fi
\fi
}
\newcommand{\hatcurLBigeccen}[1]{\ifnum#1=13 %
\hatcurLBigeccenxxxxxA
\else
\ifnum#1=14 %
\hatcurLBigeccenxxxxxB
\else
??????\fi
\fi
}
\newcommand{\hatcurLBiieccen}[1]{\ifnum#1=13 %
\hatcurLBiieccenxxxxxA
\else
\ifnum#1=14 %
\hatcurLBiieccenxxxxxB
\else
??????\fi
\fi
}
\newcommand{\hatcurLBiIeccen}[1]{\ifnum#1=13 %
\hatcurLBiIeccenxxxxxA
\else
\ifnum#1=14 %
\hatcurLBiIeccenxxxxxB
\else
??????\fi
\fi
}
\newcommand{\hatcurLBiigeccen}[1]{\ifnum#1=13 %
\hatcurLBiigeccenxxxxxA
\else
\ifnum#1=14 %
\hatcurLBiigeccenxxxxxB
\else
??????\fi
\fi
}
\newcommand{\hatcurLBiiieccen}[1]{\ifnum#1=13 %
\hatcurLBiiieccenxxxxxA
\else
\ifnum#1=14 %
\hatcurLBiiieccenxxxxxB
\else
??????\fi
\fi
}
\newcommand{\hatcurLBiiIeccen}[1]{\ifnum#1=13 %
\hatcurLBiiIeccenxxxxxA
\else
\ifnum#1=14 %
\hatcurLBiiIeccenxxxxxB
\else
??????\fi
\fi
}
\newcommand{\hatcurLBiikepeccen}[1]{\ifnum#1=14 %
\hatcurLBiikepeccenxxxxxB
\else
??????\fi
}
\newcommand{\hatcurLBiireccen}[1]{\ifnum#1=13 %
\hatcurLBiireccenxxxxxA
\else
\ifnum#1=14 %
\hatcurLBiireccenxxxxxB
\else
??????\fi
\fi
}
\newcommand{\hatcurLBiiReccen}[1]{\ifnum#1=13 %
\hatcurLBiiReccenxxxxxA
\else
\ifnum#1=14 %
\hatcurLBiiReccenxxxxxB
\else
??????\fi
\fi
}
\newcommand{\hatcurLBiizeccen}[1]{\ifnum#1=13 %
\hatcurLBiizeccenxxxxxA
\else
\ifnum#1=14 %
\hatcurLBiizeccenxxxxxB
\else
??????\fi
\fi
}
\newcommand{\hatcurLBikepeccen}[1]{\ifnum#1=14 %
\hatcurLBikepeccenxxxxxB
\else
??????\fi
}
\newcommand{\hatcurLBireccen}[1]{\ifnum#1=13 %
\hatcurLBireccenxxxxxA
\else
\ifnum#1=14 %
\hatcurLBireccenxxxxxB
\else
??????\fi
\fi
}
\newcommand{\hatcurLBiReccen}[1]{\ifnum#1=13 %
\hatcurLBiReccenxxxxxA
\else
\ifnum#1=14 %
\hatcurLBiReccenxxxxxB
\else
??????\fi
\fi
}
\newcommand{\hatcurLBizeccen}[1]{\ifnum#1=13 %
\hatcurLBizeccenxxxxxA
\else
\ifnum#1=14 %
\hatcurLBizeccenxxxxxB
\else
??????\fi
\fi
}
\newcommand{\hatcurLCbsqeccen}[1]{\ifnum#1=13 %
\hatcurLCbsqeccenxxxxxA
\else
\ifnum#1=14 %
\hatcurLCbsqeccenxxxxxB
\else
??????\fi
\fi
}
\newcommand{\hatcurLCdipeccen}[1]{\ifnum#1=13 %
\hatcurLCdipeccenxxxxxA
\else
\ifnum#1=14 %
\hatcurLCdipeccenxxxxxB
\else
??????\fi
\fi
}
\newcommand{\hatcurLCdureccen}[1]{\ifnum#1=13 %
\hatcurLCdureccenxxxxxA
\else
\ifnum#1=14 %
\hatcurLCdureccenxxxxxB
\else
??????\fi
\fi
}
\newcommand{\hatcurLCdurhreccen}[1]{\ifnum#1=13 %
\hatcurLCdurhreccenxxxxxA
\else
\ifnum#1=14 %
\hatcurLCdurhreccenxxxxxB
\else
??????\fi
\fi
}
\newcommand{\hatcurLCdurhrshorteccen}[1]{\ifnum#1=13 %
\hatcurLCdurhrshorteccenxxxxxA
\else
\ifnum#1=14 %
\hatcurLCdurhrshorteccenxxxxxB
\else
??????\fi
\fi
}
\newcommand{\hatcurLCdurshorteccen}[1]{\ifnum#1=13 %
\hatcurLCdurshorteccenxxxxxA
\else
\ifnum#1=14 %
\hatcurLCdurshorteccenxxxxxB
\else
??????\fi
\fi
}
\newcommand{\hatcurLChatnetmeccen}[1]{\ifnum#1=13 %
\hatcurLChatnetmeccenxxxxxA
\else
\ifnum#1=14 %
\hatcurLChatnetmeccenxxxxxB
\else
??????\fi
\fi
}
\newcommand{\hatcurLCiblendeccen}[1]{\ifnum#1=13 %
\hatcurLCiblendeccenxxxxxA
\else
\ifnum#1=14 %
\hatcurLCiblendeccenxxxxxB
\else
??????\fi
\fi
}
\newcommand{\hatcurLCimpeccen}[1]{\ifnum#1=13 %
\hatcurLCimpeccenxxxxxA
\else
\ifnum#1=14 %
\hatcurLCimpeccenxxxxxB
\else
??????\fi
\fi
}
\newcommand{\hatcurLCingdureccen}[1]{\ifnum#1=13 %
\hatcurLCingdureccenxxxxxA
\else
\ifnum#1=14 %
\hatcurLCingdureccenxxxxxB
\else
??????\fi
\fi
}
\newcommand{\hatcurLCPeccen}[1]{\ifnum#1=13 %
\hatcurLCPeccenxxxxxA
\else
\ifnum#1=14 %
\hatcurLCPeccenxxxxxB
\else
??????\fi
\fi
}
\newcommand{\hatcurLCPprececcen}[1]{\ifnum#1=13 %
\hatcurLCPprececcenxxxxxA
\else
\ifnum#1=14 %
\hatcurLCPprececcenxxxxxB
\else
??????\fi
\fi
}
\newcommand{\hatcurLCPshorteccen}[1]{\ifnum#1=13 %
\hatcurLCPshorteccenxxxxxA
\else
\ifnum#1=14 %
\hatcurLCPshorteccenxxxxxB
\else
??????\fi
\fi
}
\newcommand{\hatcurLCqeccen}[1]{\ifnum#1=13 %
\hatcurLCqeccenxxxxxA
\else
\ifnum#1=14 %
\hatcurLCqeccenxxxxxB
\else
??????\fi
\fi
}
\newcommand{\hatcurLCqshorteccen}[1]{\ifnum#1=13 %
\hatcurLCqshorteccenxxxxxA
\else
\ifnum#1=14 %
\hatcurLCqshorteccenxxxxxB
\else
??????\fi
\fi
}
\newcommand{\hatcurLCrprstareccen}[1]{\ifnum#1=13 %
\hatcurLCrprstareccenxxxxxA
\else
\ifnum#1=14 %
\hatcurLCrprstareccenxxxxxB
\else
??????\fi
\fi
}
\newcommand{\hatcurLCTeccen}[1]{\ifnum#1=13 %
\hatcurLCTeccenxxxxxA
\else
\ifnum#1=14 %
\hatcurLCTeccenxxxxxB
\else
??????\fi
\fi
}
\newcommand{\hatcurLCTAeccen}[1]{\ifnum#1=13 %
\hatcurLCTAeccenxxxxxA
\else
\ifnum#1=14 %
\hatcurLCTAeccenxxxxxB
\else
??????\fi
\fi
}
\newcommand{\hatcurLCTBeccen}[1]{\ifnum#1=13 %
\hatcurLCTBeccenxxxxxA
\else
\ifnum#1=14 %
\hatcurLCTBeccenxxxxxB
\else
??????\fi
\fi
}
\newcommand{\hatcurLCzetaeccen}[1]{\ifnum#1=13 %
\hatcurLCzetaeccenxxxxxA
\else
\ifnum#1=14 %
\hatcurLCzetaeccenxxxxxB
\else
??????\fi
\fi
}
\newcommand{\hatcurPPaequiveccen}[1]{\ifnum#1=13 %
\hatcurPPaequiveccenxxxxxA
\else
\ifnum#1=14 %
\hatcurPPaequiveccenxxxxxB
\else
??????\fi
\fi
}
\newcommand{\hatcurPPareccen}[1]{\ifnum#1=13 %
\hatcurPPareccenxxxxxA
\else
\ifnum#1=14 %
\hatcurPPareccenxxxxxB
\else
??????\fi
\fi
}
\newcommand{\hatcurPPareleccen}[1]{\ifnum#1=13 %
\hatcurPPareleccenxxxxxA
\else
\ifnum#1=14 %
\hatcurPPareleccenxxxxxB
\else
??????\fi
\fi
}
\newcommand{\hatcurPPfluxapeccen}[1]{\ifnum#1=13 %
\hatcurPPfluxapeccenxxxxxA
\else
\ifnum#1=14 %
\hatcurPPfluxapeccenxxxxxB
\else
??????\fi
\fi
}
\newcommand{\hatcurPPfluxapdimeccen}[1]{\ifnum#1=13 %
\hatcurPPfluxapdimeccenxxxxxA
\else
\ifnum#1=14 %
\hatcurPPfluxapdimeccenxxxxxB
\else
??????\fi
\fi
}
\newcommand{\hatcurPPfluxavgeccen}[1]{\ifnum#1=13 %
\hatcurPPfluxavgeccenxxxxxA
\else
\ifnum#1=14 %
\hatcurPPfluxavgeccenxxxxxB
\else
??????\fi
\fi
}
\newcommand{\hatcurPPfluxavgdimeccen}[1]{\ifnum#1=13 %
\hatcurPPfluxavgdimeccenxxxxxA
\else
\ifnum#1=14 %
\hatcurPPfluxavgdimeccenxxxxxB
\else
??????\fi
\fi
}
\newcommand{\hatcurPPfluxavglogeccen}[1]{\ifnum#1=13 %
\hatcurPPfluxavglogeccenxxxxxA
\else
\ifnum#1=14 %
\hatcurPPfluxavglogeccenxxxxxB
\else
??????\fi
\fi
}
\newcommand{\hatcurPPfluxperieccen}[1]{\ifnum#1=13 %
\hatcurPPfluxperieccenxxxxxA
\else
\ifnum#1=14 %
\hatcurPPfluxperieccenxxxxxB
\else
??????\fi
\fi
}
\newcommand{\hatcurPPfluxperidimeccen}[1]{\ifnum#1=13 %
\hatcurPPfluxperidimeccenxxxxxA
\else
\ifnum#1=14 %
\hatcurPPfluxperidimeccenxxxxxB
\else
??????\fi
\fi
}
\newcommand{\hatcurPPgeccen}[1]{\ifnum#1=13 %
\hatcurPPgeccenxxxxxA
\else
\ifnum#1=14 %
\hatcurPPgeccenxxxxxB
\else
??????\fi
\fi
}
\newcommand{\hatcurPPieccen}[1]{\ifnum#1=13 %
\hatcurPPieccenxxxxxA
\else
\ifnum#1=14 %
\hatcurPPieccenxxxxxB
\else
??????\fi
\fi
}
\newcommand{\hatcurPPloggeccen}[1]{\ifnum#1=13 %
\hatcurPPloggeccenxxxxxA
\else
\ifnum#1=14 %
\hatcurPPloggeccenxxxxxB
\else
??????\fi
\fi
}
\newcommand{\hatcurPPmeccen}[1]{\ifnum#1=13 %
\hatcurPPmeccenxxxxxA
\else
\ifnum#1=14 %
\hatcurPPmeccenxxxxxB
\else
??????\fi
\fi
}
\newcommand{\hatcurPPmeeccen}[1]{\ifnum#1=13 %
\hatcurPPmeeccenxxxxxA
\else
\ifnum#1=14 %
\hatcurPPmeeccenxxxxxB
\else
??????\fi
\fi
}
\newcommand{\hatcurPPmelongeccen}[1]{\ifnum#1=13 %
\hatcurPPmelongeccenxxxxxA
\else
\ifnum#1=14 %
\hatcurPPmelongeccenxxxxxB
\else
??????\fi
\fi
}
\newcommand{\hatcurPPmeshorteccen}[1]{\ifnum#1=13 %
\hatcurPPmeshorteccenxxxxxA
\else
\ifnum#1=14 %
\hatcurPPmeshorteccenxxxxxB
\else
??????\fi
\fi
}
\newcommand{\hatcurPPmlongeccen}[1]{\ifnum#1=13 %
\hatcurPPmlongeccenxxxxxA
\else
\ifnum#1=14 %
\hatcurPPmlongeccenxxxxxB
\else
??????\fi
\fi
}
\newcommand{\hatcurPPmrcorreccen}[1]{\ifnum#1=13 %
\hatcurPPmrcorreccenxxxxxA
\else
\ifnum#1=14 %
\hatcurPPmrcorreccenxxxxxB
\else
??????\fi
\fi
}
\newcommand{\hatcurPPmshorteccen}[1]{\ifnum#1=13 %
\hatcurPPmshorteccenxxxxxA
\else
\ifnum#1=14 %
\hatcurPPmshorteccenxxxxxB
\else
??????\fi
\fi
}
\newcommand{\hatcurPPperieccen}[1]{\ifnum#1=13 %
\hatcurPPperieccenxxxxxA
\else
\ifnum#1=14 %
\hatcurPPperieccenxxxxxB
\else
??????\fi
\fi
}
\newcommand{\hatcurPPphiconjeccen}[1]{\ifnum#1=13 %
\hatcurPPphiconjeccenxxxxxA
\else
\ifnum#1=14 %
\hatcurPPphiconjeccenxxxxxB
\else
??????\fi
\fi
}
\newcommand{\hatcurPPreccen}[1]{\ifnum#1=13 %
\hatcurPPreccenxxxxxA
\else
\ifnum#1=14 %
\hatcurPPreccenxxxxxB
\else
??????\fi
\fi
}
\newcommand{\hatcurPPreeccen}[1]{\ifnum#1=13 %
\hatcurPPreeccenxxxxxA
\else
\ifnum#1=14 %
\hatcurPPreeccenxxxxxB
\else
??????\fi
\fi
}
\newcommand{\hatcurPPrelongeccen}[1]{\ifnum#1=13 %
\hatcurPPrelongeccenxxxxxA
\else
\ifnum#1=14 %
\hatcurPPrelongeccenxxxxxB
\else
??????\fi
\fi
}
\newcommand{\hatcurPPreshorteccen}[1]{\ifnum#1=13 %
\hatcurPPreshorteccenxxxxxA
\else
\ifnum#1=14 %
\hatcurPPreshorteccenxxxxxB
\else
??????\fi
\fi
}
\newcommand{\hatcurPPrhoeccen}[1]{\ifnum#1=13 %
\hatcurPPrhoeccenxxxxxA
\else
\ifnum#1=14 %
\hatcurPPrhoeccenxxxxxB
\else
??????\fi
\fi
}
\newcommand{\hatcurPPrlongeccen}[1]{\ifnum#1=13 %
\hatcurPPrlongeccenxxxxxA
\else
\ifnum#1=14 %
\hatcurPPrlongeccenxxxxxB
\else
??????\fi
\fi
}
\newcommand{\hatcurPPrshorteccen}[1]{\ifnum#1=13 %
\hatcurPPrshorteccenxxxxxA
\else
\ifnum#1=14 %
\hatcurPPrshorteccenxxxxxB
\else
??????\fi
\fi
}
\newcommand{\hatcurPPtcirceccen}[1]{\ifnum#1=13 %
\hatcurPPtcirceccenxxxxxA
\else
\ifnum#1=14 %
\hatcurPPtcirceccenxxxxxB
\else
??????\fi
\fi
}
\newcommand{\hatcurPPteffeccen}[1]{\ifnum#1=13 %
\hatcurPPteffeccenxxxxxA
\else
\ifnum#1=14 %
\hatcurPPteffeccenxxxxxB
\else
??????\fi
\fi
}
\newcommand{\hatcurPPthetaeccen}[1]{\ifnum#1=13 %
\hatcurPPthetaeccenxxxxxA
\else
\ifnum#1=14 %
\hatcurPPthetaeccenxxxxxB
\else
??????\fi
\fi
}
\newcommand{\hatcurPPtinfalleccen}[1]{\ifnum#1=13 %
\hatcurPPtinfalleccenxxxxxA
\else
\ifnum#1=14 %
\hatcurPPtinfalleccenxxxxxB
\else
??????\fi
\fi
}
\newcommand{\hatcurRVecceneccen}[1]{\ifnum#1=13 %
\hatcurRVecceneccenxxxxxA
\else
\ifnum#1=14 %
\hatcurRVecceneccenxxxxxB
\else
??????\fi
\fi
}
\newcommand{\hatcurRVeccentwosiglimeccen}[1]{\ifnum#1=13 %
\hatcurRVeccentwosiglimeccenxxxxxA
\else
\ifnum#1=14 %
\hatcurRVeccentwosiglimeccenxxxxxB
\else
??????\fi
\fi
}
\newcommand{\hatcurRVfitrmsAeccen}[1]{\ifnum#1=13 %
\hatcurRVfitrmsAeccenxxxxxA
\else
\ifnum#1=14 %
\hatcurRVfitrmsAeccenxxxxxB
\else
??????\fi
\fi
}
\newcommand{\hatcurRVfitrmsBeccen}[1]{\ifnum#1=13 %
\hatcurRVfitrmsBeccenxxxxxA
\else
\ifnum#1=14 %
\hatcurRVfitrmsBeccenxxxxxB
\else
??????\fi
\fi
}
\newcommand{\hatcurRVfitrmsCeccen}[1]{\ifnum#1=13 %
\hatcurRVfitrmsCeccenxxxxxA
\else
??????\fi
}
\newcommand{\hatcurRVfitrmsDeccen}[1]{\ifnum#1=13 %
\hatcurRVfitrmsDeccenxxxxxA
\else
??????\fi
}
\newcommand{\hatcurRVgammaAeccen}[1]{\ifnum#1=13 %
\hatcurRVgammaAeccenxxxxxA
\else
\ifnum#1=14 %
\hatcurRVgammaAeccenxxxxxB
\else
??????\fi
\fi
}
\newcommand{\hatcurRVgammaBeccen}[1]{\ifnum#1=13 %
\hatcurRVgammaBeccenxxxxxA
\else
\ifnum#1=14 %
\hatcurRVgammaBeccenxxxxxB
\else
??????\fi
\fi
}
\newcommand{\hatcurRVgammaCeccen}[1]{\ifnum#1=13 %
\hatcurRVgammaCeccenxxxxxA
\else
??????\fi
}
\newcommand{\hatcurRVgammaDeccen}[1]{\ifnum#1=13 %
\hatcurRVgammaDeccenxxxxxA
\else
??????\fi
}
\newcommand{\hatcurRVheccen}[1]{\ifnum#1=13 %
\hatcurRVheccenxxxxxA
\else
\ifnum#1=14 %
\hatcurRVheccenxxxxxB
\else
??????\fi
\fi
}
\newcommand{\hatcurRVjitterAeccen}[1]{\ifnum#1=13 %
\hatcurRVjitterAeccenxxxxxA
\else
\ifnum#1=14 %
\hatcurRVjitterAeccenxxxxxB
\else
??????\fi
\fi
}
\newcommand{\hatcurRVjitterBeccen}[1]{\ifnum#1=13 %
\hatcurRVjitterBeccenxxxxxA
\else
\ifnum#1=14 %
\hatcurRVjitterBeccenxxxxxB
\else
??????\fi
\fi
}
\newcommand{\hatcurRVjitterCeccen}[1]{\ifnum#1=13 %
\hatcurRVjitterCeccenxxxxxA
\else
??????\fi
}
\newcommand{\hatcurRVjitterDeccen}[1]{\ifnum#1=13 %
\hatcurRVjitterDeccenxxxxxA
\else
??????\fi
}
\newcommand{\hatcurRVkeccen}[1]{\ifnum#1=13 %
\hatcurRVkeccenxxxxxA
\else
\ifnum#1=14 %
\hatcurRVkeccenxxxxxB
\else
??????\fi
\fi
}
\newcommand{\hatcurRVKeccen}[1]{\ifnum#1=13 %
\hatcurRVKeccenxxxxxA
\else
\ifnum#1=14 %
\hatcurRVKeccenxxxxxB
\else
??????\fi
\fi
}
\newcommand{\hatcurRVomegaeccen}[1]{\ifnum#1=13 %
\hatcurRVomegaeccenxxxxxA
\else
\ifnum#1=14 %
\hatcurRVomegaeccenxxxxxB
\else
??????\fi
\fi
}
\newcommand{\hatcurRVrheccen}[1]{\ifnum#1=13 %
\hatcurRVrheccenxxxxxA
\else
\ifnum#1=14 %
\hatcurRVrheccenxxxxxB
\else
??????\fi
\fi
}
\newcommand{\hatcurRVrkeccen}[1]{\ifnum#1=13 %
\hatcurRVrkeccenxxxxxA
\else
\ifnum#1=14 %
\hatcurRVrkeccenxxxxxB
\else
??????\fi
\fi
}
\newcommand{\hatcurRVtroneeccen}[1]{\ifnum#1=13 %
\hatcurRVtroneeccenxxxxxA
\else
\ifnum#1=14 %
\hatcurRVtroneeccenxxxxxB
\else
??????\fi
\fi
}
\newcommand{\hatcurRVtrtwoeccen}[1]{\ifnum#1=13 %
\hatcurRVtrtwoeccenxxxxxA
\else
\ifnum#1=14 %
\hatcurRVtrtwoeccenxxxxxB
\else
??????\fi
\fi
}
\newcommand{\hatcurSMEiiloggeccen}[1]{\ifnum#1=13 %
\hatcurSMEiiloggeccenxxxxxA
\else
\ifnum#1=14 %
\hatcurSMEiiloggeccenxxxxxB
\else
??????\fi
\fi
}
\newcommand{\hatcurSMEiiteffeccen}[1]{\ifnum#1=13 %
\hatcurSMEiiteffeccenxxxxxA
\else
\ifnum#1=14 %
\hatcurSMEiiteffeccenxxxxxB
\else
??????\fi
\fi
}
\newcommand{\hatcurSMEiivsineccen}[1]{\ifnum#1=13 %
\hatcurSMEiivsineccenxxxxxA
\else
\ifnum#1=14 %
\hatcurSMEiivsineccenxxxxxB
\else
??????\fi
\fi
}
\newcommand{\hatcurSMEiizfeheccen}[1]{\ifnum#1=13 %
\hatcurSMEiizfeheccenxxxxxA
\else
\ifnum#1=14 %
\hatcurSMEiizfeheccenxxxxxB
\else
??????\fi
\fi
}
\newcommand{\hatcurSMEiizfehshorteccen}[1]{\ifnum#1=13 %
\hatcurSMEiizfehshorteccenxxxxxA
\else
\ifnum#1=14 %
\hatcurSMEiizfehshorteccenxxxxxB
\else
??????\fi
\fi
}
\newcommand{\hatcurSMEiloggeccen}[1]{\ifnum#1=13 %
\hatcurSMEiloggeccenxxxxxA
\else
\ifnum#1=14 %
\hatcurSMEiloggeccenxxxxxB
\else
??????\fi
\fi
}
\newcommand{\hatcurSMEiteffeccen}[1]{\ifnum#1=13 %
\hatcurSMEiteffeccenxxxxxA
\else
\ifnum#1=14 %
\hatcurSMEiteffeccenxxxxxB
\else
??????\fi
\fi
}
\newcommand{\hatcurSMEivmaceccen}[1]{\ifnum#1=13 %
\hatcurSMEivmaceccenxxxxxA
\else
\ifnum#1=14 %
\hatcurSMEivmaceccenxxxxxB
\else
??????\fi
\fi
}
\newcommand{\hatcurSMEivmiceccen}[1]{\ifnum#1=13 %
\hatcurSMEivmiceccenxxxxxA
\else
\ifnum#1=14 %
\hatcurSMEivmiceccenxxxxxB
\else
??????\fi
\fi
}
\newcommand{\hatcurSMEivsineccen}[1]{\ifnum#1=13 %
\hatcurSMEivsineccenxxxxxA
\else
\ifnum#1=14 %
\hatcurSMEivsineccenxxxxxB
\else
??????\fi
\fi
}
\newcommand{\hatcurSMEizfeheccen}[1]{\ifnum#1=13 %
\hatcurSMEizfeheccenxxxxxA
\else
\ifnum#1=14 %
\hatcurSMEizfeheccenxxxxxB
\else
??????\fi
\fi
}
\newcommand{\hatcurSMEizfehshorteccen}[1]{\ifnum#1=13 %
\hatcurSMEizfehshorteccenxxxxxA
\else
\ifnum#1=14 %
\hatcurSMEizfehshorteccenxxxxxB
\else
??????\fi
\fi
}
\newcommand{\hatcurXAveccen}[1]{\ifnum#1=13 %
\hatcurXAveccenxxxxxA
\else
\ifnum#1=14 %
\hatcurXAveccenxxxxxB
\else
??????\fi
\fi
}
\newcommand{\hatcurXdisteccen}[1]{\ifnum#1=13 %
\hatcurXdisteccenxxxxxA
\else
\ifnum#1=14 %
\hatcurXdisteccenxxxxxB
\else
??????\fi
\fi
}
\newcommand{\hatcurXdistredeccen}[1]{\ifnum#1=13 %
\hatcurXdistredeccenxxxxxA
\else
\ifnum#1=14 %
\hatcurXdistredeccenxxxxxB
\else
??????\fi
\fi
}
\newcommand{\hatcurXEBVeccen}[1]{\ifnum#1=13 %
\hatcurXEBVeccenxxxxxA
\else
\ifnum#1=14 %
\hatcurXEBVeccenxxxxxB
\else
??????\fi
\fi
}
\newcommand{\hatcurXjhisoredeccen}[1]{\ifnum#1=13 %
\hatcurXjhisoredeccenxxxxxA
\else
\ifnum#1=14 %
\hatcurXjhisoredeccenxxxxxB
\else
??????\fi
\fi
}
\newcommand{\hatcurXjkisoredeccen}[1]{\ifnum#1=13 %
\hatcurXjkisoredeccenxxxxxA
\else
\ifnum#1=14 %
\hatcurXjkisoredeccenxxxxxB
\else
??????\fi
\fi
}
\newcommand{\hatcurXmhisoredeccen}[1]{\ifnum#1=13 %
\hatcurXmhisoredeccenxxxxxA
\else
\ifnum#1=14 %
\hatcurXmhisoredeccenxxxxxB
\else
??????\fi
\fi
}
\newcommand{\hatcurXmiisoredeccen}[1]{\ifnum#1=13 %
\hatcurXmiisoredeccenxxxxxA
\else
\ifnum#1=14 %
\hatcurXmiisoredeccenxxxxxB
\else
??????\fi
\fi
}
\newcommand{\hatcurXmjisoredeccen}[1]{\ifnum#1=13 %
\hatcurXmjisoredeccenxxxxxA
\else
\ifnum#1=14 %
\hatcurXmjisoredeccenxxxxxB
\else
??????\fi
\fi
}
\newcommand{\hatcurXmkisoredeccen}[1]{\ifnum#1=13 %
\hatcurXmkisoredeccenxxxxxA
\else
\ifnum#1=14 %
\hatcurXmkisoredeccenxxxxxB
\else
??????\fi
\fi
}
\newcommand{\hatcurXmvisoredeccen}[1]{\ifnum#1=13 %
\hatcurXmvisoredeccenxxxxxA
\else
\ifnum#1=14 %
\hatcurXmvisoredeccenxxxxxB
\else
??????\fi
\fi
}
\newcommand{\hatcurXsecdureccen}[1]{\ifnum#1=13 %
\hatcurXsecdureccenxxxxxA
\else
\ifnum#1=14 %
\hatcurXsecdureccenxxxxxB
\else
??????\fi
\fi
}
\newcommand{\hatcurXsecingdureccen}[1]{\ifnum#1=13 %
\hatcurXsecingdureccenxxxxxA
\else
\ifnum#1=14 %
\hatcurXsecingdureccenxxxxxB
\else
??????\fi
\fi
}
\newcommand{\hatcurXsecondaryeccen}[1]{\ifnum#1=13 %
\hatcurXsecondaryeccenxxxxxA
\else
\ifnum#1=14 %
\hatcurXsecondaryeccenxxxxxB
\else
??????\fi
\fi
}
\newcommand{\hatcurXsecphaseeccen}[1]{\ifnum#1=13 %
\hatcurXsecphaseeccenxxxxxA
\else
\ifnum#1=14 %
\hatcurXsecphaseeccenxxxxxB
\else
??????\fi
\fi
}
\newcommand{\hatcurXviisoredeccen}[1]{\ifnum#1=13 %
\hatcurXviisoredeccenxxxxxA
\else
\ifnum#1=14 %
\hatcurXviisoredeccenxxxxxB
\else
??????\fi
\fi
}
\newcommand{\hatcurXvkisoredeccen}[1]{\ifnum#1=13 %
\hatcurXvkisoredeccenxxxxxA
\else
\ifnum#1=14 %
\hatcurXvkisoredeccenxxxxxB
\else
??????\fi
\fi
}

\newcommand{\hatcurxxxxxA}{HATS-13}
\newcommand{\hatcurbxxxxxA}{HATS-13b}
\newcommand{\hatcurcxxxxxA}{HATS-13c}

\newcommand{\hatcurplanetnumxxxxxA}{13}

\newcommand{\hatcurRVgammaabsxxxxxA}{***TBD***}                           % Absolute Gamma velocity

\newcommand{\hatcurRVgammarelxxxxxA}{***TBD***}                           % Relative Gamma velocity. Typically that of the Keck RVs.

\newcommand{\hatcurCCtassvixxxxxA}{***TBD***}                  % TASS V-I

\newcommand{\hatcurSMEversionxxxxxA}{ii}                                       % Final SME version:i or ii?

\newcommand{\hatcurisoshortxxxxxA}{YY}
\newcommand{\hatcurisofullxxxxxA}{Yonsei-Yale (YY)}
\newcommand{\hatcurisocitexxxxxA}{yi:2001}
\newcommand{\hatcurlumindxxxxxA}{\arstar}
\newcommand{\hatcurjhkfilsetxxxxxA}{ESO}
\newcommand{\hatcurSMEteffxxxxxA}{\ifthenelse{\equal{\hatcurSMEversionxxxxxA}{i}}{\hatcurSMEiteff{\hatcurplanetnumxxxxxA}}{\hatcurSMEiiteff{\hatcurplanetnumxxxxxA}}}
\newcommand{\hatcurSMEzfehxxxxxA}{\ifthenelse{\equal{\hatcurSMEversionxxxxxA}{i}}{\hatcurSMEizfeh{\hatcurplanetnumxxxxxA}}{\hatcurSMEiizfeh{\hatcurplanetnumxxxxxA}}}
\newcommand{\hatcurSMEzfehshortxxxxxA}{\ifthenelse{\equal{\hatcurSMEversionxxxxxA}{i}}{\hatcurSMEizfehshort{\hatcurplanetnumxxxxxA}}{\hatcurSMEiizfehshort{\hatcurplanetnumxxxxxA}}}
\newcommand{\hatcurSMEloggxxxxxA}{\ifthenelse{\equal{\hatcurSMEversionxxxxxA}{i}}{\hatcurSMEilogg{\hatcurplanetnumxxxxxA}}{\hatcurSMEiilogg{\hatcurplanetnumxxxxxA}}}
\newcommand{\hatcurSMEvsinxxxxxA}{\ifthenelse{\equal{\hatcurSMEversionxxxxxA}{i}}{\hatcurSMEivsin{\hatcurplanetnumxxxxxA}}{\hatcurSMEiivsin{\hatcurplanetnumxxxxxA}}}
\newcommand{\hatcurSMEvmacxxxxxA}{\ifthenelse{\equal{\hatcurSMEversionxxxxxA}{i}}{\hatcurSMEivmac{\hatcurplanetnumxxxxxA}}{\hatcurSMEiivmac{\hatcurplanetnumxxxxxA}}}
\newcommand{\hatcurSMEvmicxxxxxA}{\ifthenelse{\equal{\hatcurSMEversionxxxxxA}{i}}{\hatcurSMEivmic{\hatcurplanetnumxxxxxA}}{\hatcurSMEiivmic{\hatcurplanetnumxxxxxA}}}

\newcommand{\hatcurxxxxxB}{HATS-14}
\newcommand{\hatcurbxxxxxB}{HATS-14b}
\newcommand{\hatcurcxxxxxB}{HATS-14c}

\newcommand{\hatcurplanetnumxxxxxB}{14}

\newcommand{\hatcurRVgammaabsxxxxxB}{***TBD***}                           % Absolute Gamma velocity

\newcommand{\hatcurRVgammarelxxxxxB}{***TBD***}                           % Relative Gamma velocity. Typically that of the Keck RVs.

\newcommand{\hatcurCCtassvixxxxxB}{***TBD***}                  % TASS V-I

\newcommand{\hatcurSMEversionxxxxxB}{ii}                                       % Final SME version:i or ii?

\newcommand{\hatcurisoshortxxxxxB}{YY}
\newcommand{\hatcurisofullxxxxxB}{Yonsei-Yale (YY)}
\newcommand{\hatcurisocitexxxxxB}{yi:2001}
\newcommand{\hatcurlumindxxxxxB}{\arstar}
\newcommand{\hatcurjhkfilsetxxxxxB}{ESO}
\newcommand{\hatcurSMEteffxxxxxB}{\ifthenelse{\equal{\hatcurSMEversionxxxxxB}{i}}{\hatcurSMEiteff{\hatcurplanetnumxxxxxB}}{\hatcurSMEiiteff{\hatcurplanetnumxxxxxB}}}
\newcommand{\hatcurSMEzfehxxxxxB}{\ifthenelse{\equal{\hatcurSMEversionxxxxxB}{i}}{\hatcurSMEizfeh{\hatcurplanetnumxxxxxB}}{\hatcurSMEiizfeh{\hatcurplanetnumxxxxxB}}}
\newcommand{\hatcurSMEzfehshortxxxxxB}{\ifthenelse{\equal{\hatcurSMEversionxxxxxB}{i}}{\hatcurSMEizfehshort{\hatcurplanetnumxxxxxB}}{\hatcurSMEiizfehshort{\hatcurplanetnumxxxxxB}}}
\newcommand{\hatcurSMEloggxxxxxB}{\ifthenelse{\equal{\hatcurSMEversionxxxxxB}{i}}{\hatcurSMEilogg{\hatcurplanetnumxxxxxB}}{\hatcurSMEiilogg{\hatcurplanetnumxxxxxB}}}
\newcommand{\hatcurSMEvsinxxxxxB}{\ifthenelse{\equal{\hatcurSMEversionxxxxxB}{i}}{\hatcurSMEivsin{\hatcurplanetnumxxxxxB}}{\hatcurSMEiivsin{\hatcurplanetnumxxxxxB}}}
\newcommand{\hatcurSMEvmacxxxxxB}{\ifthenelse{\equal{\hatcurSMEversionxxxxxB}{i}}{\hatcurSMEivmac{\hatcurplanetnumxxxxxB}}{\hatcurSMEiivmac{\hatcurplanetnumxxxxxB}}}
\newcommand{\hatcurSMEvmicxxxxxB}{\ifthenelse{\equal{\hatcurSMEversionxxxxxB}{i}}{\hatcurSMEivmic{\hatcurplanetnumxxxxxB}}{\hatcurSMEiivmic{\hatcurplanetnumxxxxxB}}}

\newcommand{\hatcur}[1]{\ifnum#1=13 %
\hatcurxxxxxA
\else
\ifnum#1=14 %
\hatcurxxxxxB
\else
??????\fi
\fi
}
\newcommand{\hatcurb}[1]{\ifnum#1=13 %
\hatcurbxxxxxA
\else
\ifnum#1=14 %
\hatcurbxxxxxB
\else
??????\fi
\fi
}
\newcommand{\hatcurc}[1]{\ifnum#1=13 %
\hatcurcxxxxxA
\else
\ifnum#1=14 %
\hatcurcxxxxxB
\else
??????\fi
\fi
}
\newcommand{\hatcurCCtassvi}[1]{\ifnum#1=13 %
\hatcurCCtassvixxxxxA
\else
\ifnum#1=14 %
\hatcurCCtassvixxxxxB
\else
??????\fi
\fi
}
\newcommand{\hatcurisocite}[1]{\ifnum#1=13 %
\hatcurisocitexxxxxA
\else
\ifnum#1=14 %
\hatcurisocitexxxxxB
\else
??????\fi
\fi
}
\newcommand{\hatcurisofull}[1]{\ifnum#1=13 %
\hatcurisofullxxxxxA
\else
\ifnum#1=14 %
\hatcurisofullxxxxxB
\else
??????\fi
\fi
}
\newcommand{\hatcurisoshort}[1]{\ifnum#1=13 %
\hatcurisoshortxxxxxA
\else
\ifnum#1=14 %
\hatcurisoshortxxxxxB
\else
??????\fi
\fi
}
\newcommand{\hatcurjhkfilset}[1]{\ifnum#1=13 %
\hatcurjhkfilsetxxxxxA
\else
\ifnum#1=14 %
\hatcurjhkfilsetxxxxxB
\else
??????\fi
\fi
}
\newcommand{\hatcurlumind}[1]{\ifnum#1=13 %
\hatcurlumindxxxxxA
\else
\ifnum#1=14 %
\hatcurlumindxxxxxB
\else
??????\fi
\fi
}
\newcommand{\hatcurplanetnum}[1]{\ifnum#1=13 %
\hatcurplanetnumxxxxxA
\else
\ifnum#1=14 %
\hatcurplanetnumxxxxxB
\else
??????\fi
\fi
}
\newcommand{\hatcurRVgammaabs}[1]{\ifnum#1=13 %
\hatcurRVgammaabsxxxxxA
\else
\ifnum#1=14 %
\hatcurRVgammaabsxxxxxB
\else
??????\fi
\fi
}
\newcommand{\hatcurRVgammarel}[1]{\ifnum#1=13 %
\hatcurRVgammarelxxxxxA
\else
\ifnum#1=14 %
\hatcurRVgammarelxxxxxB
\else
??????\fi
\fi
}
\newcommand{\hatcurSMElogg}[1]{\ifnum#1=13 %
\hatcurSMEloggxxxxxA
\else
\ifnum#1=14 %
\hatcurSMEloggxxxxxB
\else
??????\fi
\fi
}
\newcommand{\hatcurSMEteff}[1]{\ifnum#1=13 %
\hatcurSMEteffxxxxxA
\else
\ifnum#1=14 %
\hatcurSMEteffxxxxxB
\else
??????\fi
\fi
}
\newcommand{\hatcurSMEversion}[1]{\ifnum#1=13 %
\hatcurSMEversionxxxxxA
\else
\ifnum#1=14 %
\hatcurSMEversionxxxxxB
\else
??????\fi
\fi
}
\newcommand{\hatcurSMEvmac}[1]{\ifnum#1=13 %
\hatcurSMEvmacxxxxxA
\else
\ifnum#1=14 %
\hatcurSMEvmacxxxxxB
\else
??????\fi
\fi
}
\newcommand{\hatcurSMEvmic}[1]{\ifnum#1=13 %
\hatcurSMEvmicxxxxxA
\else
\ifnum#1=14 %
\hatcurSMEvmicxxxxxB
\else
??????\fi
\fi
}
\newcommand{\hatcurSMEvsin}[1]{\ifnum#1=13 %
\hatcurSMEvsinxxxxxA
\else
\ifnum#1=14 %
\hatcurSMEvsinxxxxxB
\else
??????\fi
\fi
}
\newcommand{\hatcurSMEzfeh}[1]{\ifnum#1=13 %
\hatcurSMEzfehxxxxxA
\else
\ifnum#1=14 %
\hatcurSMEzfehxxxxxB
\else
??????\fi
\fi
}
\newcommand{\hatcurSMEzfehshort}[1]{\ifnum#1=13 %
\hatcurSMEzfehshortxxxxxA
\else
\ifnum#1=14 %
\hatcurSMEzfehshortxxxxxB
\else
??????\fi
\fi
}

\newcounter{planetcounter}

\newboolean{emulateapj}
\setboolean{emulateapj}{true}

\newboolean{rvtablelong}
\setboolean{rvtablelong}{true}

\newboolean{astroph}
\setboolean{astroph}{true}
\begin{document}

\title{\hatcur{13}\lowercase{b} and \hatcur{14}\lowercase{b}: two
transiting hot Jupiters \\ from the HATSouth survey\thanks{The
HATSouth network is operated by a collaboration consisting of
Princeton University (PU), the Max Planck Institute f\"ur
Astronomie (MPIA), the Australian National University (ANU), and
the Pontificia Universidad Cat\'olica de Chile (PUC).  The station
at Las Campanas Observatory (LCO) of the Carnegie Institute is
operated by PU in conjunction with PUC, the station at the High
Energy Spectroscopic Survey (H.E.S.S.) site is operated in
conjunction with MPIA, and the station at Siding Spring
Observatory (SSO) is operated jointly with ANU.
Based in part on observations made with ($i$) the Subaru
Telescope, which is operated by the National Astronomical
Observatory of Japan; ($ii$) the MPG~2.2\,m and the ($iii$)
Euler~1.2\,m Telescopes at the ESO Observatory in La Silla; ($iv$)
the CTIO~0.9\,m Telescope at the Observatory of Cerro Tololo.
}}
   \subtitle{}
\titlerunning{\hatcur{13}\lowercase{b} and \hatcur{14}\lowercase{b}}

   \author{
          L. Mancini \inst{1}
          \and
          J.~D. Hartman \inst{2}
          \and
          K. Penev \inst{2}
          \and
          G.\'A. Bakos \inst{2}
          \and
          R. Brahm \inst{3,4}
          \and
          S. Ciceri \inst{1}
          \and
          Th. Henning \inst{1}
          \and
          Z. Csubry \inst{2}
          \and
          D. Bayliss \inst{5}
          \and
          G. Zhou \inst{6}
          \and
          M. Rabus \inst{3,1}
          \and
          M. de Val-Borro \inst{2}
          \and
          N. Espinoza \inst{3,4}
          \and
          A. Jord\'an \inst{3,4}
          \and
          V. Suc \inst{3}
          \and
          W. Bhatti \inst{2}
          \and
          B. Schmidt \inst{6}
          \and
          B. Sato \inst{7}
          \and
          T.~G. Tan \inst{8}
          \and
          D.~J. Wright\inst{2}
          \and
          C.~G. Tinney \inst{2}
          \and
          B.~C. Addison \inst{9}
          \and
          R.~W. Noyes \inst{10}
          \and
          J. L\'az\'ar \inst{11}
          \and
          I. Papp \inst{11}
          \and
          P. S\'ari \inst{11}
          }
{
       \institute{
       % 1
       Max Planck Institute for Astronomy, K\"{o}nigstuhl 17, 69117 -- Heidelberg, Germany \\
       \email{mancini@mpia.de}
       \and
       % 2
       Department of Astrophysical Sciences, Princeton University, Princeton, NJ 08544, USA
       \and
       % 3
       Instituto de Astrof\'{i}sica, Pontificia Universidad Cat\'{o}lica de Chile, Av. Vicu\~{n}a Mackenna 4860, 7820436 Macul, Santiago, Chile %
       \and
       % 4
       Millennium Institute of Astrophysics, Av. Vicu\~{n}a Mackenna 4860, 7820436 Macul, Santiago, Chile %
       \and
       % 5
       Observatoire Astronomique de l'Universit\'{e} de Gen\`{e}ve, 51 ch. des Maillettes, 1290 Versoix, Switzerland
       \and
       % 6
       The Australian National University, Canberra, Australia
       \and
       % 7
       Department of Earth and Planetary Sciences, Tokyo Institute of Technology, 2-12-1 Ookayama, Meguro-ku, Tokyo 152-8551, Japan %
       \and
       % 8
       Perth Exoplanet Survey Telescope, Perth, Australia
       \and
       % 9
       Exoplanetary Science Group, School of Physics, University of New South Wales, Sydney, NSW 2052, Australia
       \and
       % 10
       Harvard-Smithsonian Center for Astrophysics, Cambridge, MA 02138 USA %
       \and
       % 11
       Hungarian Astronomical Association, Budapest, Hungary %
}}

%   \date{Received ; Accepted}
 \abstract
{We report the discovery of HATS-13b and HATS-14b, two hot-Jupiter
transiting planets discovered by the HATSouth survey. The host
stars are quite similar to each other (\emph{HATS-13:}
$V=13.9$\,mag, $M_{\star}=0.96\,M_{\sun}$,
$R_{\star}=0.89\,R_{\sun}$, $T_{\mathrm{eff}} \approx 5500$\,K,
[Fe/H]$=0.05$; \emph{HATS-14:} $V=13.8$\,mag,
$M_{\star}=0.97\,M_{\sun}$, $R_{\star}=0.93\,R_{\sun}$,
$T_{\mathrm{eff}} \approx 5350$\,K, [Fe/H]$=0.33$) and both the
planets orbit around them with a period of $\sim 3$\,days and a
separation of $\sim 0.04$\,au. However, even though they are
irradiated in a similar way, the physical characteristics of the
two planets are very different. HATS-13b, with a mass of
$M_{\mathrm{p}}=0.543 \pm 0.072\,M_{\mathrm{J}}$ and a radius of
$R_{\mathrm{p}}=1.212 \pm 0.035\,R_{\mathrm{J}}$, appears as an
inflated planet, while HATS-14b, having a mass of
$M_{\mathrm{p}}=1.071 \pm 0.070\,M_{\mathrm{J}}$ and a radius of
$R_{\mathrm{p}}=1.039 \pm 0.032\,R_{\mathrm{J}}$, is only slightly
larger in radius than Jupiter.}

 % 5 {} token are mandatory
%{a}
%{a}
%{a}
%{a}
%{a}


\maketitle

% Sect. 1
% #####################################################################
%% Introduction
\section{Introduction}
\label{sec:introduction}
%++++++++++++++++++++++++++++++++++++++++++++++++++++++++++++++++++++++
\begin{comment}

    % The suggestions and structure below can be ignored.
    % This is inherited from HAT-P-10b.

    % 1 General: why transiting planets are important. Brief review of
    % recent findings.

    % 2 Transit search projects, current census of planets, range of
    % planetary properties.

    % 3 Summary of progress made in theory, with special attention to
    % aspects of the current system. E.g if \hatcurb is a hot Saturn,
    % then review of hot Saturns.

    % 4 Summary of HATSouth.
\end{comment}
%++++++++++++++++++++++++++++++++++++++++++++++++++++++++++++++++++++++
%% EOF introduction

After 20 years from when the human knowledge crossed the borders
of the solar system and found a planet orbiting another main
sequence star \citep{mayor:1995}, we can now count more than 1800
exoplanets in our Galaxy, and marvel how physically varied and
intriguing most of them are. The first class of \emph{unexpected}
planets with which we faced is composed by the so-called \emph{hot
Jupiters}, i.e. giant gaseous planets in close orbits
%(between $\sim 0.01$ and $0.4$\,au)
around their host stars, able to perform a complete orbit in a
relatively short time ($\sim 0.1-10$\,days). Even though they are
rarer than small-size rocky and Neptunian planets
\citep{fressin:2013,dressing:2013,petigura:2013}, there are
numerous reasons that make them very interesting to study,
especially those that transit their parent stars. Indeed, since
hot Jupiters are more massive and larger than rocky planets, it is
possible to measure their physical parameters with a much better
accuracy: \emph{in primis} mass and radius, but also their
spin-orbit alignment (from the Rossiter-McLaughlin effect), their
thermal flux and reflected light (from \emph{occultations} and
\emph{phase curve}), the chemical composition of their atmosphere
(from \emph{emission} and \emph{transmission spectra}), etc.
However, although all these parameters are accessible, even with
moderate-sized ground-based telescopes, there are various aspects
of the hot-Jupiter population that were not well understood. We
did not find, for example, any convincing way to group them in
classes based on some of their features (e.g.,
\citealp{hansen:2007,fortney:2008,schlaufman:2010,madhusudhan:2012}).
It is also very puzzling to determine what are the physical
mechanisms that regulate the formation, accretion, evolution and
cause the migration of giant planets from the snow line ($\sim
3$\,au) up to roughly $0.01$\,au from their host stars. In this
context, several scaling laws have been suggested between their
parameters (e.g.,
\citealp{southworth:2007,knutson:2010,hartman:2010}), but none
seems to be generally apply to all planets.

Answering the above questions is possible only by enlarging the
sample at our disposal, in particular the regions of the parameter
space which are currently deserted because of observational
biases. In the last three \emph{lustra}, ground-based transit
surveys have played a major role in exoplanet detection and thus
in the growth of our scientific knowledge about planetary systems.
In a fair competition with other teams (e.g., HATNet:
\citealp{bakos:2004}; WASP: \citealp{pollacco:2006}; KELT:
\citealp{pepper:2007}; MEARTH: \citealp{charbonneau:2009}; QES:
\citealp{alsubai:2013}; APACHE: \citealp{sozzetti:2013}; NGTS:
\citealp{wheatley:2013}), we are undertaking the HATSouth project,
which consists of the monitoring of millions of stars in the
southern sky to look for new exoplanet transit signals. Our survey
is carried out by a network of 6 telescope systems, employing 24
astrographs, distributed over three continents (South America,
Africa, and Australia), thus increasing the sensitive to
long-period ($>10$\,days) planets \citep{bakos:2013}.

Here we present two new transiting extrasolar planets: HATS-13b
and HATS-14b. The paper is organized as follows: in
\refsecl{sec:obs} we summarize the detection of the photometric
transit signal and the subsequent spectroscopic and photometric
observations of each star to confirm the planets. In
\refsecl{sec:analysis} we analyze the data to rule out false
positive scenarios, and to determine the stellar and planetary
parameters. Our findings are summarized and discussed in
\refsecl{sec:discussion}.

% #####################################################################
\section{Observations}
\label{sec:obs}
% #####################################################################

% =====================================================================
%% Photometric detection
\subsection{Photometric detection}
\label{sec:detection}
% =====================================================================

The \emph{modus operandi} of the HATSouth survey is
comprehensively described in \citet{bakos:2013}. In brief,
HATSouth is a network of completely automated wide-field
telescopes, consisting in six homogeneous units located at three
different places in the southern hemisphere, i.e. Las Campanas
Observatory (LCO) in Chile, the H.E.S.S. site in Namibia, and
Siding Spring Observatory (SSO) in Australia. Each unit is
equipped with four 18\,cm $f/2.8$ Takahasi astrographs, each
working in pairs with Apogee U16M Alta 4k$\times$4k CCD cameras,
with a total mosaic field-of-view on the sky of $8^{\circ} \times
8^{\circ}$ at a scale of $3.7$\,arcsec\,pixel$^{-1}$. Observations
are performed through a Sloan-$r$ filter with an exposure time of
4\,minutes. Scientific images are automatically calibrated and
light curves are extracted by aperture photometry. They are then
treated with decorrelation and detrending
algorithms\footnote{External Parameter Decorrelation (EPD;
\citealp{bakos:2010}); Trend Filtering Algorithm (TFA;
\citealp{kovacs:2005}).} and finally run through with BLS
(Box-fitting Least Squares; \citealp{kovacs:2002}) to find
periodic signals by transiting exoplanets.

\begin{figure*}
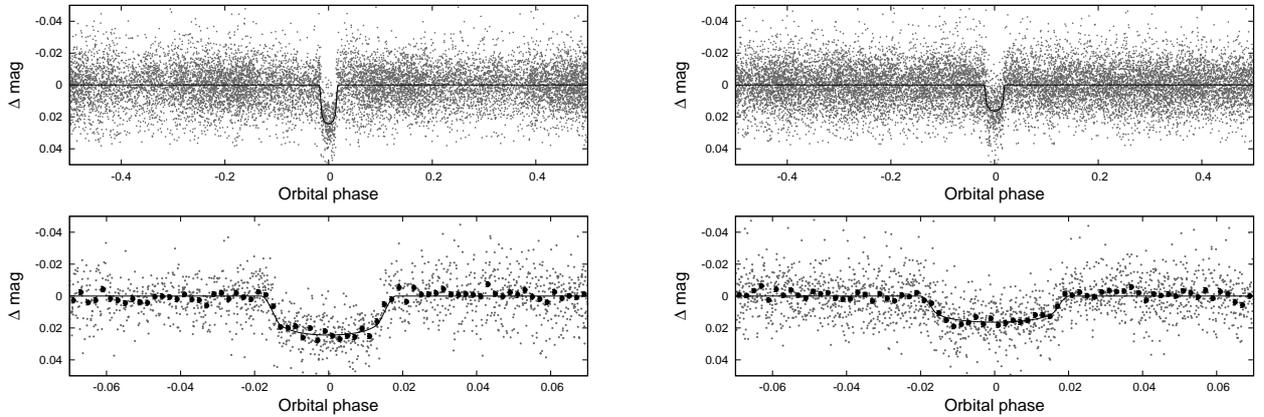
%
\centering
{{\includegraphics[width=8cm]{\hatcurhtr{13}-hs.eps} }}%
\qquad
{{\includegraphics[width=8cm]{\hatcurhtr{14}-hs.eps} }}%
\caption{Phase-folded unbinned HATSouth light curves for HATS-13
(\emph{left}) and HATS-14 (\emph{right}). In each case we show two
panels. The top panel shows the full light curve, while the bottom
panel shows the light curve zoomed-in on the transit. The solid
lines show the model fits to the light curves. The dark filled
circles in the bottom panels show the light curves binned in phase
with a bin
size of 0.002.}%
\label{fig:hatsouth}%
\end{figure*}

The stars HATS-13 (aka \hatcurCCtwomass{13}; $\alpha =
\hatcurCCra{13}$, $\delta = \hatcurCCdec{13}$; J2000) and HATS-14
(aka \hatcurCCtwomass{14}; $\alpha = \hatcurCCra{14}$, $\delta =
\hatcurCCdec{14}$; J2000) are two moderately bright
($V=13.89$\,mag and $V=13.79$\,mag, respectively) stars. They were
monitored between Nov 2009 and September 2010 by three of the
HATSouth units, which collected more than 10\,000 images for both
of them. Details of the observations are reported in
Table\,\ref{tab:photobs}. The corresponding light curves, folded
with a period of $P \sim 3.04$ and $2.77$\,days, respectively, are
plotted in Fig.\,\ref{fig:hatsouth}, both clearly showing
transiting-planet signals with depths of $\sim 2\%$ and $\sim
1\%$, respectively.

\begin{table*}
\caption{Summary of photometric observations} %
\label{tab:photobs} %
\centering     %
\tiny          %
\setlength{\tabcolsep}{8pt}
\begin{tabular}{llrrcr}
\hline\hline %
Instrument/Field$^{\mathrm{a}}$ & UT Date(s) & \# Images & Cadence$^{\mathrm{b}}$ & Filter & Precision$^{\mathrm{c}}$ \\
                                &            &           & (sec)~~~               &        & (mmag)~~             \\
\hline \\%
\multicolumn{2}{l}{\textbf{HATS-13}}              \\ [2pt]%
~~~~HS-2/G582 & 2009 Nov--2010 Sep & 2486 & 288~~~~ & $r$ & 12.6 ~~~~\\
~~~~HS-4/G582 & 2009 Sep--2010 Sep & 8565 & 288~~~~ & $r$ & 12.2 ~~~~\\
~~~~HS-6/G582 & 2010 Apr--2010 Sep &  356 & 265~~~~ & $r$ & 13.1 ~~~~\\
~~~~CTIO~0.9m$^{\mathrm{d}}$ & 2012 Aug 26        &   68 & 237~~~~ & $z$ &  2.9 ~~~~\\
~~~~GROND$^{\mathrm{d}}$     & 2012 Oct 17        &   82 &  87~~~~ & $g$ &  1.4 ~~~~\\
~~~~GROND$^{\mathrm{d}}$     & 2012 Oct 17        &   83 &  87~~~~ & $r$ &  1.2 ~~~~\\
~~~~GROND$^{\mathrm{d}}$     & 2012 Oct 17        &   82 &  87~~~~ & $i$ &  1.7 ~~~~\\
~~~~GROND$^{\mathrm{d}}$     & 2012 Oct 17        &   82 &  87~~~~ & $z$ &  1.6 ~~~~\\
~~~~PEST      & 2013 May 3         &   99 & 130~~~~ & $R$ &  4.7 ~~~~\\
~~~~PEST      & 2013 Jun 30        &  189 & 130~~~~ & $R$ &  4.7 ~~~~\\ [6pt] %
\multicolumn{2}{l}{\textbf{HATS-14}}              \\ [2pt]%
~~~~HS-2/G582 & 2009 Nov--2010 Sep & 4866 & 284~~~~ & $r$ & 11.5 ~~~~\\
~~~~HS-4/G582 & 2009 Sep--2010 Sep & 8889 & 288~~~~ & $r$ & 12.6 ~~~~\\
~~~~HS-6/G582 & 2010 Aug--2010 Sep & 200  & 290~~~~ & $r$ & 11.6 ~~~~\\
~~~~PEST      & 2013 Jun 06        & 131  & 131~~~~ & $R$ &  4.8 ~~~~\\
~~~~GROND$^{\mathrm{d}}$     & 2013 Jun 12        & 114  & 192~~~~ & $g$ &  1.6 ~~~~\\
~~~~GROND$^{\mathrm{d}}$     & 2013 Jun 12        & 114  & 192~~~~ & $r$ &  1.8 ~~~~\\
~~~~GROND$^{\mathrm{d}}$     & 2013 Jun 12        & 114  & 192~~~~ & $i$ &  2.6 ~~~~\\
~~~~GROND$^{\mathrm{d}}$     & 2013 Jun 12        & 114  & 192~~~~ & $z$ &  2.0 ~~~~\\
\hline
\end{tabular}
\tablefoot{\\
    \tiny{
    $^{\mathrm{a}}$
    For HATSouth data we list the HATSouth unit and field name from
    which the observations are taken. HS-1 and -2 are located at Las
    Campanas Observatory in Chile, HS-3 and -4 are located at the
    H.E.S.S. site in Namibia, and HS-5 and -6 are located at Siding
    Spring Observatory in Australia. Each field corresponds to one of
    838 fixed pointings used to cover the full 4$\pi$ celestial
    sphere. All data from a given HATSouth field are reduced together,
    while detrending through External Parameter Decorrelation (EPD) is
    done independently for each unique field+unit combination.
    \\ [2pt]%
    $^{\mathrm{b}}$
    The median time between consecutive images rounded to the nearest
    second. Due to weather, the day--night cycle, guiding and focus
    corrections, and other factors, the cadence is only approximately
    uniform over short timescales.
    \\ [2pt]%
    $^{\mathrm{c}}$
    The RMS of the residuals from the best-fit model.
        \\ [2pt]%
    $^{\mathrm{d}}$
    The \emph{telescope-defocussing} technique \citep{southworth:2009} was used for this transit observation.}}
\end{table*}

% =====================================================================
\subsection{Spectroscopic Observations}
\label{sec:obsspec}
% =====================================================================
After being selected as \emph{HATSouth planet} candidates, HATS-13
and HATS-14 underwent spectral reconnaissance through low- and
medium-resolution observations with the Wide Field Spectrograph
(WiFeS; \citealp{dopita:2007}) mounted on the ANU 2.3\,m telescope
at SSO. This first step is very useful in the planet confirmation
process because it can immediately rule out possible false
positive cases, mainly caused by giant stars, F-M binary systems
and blending with faint eclipsing-binary systems.

Using WiFES, we identified both the targets as dwarf stars.
HATS-13 and HATS-14 were then accurately monitored with an array
of telescopes equipped with high-resolution spectrographs,
covering wide ranges of optical wavelengths, to look for possible
radial-velocity (RV) variations compatible with the presence of
planetary companions.

Four and five spectra were observed in May 2012 for HATS-13 and
HATS-14, respectively, with CYCLOPS mounted on the 3.9\,m
Anglo-Australian Telescope at SSO. A better RV accuracy was
achieved between May and November 2012 thanks to FEROS
\citep{kaufer:1998} on the MPG 2.2\,m telescope at the ESO
Observatory in La Silla and Coralie \citep{queloz:2001} on the
Euler 1.2\,m telescope, also located in La Silla. In total, with
these two instruments, we collected 32 and 31 spectra for HATS-13
and HATS-14, respectively, with an average precision of some tens
of meters per second. Information about these spectropic
observations are summarized in Table\,\ref{tab:specobs}, yet we
did not use all the spectra in the analysis, as some of them were
discarded due to high-sky contamination. Additional details about
the instruments and data-reduction processes are exhaustively
discussed in previous works of the HATS team and we refer the
reader to those (i.e.
\citealp{penev:2013,mohler:2013,bayliss:2013}). In particular,
Coralie and FEROS spectra were reduced using the new procedure
described in \citet{jordan:2014} and \citet{brahm:2015}.

To better characterize the periodic signal of the RV variation of
HATS-13, it was necessary to observe this target with higher RV
precision. On September 2012, we used the High Dispersion
Spectrograph (HDS; \citealp{noguchi:2002}) on the Subaru telescope
at the Observatory of Mauna Kea, Hawaii. Observations were spread
over four nights and performed in a way similar to those for
HATS-5 \citep{zhou:2014}, i.e. using a $0^{\prime\prime}.6 \times
2^{\prime\prime}.0$ slit and a setup which guaranteed a
wavelength-range coverage of $3500-6200$\,\AA, with a resolution
of $R=60\,000$. Ten spectra were taken using an I$_2$ cell and
another three without it (Table\,\ref{tab:specobs}). All of HDS
observations were reduced following \citet{sato:2002,sato:2012}.

\begin{figure*}
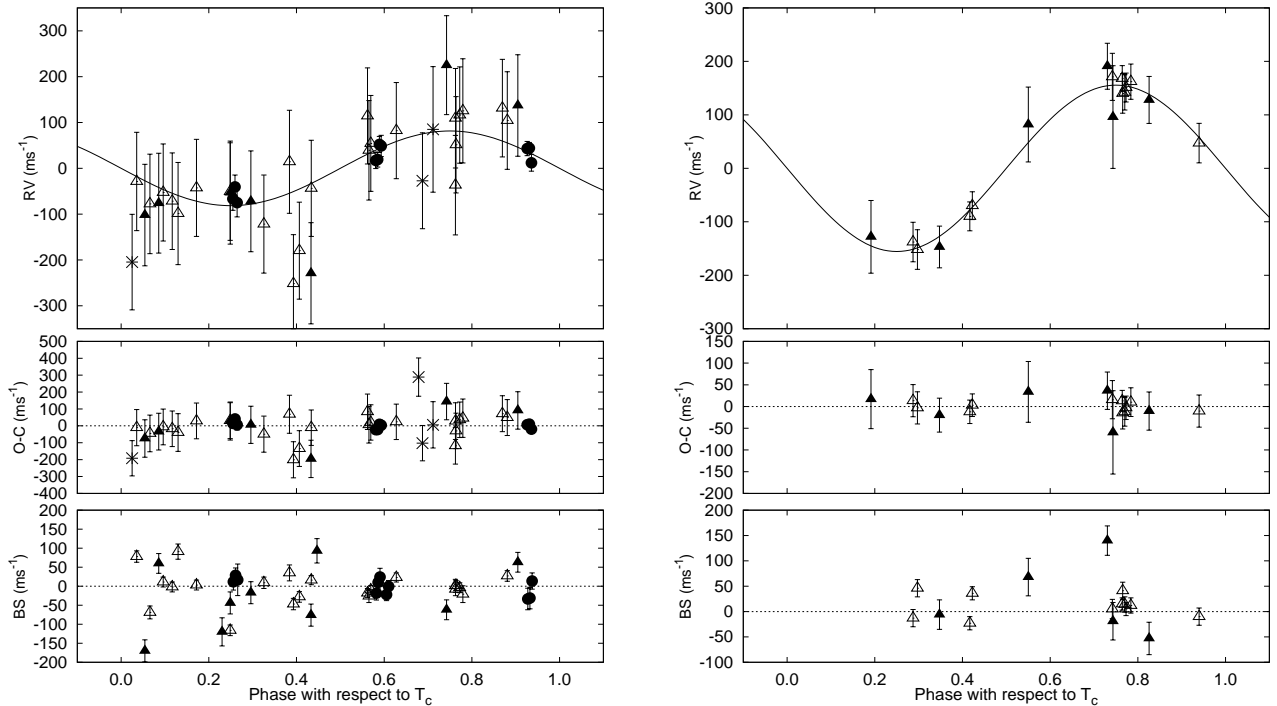
%
\centering
{{\includegraphics[width=8cm]{\hatcurhtr{13}-rv.eps} }}%
\qquad
{{\includegraphics[width=8cm]{\hatcurhtr{14}-rv.eps} }}%
\caption{Phased high-precision RV measurements for
\hbox{\hatcur{13}{}} (\emph{left}), and \hbox{\hatcur{14}{}}
(\emph{right}) from HDS (filled circles), FEROS (open triangles),
Coralie (filled triangles), and CYCLOPS (stars). In each case we
show three panels. The top panel shows the phased measurements
together with our best-fit model (see
Table\,\ref{tab:planetparam}) for each system. Zero-phase
corresponds to the time of mid-transit. The center-of-mass
velocity has been subtracted. The second panel shows the velocity
O--C residuals from the best fit. The error bars include the
jitter terms listed in Table\,\ref{tab:planetparam} added in
quadrature to the formal errors for each instrument. The third
panel shows the bisector spans (BS), with the mean value
subtracted. Note the different vertical scales of the panels.}
\label{fig:rvbis}%
\end{figure*}

All the RV measurements, extracted from the spectra here
discussed, are listed in Table\,\ref{tab:rvs1} and \ref{tab:rvs2}.
Phased RV and BS measurements are shown for each system in
Fig.\,\ref{fig:rvbis}.

\begin{table*}
\caption{Summary of spectroscopy observations} %
\label{tab:specobs} %
\centering     %
\tiny          %
\setlength{\tabcolsep}{8pt}
\begin{tabular}{llrrrrr}
\hline\hline %
Telescope/Instrument & UT Date(s) & \# Spec. & Res.~~~~                        & S/N Range$^{\mathrm{a}}$ & $\gamma_{\rm RV}$$^{\mathrm{b}}~~$ & RV Precision$^{\mathrm{c}}$ \\ %
                     &            &          & $\Delta \lambda$/$\lambda$/1000 &                          & (\kms)                             & (\ms)~~~~~                  \\ %
\hline \\%
\multicolumn{2}{l}{\textbf{HATS-13}}              \\ [2pt]%
ANU~2.3\,m/WiFeS        & 2012 Apr 10     & 1  & 3  & 90     & $\cdots$ & $\cdots$ \\
ANU~2.3\,m/WiFeS        & 2012 Apr 11-12  & 2  & 7  & 30--40 & 25.0     & 1000     \\
AAT~3.9\,m/CYCLOPS      & 2012 May 5-11   & 4  & 70 & 10--20    & 25.8     & 190      \\
Euler~1.2\,m/Coralie    & 2012 Jun--Nov   & 9  & 60 & 12--17 & 25.8     & 100      \\
MPG~2.2\,m/FEROS        & 2012 May--Oct   & 23 & 48 & 29--72 & 25.8     & 68       \\
Subaru 8\,m/HDS         & 2012 Sep 19     & 3  & 60 & 17--32 & $\cdots$ & $\cdots$ \\
Subaru 8\,m/HDS+I$_{2}$ & 2012 Sep 20--22 & 10 & 60 & 15--27 & $\cdots$ & 21       \\ [6pt] %
\multicolumn{2}{l}{\textbf{HATS-14}}              \\ [2pt]%
ANU~2.3\,m/WiFeS        & 2012 Apr 10        & 1  & 3  & 70     & $\cdots$ & $\cdots$ \\
ANU~2.3\,m/WiFeS        & 2012 Apr 11-13     & 3  & 7  & 23--31 & 28.5     & 380      \\
AAT~3.9\,m/CYCLOPS      & 2012 May 5-11      & 5  & 70 & 15--25    & 29.9     & 110      \\
Euler~1.2\,m/Coralie    & 2012 Jun--Nov      & 14 & 60 & 9--16  & 30.2     & 37       \\
MPG~2.2\,m/FEROS        & 2012 May--2013 Jul & 17 & 48 & 25--68 & 30.2     & 12       \\
\hline %
\end{tabular}
\tablefoot{\\
    \tiny{
    $^{\mathrm{a}}$
    S/N per resolution element near 5180\,\AA.
    \\ [2pt]%
    $^{\mathrm{b}}$
    For Coralie, FEROS and CYCLOPS this is the systemic RV from fitting an orbit to the observations in \ref{sec:globmod}.
    For WiFeS it is the mean of the observations, and for the du~Pont Echelle it is the measured RV of the single observation.
    We do not provide this quantity for instruments for which only relative RVs are measured, or for the lower resolution WiFeS
    observations which were only used to measure stellar atmospheric parameters.
    \\ [2pt]%
    $^{\mathrm{c}}$
    For High-precision RV observations included in the orbit
    determination this is the RV residuals from the best-fit orbit,
    for other instruments used for reconnaissance spectroscopy this is
    an estimate of the precision, or the measured standard
    deviation.}}
\end{table*}

% =====================================================================
\subsection{Photometric follow-up observations}
\label{sec:phot}
% =====================================================================
High-quality photometric follow-up observations of additional
transit events of the two targets were subsequently performed with
larger telescopes than the HATSouth units. This is also an
important step because it allowed us to have a precise light-curve
anatomy of the planetary transits (depth, duration and sharpness)
and -- by constraining the eccentricity via RV variations --
measure the mean density of the parent stars with high accuracy
and with no systematic errors \citep{seager:2003}. As we will see
in Sect.\,\ref{sec:stelparam} the knowledge of the stellar mean
density is a very useful constraint for the determination of the
other physical parameters of the two systems.

Concerning HATS-13, two complete and two incomplete transits were
observed using the MPG 2.2\,m, the CTIO 0.9\,m, and the PEST
0.3\,m telescopes. Two complete transit events were successfully
monitored for HATS-14 with the MPG 2.2\,m and PEST telescopes.
Relevant information about these observations (i.e. dates,
cadence, filter, precision) are reported in
Table\,\ref{tab:photobs}. In particular, the MPG 2.2\,m telescope
is equipped with GROND, a multi-imaging camera, able to observe a
field-of-view (FOV) of $5.4^{\prime} \times 5.4^{\prime}$ in four
different filters (similar to Sloan $g,r,i,z$) simultaneously
\citep{greiner:2008}. Details of the GROND camera and data
reduction are reported in \citet{penev:2013} and
\citet{mohler:2013}, while studies of the accuracy and S/N
expectations for this instrument were done by \citet{pierini:2012}
and \citet{mancini:2014}. The PEST telescope and data reduction
method are discussed in \citet{bayliss:2013}. The same information
for the CTIO 0.9\,m telescope have been reported by
\citet{hartman:2014}.

\begin{figure*}
\centering
\includegraphics[width=18cm]{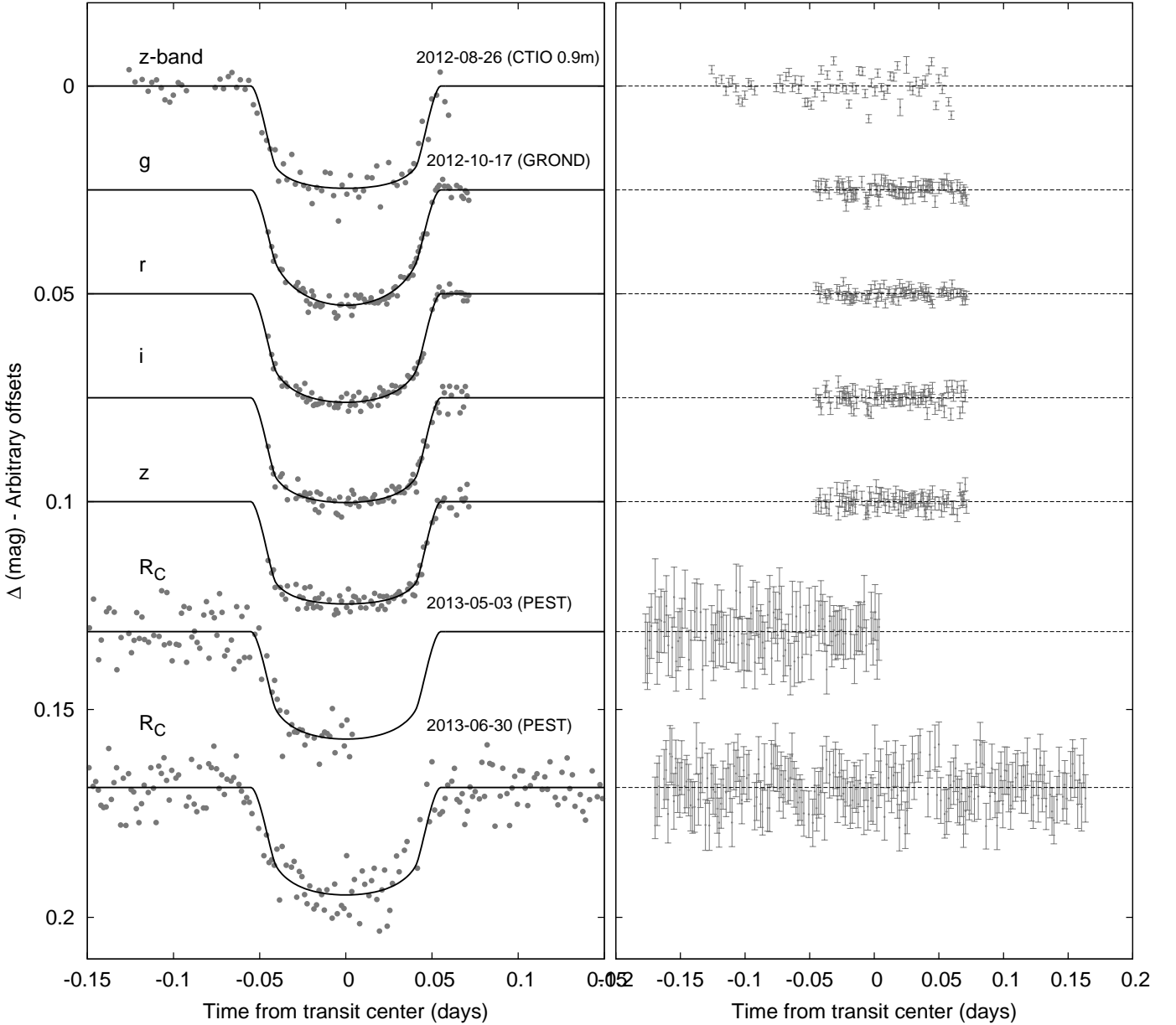}
\caption{\emph{Left panel}: Unbinned transit light curves for
\hatcur{13}. The light curves have been corrected for quadratic
trends in time fitted simultaneously with the transit model. The
dates of the events, filters and instruments used are indicated.
Light curves following the first are displaced vertically for
clarity.  Our best fit from the global modeling described in
\refsecl{sec:globmod} is shown by the solid lines. \emph{Right
panel}: residuals from the fits are displayed in the same order as
the left curves. The error bars represent the photon and
background shot noise, plus the readout noise.} %
\label{fig:lc13}
\end{figure*}

\begin{figure*}
\centering
\includegraphics[width=18cm]{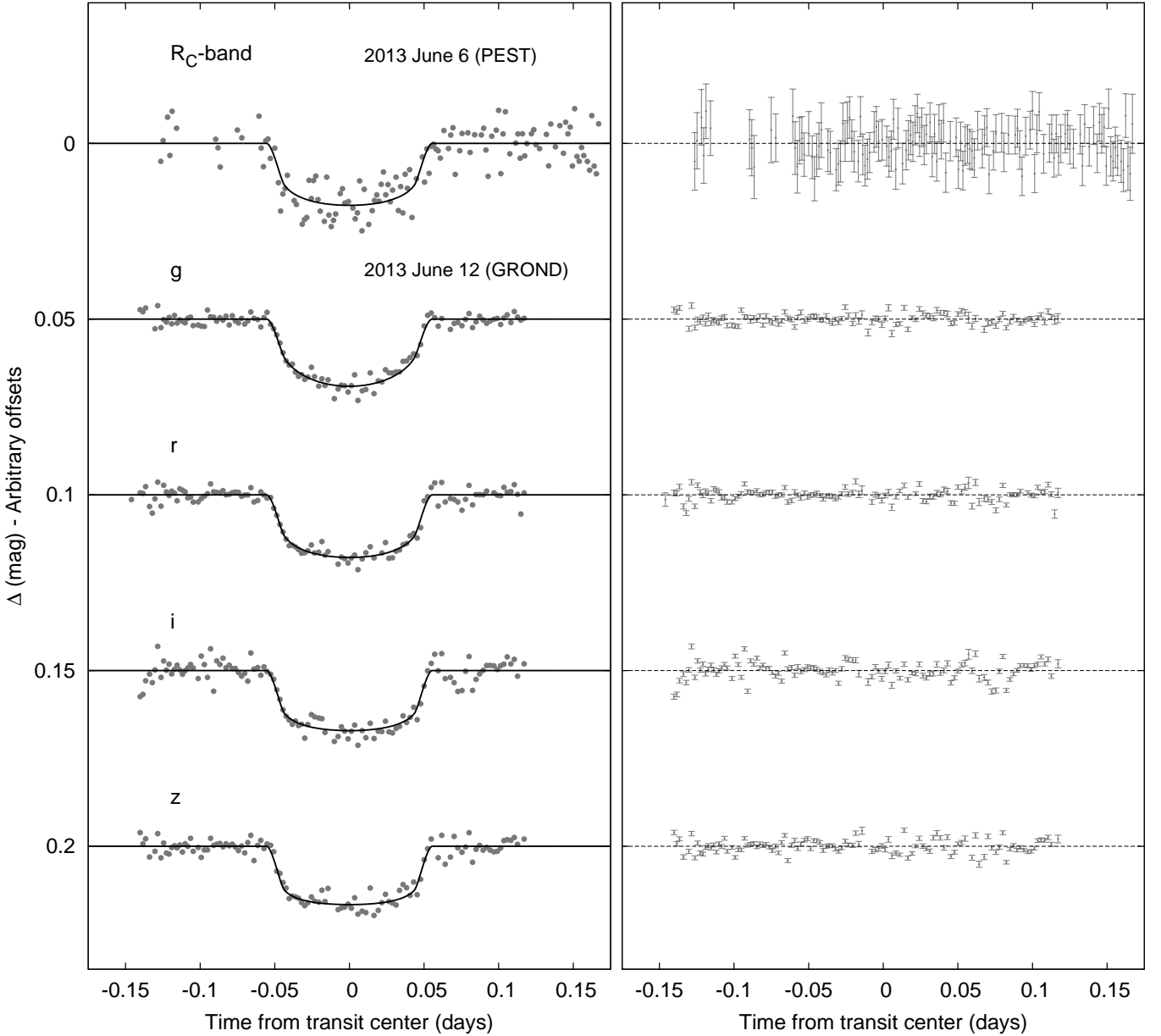}
\caption{Similar to \reffigl{fig:lc13}; here we show the follow-up
light curves for \hatcur{14}.} %
\label{fig:lc14}
\end{figure*}

The light curves for HATS-13 and HATS-14 are shown in
Fig.\,\ref{fig:lc13} and Fig.\,\ref{fig:lc14}, respectively. The
corresponding data, including those from the HATS units, are given
in Table 3.


% =====================================================================
\section{Analysis}
\label{sec:analysis}
% =====================================================================
Based on the data previously presented, this section is dedicated
to the derivation of the physical parameters of the HATS-13 and
HATS-14 planet hosts.

% =====================================================================
\subsection{Properties of the parent stars}
\label{sec:stelparam}
% =====================================================================
We used 17 and 14 high-resolution FEROS spectra to determine the
\emph{atmospheric} properties (metallicity, effective temperature
and surface gravity) of the stars HATS-13 and HATS-14,
respectively. This was accomplished by using the new routine ZASPE
(Zonal Atmospherical Stellar Parameter Estimator), which is fully
described in Brahm et al. (2015). The other principal stellar
parameters (like mass, radius, luminosity, age, etc.) and
corresponding uncertainties were estimated thanks to a Markov
chain Monte Carlo (MCMC) global analysis of our photometric and
spectroscopic data, following the methodology of
\citet{sozzetti:2007}. This is based on stellar effective
temperature $\teffstar$, which we determined with ZASPE, the
stellar mean density $\rhostar$, estimated from the light-curve
fitting (see Sect.\,\ref{sec:globmod}), and from the Yonsei-Yale
(YY; \citealp{yi:2001}) evolutionary tracks.

Spanning a range of reliable values for the metallicity, we
calculated the YY isochrones for each of the two systems over a
wide a range of ages and compared the resulting $\teffstar$ and
$\rhostar$ with those estimated from the data. The best agreement
returned the values of the other stellar parameters. In
particular, the better estimation of the stellar logarithmic
surface gravity (\loggstar$=4.524 \pm 0.017$ for HATS-13 and
\loggstar$=4.484 \pm 0.020$ for HATS-14), was used for a second
iteration of ZASPE, by fixing these values, to revise the other
atmospheric parameters.

The stellar properties that we derived are reported in
Table\,\ref{tab:stellar}, along with their 1$\sigma$
uncertainties. Model isochrones are shown in the panels of
Fig.\,\ref{fig:iso}, in which the positions of the two stars in
the $\teffstar-\rhostar$ diagram are also marked.

We found that both the stars are slightly smaller and less massive
than the Sun, with parameters listed in Table\,\ref{tab:stellar}.
In particular, with \teffstar$=5523 \pm 69$\,K, $\mstar=0.962 \pm
0.029 \, M_{\sun}$, $\rstar=0.887 \pm 0.019\, R_{\sun}$, $B-V=0.80
\pm 0.03$, $V-H=1.84 \pm 0.04$, HATS-13 is a G5\,V star, whereas
HATS-14, characterized by \teffstar$=5346 \pm 60$\,K,
$\mstar=0.967 \pm 0.024\, M_{\sun}$,
$\rstar=0.933_{-0.015}^{+0.023}\, R_{\sun}$, $B-V=0.83 \pm 0.2$,
$V-H=1.87 \pm 0.3$, has a spectral class close to the K/G
transition \citep{pecaut:2013}. The preferred metallicities are
$\feh=0.050 \pm 0.060$ and $\feh=0.330 \pm 0.060$ for HATS-13 and
HATS-14, respectively.

Table\,\ref{tab:stellar} also shows the magnitudes of the two
stars in the optical bands (taken from APASS as listed in the UCAC
4 catalog; \citealp{zacharias:2012}) and in the NIR bands (from
2MASS). We compared these values with the predicted magnitudes in
each filter from the isochrones, determining the distance of the
two stars, that is $476 \pm 12$\,pc for HATS-13 and $513 \pm
14$\,pc for HATS-14. Here the extinction was estimated by assuming
an $R_{V}=3.1$ law from \citet{cardelli:1989}.

\begin{figure*}%
\centering
{{\includegraphics[width=8cm]{\hatcurhtr{13}-iso-rho.eps} }}%
\qquad
{{\includegraphics[width=8cm]{\hatcurhtr{14}-iso-rho.eps} }}%
\caption{Model isochrones from \citet{yi:2001} for the measured
metallicities of \hatcur{13} (\emph{left panel}) and \hatcur{14}
(\emph{right panel}). In each case we show models for ages of
0.2\,Gyr and 1.0 to 14.0\,Gyr in 1.0\,Gyr increments (ages
increasing from left to right). The adopted values of $\teffstar$
and \rhostar\ are shown together with their 1$\sigma$ and
2$\sigma$ confidence ellipsoids. The initial values of \teffstar\
and \rhostar\ from the first ZASPE and \lc\ analyses are
represented with a triangle.}
\label{fig:iso}%
\end{figure*}

\begin{table*}
\caption{Stellar parameters for \hatcur{13} and \hatcur{14}} %
\label{tab:stellar} %
\centering     %
\tiny          %
\setlength{\tabcolsep}{8pt}
\begin{tabular}{lccl}
\hline\hline %
& HATS-13b & HATS-14b \\ %
~~~~~~~~~~~~~~~Parameter~~~~~~~~~~~~~~~ & Value & Value & Source \\ %
\hline \\%
\multicolumn{2}{l}{Astrometric properties and cross-identifications}                     \\ [1pt] %
~~~~2MASS-ID\dotfill          & \hatcurCCtwomass{13}  & \hatcurCCtwomass{14} & \\
~~~~GSC-ID\dotfill            & \hatcurCCgsc{13}      & \hatcurCCgsc{14}     & \\
~~~~R.A. (J2000)\dotfill      & \hatcurCCra{13}       & \hatcurCCra{14}      & 2MASS\\
~~~~Dec. (J2000)\dotfill      & \hatcurCCdec{13}      & \hatcurCCdec{14}     & 2MASS\\
~~~~$\mu_{\rm R.A.}$ (\masy)  & \hatcurCCpmra{13}     & \hatcurCCpmra{14}    & UCAC4\\
~~~~$\mu_{\rm Dec.}$ (\masy)  & \hatcurCCpmdec{13}    & \hatcurCCpmdec{14}   & UCAC4\\[3pt]
\multicolumn{2}{l}{Spectroscopic properties}                     \\ [1pt] %
% TODO: comment those that are not relevant to the paper.
~~~~$\teffstar$ (K)\dotfill         &  \hatcurSMEteff{13}      & \hatcurSMEteff{14} & ZASPE $^{\mathrm{a}}$\\
~~~~$\feh$\dotfill                  &  \hatcurSMEzfeh{13}      & \hatcurSMEzfeh{14} & ZASPE                 \\
~~~~$\vsini$ (\kms)\dotfill         &  \hatcurSMEvsin{13}      & \hatcurSMEvsin{14} & ZASPE                 \\
%~~~~$\vmac$ (\kms)\dotfill          &  ***TBD***   & ***TBD*** & Assumed              \\
%~~~~$\vmic$ (\kms)\dotfill          &  ***TBD***   & ***TBD*** & Assumed              \\
~~~~$\gamma_{\rm RV}$ (\kms)\dotfill &  $25.804 \pm 0.014$   & $30.190\pm 0.008$ & Coralie, FEROS  \\[3pt]
\multicolumn{2}{l}{Photometric properties}                     \\ [1pt] %
% TODO: comment those that are not relevant to the paper.
%       Add photometry from other sources, e.g. Tycho-2.
%~~~~$B_T$ (mag)\dotfill             &  10.494 $\pm$ 0.031\phn  & Tycho-2     \\
%~~~~$V_T$ (mag)\dotfill             &  10.038 $\pm$ 0.029\phn  & Tycho-2     \\
~~~~$B$ (mag)\dotfill               &  \hatcurCCtassmB{13}  & \hatcurCCtassmB{14} & APASS $^{\mathrm{b}}$ \\
~~~~$V$ (mag)\dotfill               &  \hatcurCCtassmv{13}  & \hatcurCCtassmv{14} & APASS $^{\mathrm{b}}$ \\
~~~~$g$ (mag)\dotfill               &  \hatcurCCtassmg{13}  & $\cdots$ & APASS $^{\mathrm{b}}$ \\
~~~~$r$ (mag)\dotfill               &  \hatcurCCtassmr{13}  & $\cdots$ & APASS $^{\mathrm{b}}$ \\
~~~~$i$ (mag)\dotfill               &  \hatcurCCtassmi{13}  & $\cdots$ & APASS $^{\mathrm{b}}$ \\
~~~~$J$ (mag)\dotfill               &  \hatcurCCtwomassJmag{13} & \hatcurCCtwomassJmag{14} & 2MASS           \\
~~~~$H$ (mag)\dotfill               &  \hatcurCCtwomassHmag{13} & \hatcurCCtwomassHmag{14} & 2MASS           \\
~~~~$K_s$ (mag)\dotfill             &  \hatcurCCtwomassKmag{13} & \hatcurCCtwomassKmag{14} & 2MASS           \\[3pt]
\multicolumn{2}{l}{Derived properties}                     \\ [1pt] %
~~~~$\mstar$ ($\msun$) \dotfill              &  \hatcurISOmlong{13} & \hatcurISOmlong{14} & YY+$\rhostar$+ZASPE $^{\mathrm{c}}$ \\
~~~~$\rstar$ ($\rsun$)\dotfill               &  \hatcurISOrlong{13} & \hatcurISOrlong{14} & YY+$\rhostar$+ZASPE                 \\
~~~~$\loggstar$ (cgs)\dotfill                &  \hatcurISOlogg{13}  & \hatcurISOlogg{14}  & YY+$\rhostar$+ZASPE                 \\
~~~~$\rhostar$ (\gcmc)\dotfill               &  \hatcurISOrho{13}   & \hatcurISOrho{14}   & YY+$\rhostar$+ZASPE $^{\mathrm{d}}$ \\
~~~~$\lstar$ ($\lsun$)\dotfill               &  \hatcurISOlum{13}   & \hatcurISOlum{14}   & YY+$\rhostar$+ZASPE                 \\
~~~~$M_V$ (mag)\dotfill                      &  \hatcurISOmv{13}    & \hatcurISOmv{14}    & YY+$\rhostar$+ZASPE                 \\
~~~~$M_K$ (mag,\hatcurjhkfilset{13})\dotfill &  \hatcurISOMK{13}    & \hatcurISOMK{14}    & YY+$\rhostar$+ZASPE                 \\
~~~~Age (Gyr)\dotfill                        &  \hatcurISOage{13}   & \hatcurISOage{14}   & YY+$\rhostar$+ZASPE                 \\
~~~~$A_{V}$ (mag)\dotfill                    &  \hatcurXAv{13}      & \hatcurXAv{14}      & YY+$\rhostar$+ZASPE                 \\
~~~~Distance (pc)\dotfill                    &  \hatcurXdistred{13} & \hatcurXdistred{14} & YY+$\rhostar$+ZASPE                 \\ [2pt] %
\hline %
\end{tabular}
\tablefoot{\\
    \tiny{
    $^{\mathrm{a}}$
    ZASPE = Zonal Atmospherical Stellar Parameter Estimator routine
    for the analysis of high-resolution spectra \citep{brahm:2015},
    applied to the FEROS spectra of \hatcur{13} and
    \hatcur{14}. These parameters rely primarily on ZASPE, but have a
    small dependence also on the iterative analysis incorporating the
    isochrone search and global modeling of the data.
    \\ [2pt]%
    $^{\mathrm{b}}$
    From APASS DR6 for \hatcur{13}, \hatcur{14} as
    listed in the UCAC 4 catalog \citep{zacharias:2012}.
    \\ [2pt]%
    $^{\mathrm{c}}$
    \hatcurisoshort{13}+\rhostar+ZASPE = Based on the \hatcurisoshort{13}
    isochrones \citep{\hatcurisocite{13}}, \rhostar\ as a luminosity
    indicator, and the ZASPE results.
    \\ [2pt]%
    $^{\mathrm{d}}$
    The parameter $\rhostar$  is primarily determined
    from the global fit to the light curves and RV data. The value
    shown here also has a slight dependence on the stellar models and
    ZASPE parameters due to restricting the posterior distribution to
    combinations of $\left[ \rhostar, \teffstar, \mathrm{Fe}/\mathrm{H} \right]$ that match to a
    \hatcurisoshort{13} stellar model.
}}
\end{table*}

% =====================================================================
\subsection{Excluding blend scenarios}
\label{sec:blend}
% =====================================================================
To rule out the possibility that either \hatcur{13} or \hatcur{14}
is a blend between an eclipsing binary and a third star
(potentially in the foreground or background of the binary), we
carried out a blend analysis following \citet{hartman:2012}. We
find that for both objects the single star with a transiting
planet model fits the light curves and broad-band photometric
color data better than a blended eclipsing binary model. For
\hatcur{13} the best-fit transiting planet model is preferred with
$2\sigma$ confidence over the best-fit blend model, while for
\hatcur{14} the best-fit transiting planet model is preferred with
$4\sigma$ confidence. Moreover, we find that any blend model that
comes close to fitting the photometric data would have been easily
detected as a composite object based on the spectroscopic data
(there would be two clear peaks in the CCFs, the RVs from the
highest peak would vary by more than 1\,\kms, as would the
bisector spans). We conclude that both \hatcur{13} and \hatcur{14}
are transiting planet systems. We cannot, however, rule out the
possibility that either object is a blend between a transiting
planet system and a third star that is fainter than the
planet-hosting star. For \hatcur{13} we find that including a
physical stellar companion with a mass greater than $0.84$\,\msun\
leads to a worse fit than not including the companion, however
even a companion up to the mass of the primary star cannot be
ruled out with greater than $5\sigma$ confidence. For \hatcur{14}
we can rule out companions with a mass greater than $0.92$\,\msun\
with greater than 5$\sigma$ confidence, while including a
companion with a mass greater than $0.5$\,\msun\ leads to a worse
fit of the data than a non-composite system. High-resolution
imaging and/or long-term RV monitoring are needed to determine if
either source has a stellar companion. For the remainder of the
paper we assume both objects are single stars with transiting
planets, however if either system has a stellar companion the true
radius and mass of the planet would be larger than what we infer
here.

% =====================================================================
\subsection{Global modelling of the data}
\label{sec:globmod}
% =====================================================================

We modeled the HATSouth photometry, the follow-up photometry, and
the high-precision RV measurements following \citet{pal:2008},
\citet{bakos:2010}, \citet{hartman:2012}. We fit
\citet{mandel:2002} transit models to the light curves, allowing
for a dilution of the HATSouth transit depth as a result of
blending from neighboring stars and over-correction by the
trend-filtering method. For the follow-up light curves we include
a quadratic trend in time in our model for each event to correct
for systematic errors in the photometry. We fit Keplerian orbits
to the RV curves allowing the zero-point for each instrument to
vary independently in the fit, and allowing for RV jitter which we
also vary as a free parameter for each instrument. We used a
Differential Evolution Markov Chain Monte Carlo procedure to
explore the fitness landscape and to determine the posterior
distribution of the parameters. One may see that for \hatcur{14}
the scatter in the Coralie and FEROS RV residuals is consistent
with the uncertainties (see Fig.\,\ref{fig:rvbis}), so our
modelling finds jitter values of $0$ for both instruments.

The resulting parameters for each system are listed in
\reftabl{tab:planetparam}. They were determined assuming circular
orbits. We have also explored non-zero eccentricities, by varying
$\sqrt{e}\cos{\omega}$ and $\sqrt{e}\sin{\omega}$ in the fitting
process, $e$ being the eccentricity and $\omega$ the argument of
the periastron. In this case, we got that $e<0.181\,(<0.142)$ at
$95\%$ confidence for HATS-3\,(HATS-4).

By inspecting \reftabl{tab:planetparam}, we can note that while
HATS-14b has mass ($M_{\mathrm{p}}=1.071 \pm 0.070\,\mjup$) and
size ($R_{\mathrm{p}}=1.039^{+0.032}_{-0.022}\,\rjup$) slightly
larger than those of Jupiter, HATS-13 is much less massive (only
$M_{\mathrm{p}}=0.543 \pm 0.072\,\mjup$), but bloated
($R_{\mathrm{p}}=1.212 \pm 0.035\,\rjup$). The above values lead
to mean densities extremely different, i.e.
$\rho_{\mathrm{p}}=0.377 \pm 0.058$\,g\,cm$^{-3}$ for HATS-13b and
$\rho_{\mathrm{p}}=1.191_{-0.140}^{+0.098}$\,g\,cm$^{-3}$  for
HATS-14b. Curiously, even though they have different physical
properties, their orbital periods (3.04 and 2.77 days) and
separation from the own host star (0.041 and 0.038 \,au) are
similar to each other.

\begin{table*}
\caption{Orbital and planetary parameters for \hatcurb{13} and
\hatcurb{14}} %
\label{tab:planetparam} %
\centering     %
\tiny          %
\setlength{\tabcolsep}{8pt}
\begin{tabular}{lcc}
\hline\hline %
& HATS-13b & HATS-14b \\ %
~~~~~~~~~~~~~~~Parameter~~~~~~~~~~~~~~~ & Value & Value \\ %
\hline \\%
\multicolumn{2}{l}{Light curve parameters}                     \\ [2pt] %
~~~$P$ (days)                               \dotfill    & $\hatcurLCP{13}$       & $\hatcurLCP{14}$       \\
~~~$T_c$ (${\rm BJD}$)      $^{\mathrm{a}}$ \dotfill    & $\hatcurLCT{13}$       & $\hatcurLCT{14}$       \\
~~~$T_{14}$ (days)          $^{\mathrm{a}}$ \dotfill    & $\hatcurLCdur{13}$     & $\hatcurLCdur{14}$     \\
~~~$T_{12} = T_{34}$ (days) $^{\mathrm{a}}$ \dotfill    & $\hatcurLCingdur{13}$  & $\hatcurLCingdur{14}$  \\
~~~$\arstar$                                \dotfill    & $\hatcurPPar{13}$      & $\hatcurPPar{14}$      \\
~~~$\zrstar$ $^{\mathrm{b}}$                \dotfill    & $\hatcurLCzeta{13}$    & $\hatcurLCzeta{14}$    \\
~~~$\rpl/\rstar$                            \dotfill    & $\hatcurLCrprstar{13}$ & $\hatcurLCrprstar{14}$ \\
~~~$b^2$                                    \dotfill    & $\hatcurLCbsq{13}$     & $\hatcurLCbsq{14}$     \\ [2pt] %
~~~$b \equiv a \cos i/\rstar$               \dotfill    & $\hatcurLCimp{13}$     & $\hatcurLCimp{14}$     \\
~~~$i$ (deg)                                \dotfill    & $\hatcurPPi{13}$       & $\hatcurPPi{14}$       \\
% If there is a periastron passage
%~~~$T_{peri}$ (days)      \dotfill   & $\hatcurPPperi{13}$           \\
%
\multicolumn{2}{l}{Limb-darkening coefficients $^{\mathrm{c}}$} \\ [2pt] %
%% TODO: comment/uncomment LD terms that were used
~~~$c_1,g$ (linear term)    \dotfill    & $\hatcurLBig{13}$  & $\hatcurLBig{14}$  \\
~~~$c_2,g$ (quadratic term) \dotfill    & $\hatcurLBiig{13}$ & $\hatcurLBiig{14}$ \\
~~~$c_1,r$                  \dotfill    & $\hatcurLBir{13}$  & $\hatcurLBir{14}$  \\
~~~$c_2,r$                  \dotfill    & $\hatcurLBiir{13}$ & $\hatcurLBiir{14}$ \\
~~~$c_1,i$                  \dotfill    & $\hatcurLBii{13}$  & $\hatcurLBii{14}$  \\
~~~$c_2,i$                  \dotfill    & $\hatcurLBiii{13}$ & $\hatcurLBiii{14}$ \\
~~~$c_1,z$                  \dotfill    & $\hatcurLBiz{13}$  & $\hatcurLBiz{14}$  \\
~~~$c_2,z$                  \dotfill    & $\hatcurLBiiz{13}$ & $\hatcurLBiiz{14}$ \\
~~~$c_1,R$                  \dotfill    & $\hatcurLBiR{13}$  & $\hatcurLBiR{14}$  \\
~~~$c_2,R$                  \dotfill    & $\hatcurLBiiR{13}$ & $\hatcurLBiiR{14}$ \\
%~~~$c_1,I$               \dotfill    & $\hatcurLBIi{13}$             \\
%~~~$c_2,I$               \dotfill    & $\hatcurLBIii{13}$            \\
%
\multicolumn{2}{l}{RV parameters}                               \\ [2pt] %
~~~$K$ (\ms)              \dotfill    & $\hatcurRVK{13}$ & $\hatcurRVK{14}$ \\
% Include if linear drift was also fitted.
%~~~$G_1$ (\ms/day)       \dotfill    & $\hatcurRVlindrift{13}$       \\
%~~~$k_{\rm RV}$\tablenotemark{c}
%                          \dotfill    & $\hatcurRVk{13}$\phs & $\hatcurRVk{14}$\phs \\
%~~~$h_{\rm RV}$\tablenotemark{c}
%                          \dotfill    & $\hatcurRVh{13}$ & $\hatcurRVh{14}$ \\
%
% Only if corrected k and h values were given based on isochrones
% and independent luminosity indicator, such as parallax.
%~~~$k_C$\tablenotemark{d} \dotfill   & $\hatcurRVkcorr{13}$          \\
%~~~$h_C$\tablenotemark{d} \dotfill   & $\hatcurRVhcorr{13}$          \\
~~~$e$ $^{\mathrm{d}}$                 \dotfill    & $\hatcurRVeccentwosiglimeccen{13}$ & $\hatcurRVeccentwosiglimeccen{14}$ \\
%
% Only give omega if the orbit was eccentric
%~~~$\omega$ (deg)         \dotfill    & $\hatcurRVomega{13}$\phn & $\hatcurRVomega{14}$\phn & $\hatcurRVomega{52}$\phn & $\hatcurRVomega{53}$\phn      \\
~~~RV jitter HDS (\ms) $^{\mathrm{e}}$ \dotfill    & \hatcurRVjitterA{13} & $\cdots$             \\
~~~RV jitter FEROS (\ms)               \dotfill    & \hatcurRVjitterB{13} & \hatcurRVjitterB{14} \\
~~~RV jitter Coralie (\ms)             \dotfill    & \hatcurRVjitterC{13} & \hatcurRVjitterA{14} \\
~~~RV jitter CYCLOPS (\ms)             \dotfill    & \hatcurRVjitterD{13} & $\cdots$             \\
\multicolumn{2}{l}{Planetary parameters}                          \\ [2pt] %
~~~$\mpl$ ($\mjup$)       \dotfill    & $\hatcurPPmlong{13}$ & $\hatcurPPmlong{14}$ \\
~~~$\rpl$ ($\rjup$)       \dotfill    & $\hatcurPPrlong{13}$ & $\hatcurPPrlong{14}$ \\
~~~$C\,(\mpl,\rpl)$ $^{\mathrm{f}}$     \dotfill    & $\hatcurPPmrcorr{13}$ & $\hatcurPPmrcorr{14}$ \\
~~~$\rhopl$ (\gcmc)       \dotfill    & $\hatcurPPrho{13}$ & $\hatcurPPrho{14}$ \\
~~~$\log g_p$ (cgs)       \dotfill    & $\hatcurPPlogg{13}$ & $\hatcurPPlogg{14}$ \\
~~~$a$ (AU)               \dotfill    & $\hatcurPParel{13}$ & $\hatcurPParel{14}$ \\
~~~$T_{\rm eq}$ (K)        \dotfill   & $\hatcurPPteff{13}$ & $\hatcurPPteff{14}$ \\
~~~$\Theta$ $^{\mathrm{g}}$ \dotfill & $\hatcurPPtheta{13}$ & $\hatcurPPtheta{14}$ \\
%
% These are given if the orbit is eccentric
%~~~$F_{per}$ (\ergscmsq) \tablenotemark{e}
%                          \dotfill    & $\hatcurPPfluxperi{13}$      \\
%~~~$F_{ap}$  (\ergscmsq) \tablenotemark{e}
%                          \dotfill    & $\hatcurPPfluxap{13}$        \\
~~~$\log_{14}\langle F \rangle$ (cgs) $^{\mathrm{h}}$ \dotfill & $\hatcurPPfluxavglog{13}$ & $\hatcurPPfluxavglog{14}$ \\ [2pt] %
\hline %
\end{tabular}
\tablefoot{\\
    \tiny{
    $^{\mathrm{a}}$
    Times are in Barycentric Julian Date calculated directly from UTC {\em without} correction for leap seconds.
    \ensuremath{T_c}: Reference epoch of mid transit that minimizes the correlation with the orbital period.
    \ensuremath{T_{14}}: total transit duration, time between first to last contact;
    \ensuremath{T_{12}=T_{34}}: ingress/egress time, time between first and second, or third and fourth contact.
    \\ [2pt]%
    $^{\mathrm{b}}$
    Reciprocal of the half duration of the transit used as a jump parameter in our MCMC analysis in place of $\arstar$. It is
    related to $\arstar$ by the expression $\zrstar = \arstar(2\pi(1+e\sin\omega))/(P\sqrt{1-b^2}\sqrt{1-e^2})$
    \citep{bakos:2010}.
    \\ [2pt]%
    $^{\mathrm{c}}$
    Values for a quadratic law, adopted from the tabulations by \cite{claret:2004} according to the spectroscopic (ZASPE)
    parameters listed in \reftabl{tab:stellar}.
    \\ [2pt]%
    $^{\mathrm{d}}$
    As discussed in \refsecl{sec:globmod} the adopted parameters for all four systems are determined assuming circular orbits. We also list
    the 95\% confidence upper limit on the eccentricity determined when $\sqrt{e}\cos\omega$ and $\sqrt{e}\sin\omega$ are allowed to
    vary in the fit.
    \\ [2pt]%
    $^{\mathrm{e}}$
    Term added in quadrature to the formal RV uncertainties for each instrument. This is treated as a free parameter in the fitting routine.
    \\ [2pt]%
    $^{\mathrm{f}}$
    Correlation coefficient between the planetary mass $\mpl$ and radius $\rpl$ estimated from the posterior parameter distribution.
    \\ [2pt]%
    $^{\mathrm{g}}$
    The Safronov number is given by $\Theta = \frac{1}{2}(V_{\rm esc}/V_{\rm orb})^2 = (a/\rpl)(\mpl / \mstar )$
    \citep[see][]{hansen:2007}.
    \\ [2pt]%
    $^{\mathrm{h}}$
    Incoming flux per unit surface area, averaged over the orbit.
}}
\end{table*}
%

% =====================================================================
\section{Discussion and conclusions}
\label{sec:discussion}
% =====================================================================

After having monitored more than 3 million stars in its almost
first five years of life, the HATSouth survey is now entering in a
phase of continuous flow of exoplanet discoveries. In this work we have presented two new
hot-Jupiter transiting planets, HATS-13b and HATS-14b, both
orbiting around slightly metal rich, mild main-sequence stars with
a period of $\sim 3$\,days. Their detection is robustly based on
extensive photometric observations and numerous RV measurements,
as we described in the previous sections.

Orbiting around similar stars at similar distances, the stellar
radiation that the two planets receive are quite similar, i.e.
$\sim 5.4$ and $\sim 6.0 \times 10^{8}$\,erg\,s$^{-1}$\,cm$^{-2}$
for HATS-13b and HATS-14b, respectively, putting them in the pL
class, according to the terminology of \citet{fortney:2008}. Based
on their equilibrium temperature and surface gravity (see
Table\,\ref{tab:planetparam}), their scale heights are $\sim740$
and $\sim230$\,km, respectively. So, HATS-13b would be a suitable
target for transmission-spectroscopy follow-up observations, but,
since it is a pL planet, we do not expect that its atmosphere
hosts a large amount of absorbing molecules in the optical
wavelength range \citep{fortney:2010}. However, past observations
of transiting gas giants reveal a wide diversity (e.g.,
\citealp{wakeford:2015}) and a more sophisticated classification
scheme for hydrogen-dominated exoplanetary atmospheres would be
necessary (see \citealp{madhusudhan:2014} and references therein).

\begin{figure*}
\centering
\includegraphics[width=18.0cm]{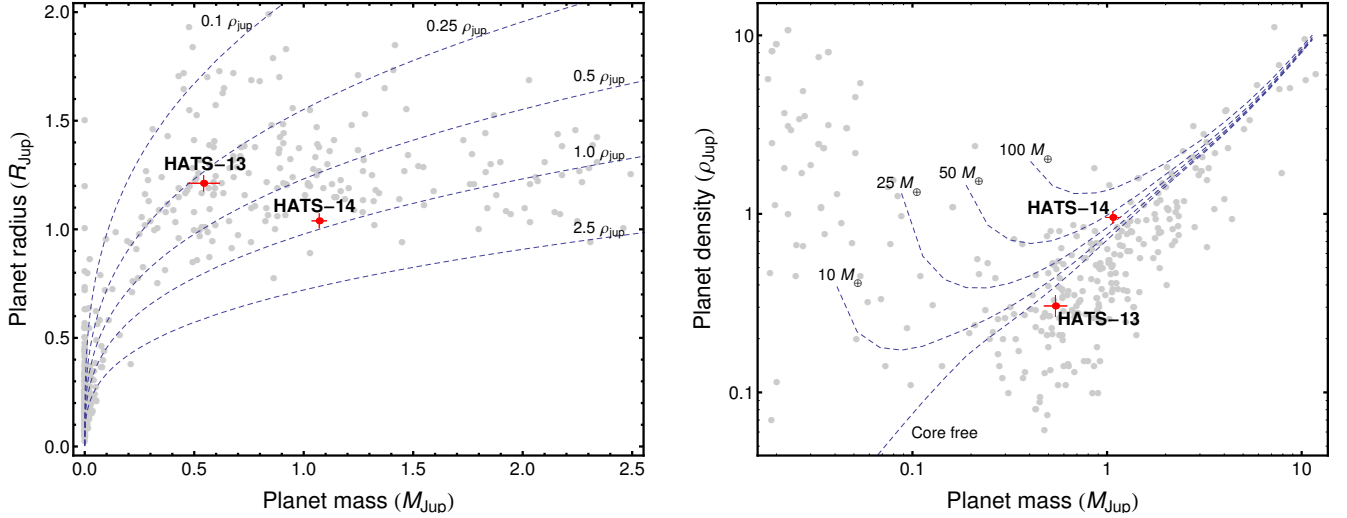}
\caption{\emph{Left panel}: Masses and radii of the known
transiting extrasolar planets. The grey points denote values taken
from TEPCat. Their error bars have been suppressed for clarity.
HATS-13b and HATS-14b are shown with red points with error bars.
Dotted lines show where density is 2.5, 1.0, 0.5, 0.25 and 0.1
$\rho_{\mathrm{J}}$. \emph{Right panel}: the mass-density diagram
of the currently known transiting exoplanets (taken from TEPCat).
Again HATS-13b and HATS-14b are shown with red points with error
bars. Four planetary models with various core masses (10, 25, 50
and 100 Earth mass) and another without a
core \citep{fortney:2007} are plotted for comparison.} %
\label{fig:diagrams}
\end{figure*}

If we look to their Safranov number, HATS-13b and HATS-14b would
belong to separate classes of planets and should have had quite
different evolution, migration and evaporation processes
\citep{hansen:2007}. Actually, even though the parent stars have
similar masses, their inferred ages differ by a factor of $\sim2$
(see Table\,\ref{tab:stellar}). Fig.\,\ref{fig:diagrams} shows the
positions of the two new HATS planets in the current planet
mass-radius plot (left panel) and planet mass-density plot (right
panel). They are shown together with those of all the other known
transiting exoplanets (data taken from the TEPCat
catalogue\footnote{The Transiting Extrasolar Planet Catalogue
(TEPCat) is available at www.astro.keele.ac.uk/jkt/tepcat/
\citep{southworth:2011}.} on March 9, 2015). It can be noted
immediately that they occupy two quite different positions in both
the diagrams. In the left panel, HATS-14b appears to be a bit out
from the population of Jupiters with masses near
$1\,M_{\mathrm{J}}$, whereas HATS-13b is in the middle of a
cluster of planets with masses around $0.5\,M_{\mathrm{J}}$ and
inflated radii. In addition to the position of the planets, the
right panel also shows 3.2\,Gyr isochrones of giant planets at
0.045\,au orbital separation from a solar analogue
\citep{fortney:2007}. The plot suggests that HATS-13b should be a
core-free planet, while HATS-14 should have a massive core of
$\sim 50\,M_{\oplus}$ (we stress that, although we cannot rule out
the possibility that HATS-14 has a stellar companion which is
diluting the transit -- see discussion in Sect.\,\ref{sec:blend}
-- our 3$\sigma$ upper limit on the radius of the planet under
this scenario is $1.11\,\rjup$).
%

%  Acknowledgements
%%%%%%%%%%%%%%%%%%%%%%%%%%%%%%%%%%%%%%%%%%%%%%%%%%%%%%%%%%%%%%%%%%%
\begin{acknowledgements}
Development of the HATSouth project was funded by NSF MRI grant
NSF/AST-0723074, operations have been supported by NASA grants
NNX09AB29G and NNX12AH91H, and follow-up observations receive
partial support from grant NSF/AST-1108686. A.J. acknowledges
support from FONDECYT project 1130857, BASAL CATA PFB-06, and
project IC120009 ``Millennium Institute of Astrophysics (MAS)'' of
the Millenium Science Initiative, Chilean Ministry of Economy.
R.B. and N.E. are supported by CONICYT- PCHA/Doctorado Nacional.
R.B. and N.E. acknowledge additional support from project IC120009
``Millenium Institute of Astrophysics (MAS)'' of the Millennium
Science Initiative, Chilean Ministry of Economy. V.S. acknowledges
support form BASAL CATA PFB-06. K.P. acknowledges support from
NASA grant NNX13AQ62G. This work is based on observations made
with telescopes at the ESO Observatory of La Silla. This paper
also uses observations obtained with facilities of the Las Cumbres
Observatory Global Telescope. Work at the Australian National
University is supported by ARC Laureate Fellowship Grant
FL0992131. We acknowledge the use of the AAVSO Photometric All-Sky
Survey (APASS), funded by the Robert Martin Ayers Sciences Fund,
and the SIMBAD database, operated at CDS, Strasbourg, France.
Operations at the MPG 2.2m Telescope are jointly performed by the
Max Planck Gesellschaft and the European Southern Observatory. The
imaging system GROND has been built by the high-energy group of
MPE in collaboration with the LSW Tautenburg and ESO. We thank
R\'{e}gis Lachaume for his technical assistance during the
observations at the MPG 2.2m Telescope. We are grateful to P.
Sackett for her help in the early phase of the HATSouth project.
The reduced light curves presented in this work will be made
available at the CDS (http://cdsweb.u-strasbg.fr/). We acknowledge
the use of the following internet-based resources: the ESO
Digitized Sky Survey; the TEPCat catalog; the SIMBAD data base
operated at CDS, Strasbourg, France; and the arXiv scientific
paper preprint service operated by Cornell University.
\end{acknowledgements}

\bibliographystyle{aa}

\begin{appendix}

%%%%%%%%%%%%%%%%%%%%%%%%%%%%%%%%%%%%%%%%%%%%%%%%%%%%%
\section{Supplementary tables}
\label{Appendix_A}
%%%%%%%%%%%%%%%%%%%%%%%%%%%%%%%%%%%%%%%%%%%%%%%%%%%%%

\begin{table*}
\caption{Light curve data for \hatcur{13} and \hatcur{14}} %
\label{tab:phfu} %
\centering     %
\tiny          %
\setlength{\tabcolsep}{8pt}
\begin{tabular}{lcrcccc}
\hline\hline %
Object$^{\mathrm{a}}$ & BJD$^{\mathrm{b}}$ & Mag$^{\mathrm{c}}$ & $\sigma_{\rm Mag}$ & Mag(orig)$^{\mathrm{d}}$ & Filter & Instrument \\
                      & ($2,400,000+$)     &                    &                    &                          &        &            \\
\hline \\%
HATS-13 & $ 55419.49174 $ & $  -0.01156 $ & $   0.00605 $ & $   0.00000 $ & $ r$ &         HS\\
HATS-13 & $ 55434.71221 $ & $  -0.00680 $ & $   0.00753 $ & $   0.00000 $ & $ r$ &         HS\\
HATS-13 & $ 55413.40415 $ & $   0.00030 $ & $   0.00605 $ & $   0.00000 $ & $ r$ &         HS\\
HATS-13 & $ 55416.44831 $ & $   0.00714 $ & $   0.00620 $ & $   0.00000 $ & $ r$ &         HS\\
HATS-13 & $ 55352.52327 $ & $   0.01857 $ & $   0.00631 $ & $   0.00000 $ & $ r$ &         HS\\
HATS-13 & $ 55422.53650 $ & $  -0.00515 $ & $   0.00599 $ & $   0.00000 $ & $ r$ &         HS\\
HATS-13 & $ 55355.56742 $ & $   0.00957 $ & $   0.00621 $ & $   0.00000 $ & $ r$ &         HS\\
HATS-13 & $ 55291.64236 $ & $   0.02224 $ & $   0.00758 $ & $   0.00000 $ & $ r$ &         HS\\
HATS-13 & $ 55373.83182 $ & $   0.01939 $ & $   0.00907 $ & $   0.00000 $ & $ r$ &         HS\\
HATS-13 & $ 55361.65578 $ & $  -0.00370 $ & $   0.00583 $ & $   0.00000 $ & $ r$ &         HS\\
\hline
\end{tabular}
\tablefoot{\\
    \tiny{This table is available in a machine-readable form in the online
    journal.  A portion is shown here for guidance regarding its form
    and content.        \\ [2pt]%
    $^{\mathrm{a}}$
    Either HATS-13, or HATS-14.
        \\ [2pt]%
    $^{\mathrm{b}}$
    Barycentric Julian Date is computed directly from the UTC time
    without correction for leap seconds.
            \\ [2pt]%
    $^{\mathrm{c}}$
    The out-of-transit level has been subtracted. For observations
    made with the HATSouth instruments (identifed by ``HS'' in the
    ``Instrument'' column) these magnitudes have been corrected for
    trends using the EPD and TFA procedures applied {\em prior} to
    fitting the transit model. This procedure may lead to an
    artificial dilution in the transit depths. For \hatcur{13} the
    transit depth is 95\% that of the true depth, with for \hatcur{14}
    it is 93\% that of the true depth. For observations made with
    follow-up instruments (anything other than ``HS'' in the
    ``Instrument'' column), the magnitudes have been corrected for a
    quadratic trend in time fit simultaneously with the transit.
            \\ [2pt]%
    $^{\mathrm{d}}$
    Raw magnitude values without correction for the quadratic trend in
    time. These are only reported for the follow-up observations.
    }}
\end{table*}

\begin{table*}
\caption{Relative radial velocities and bisector spans for
\hatcur{13}} %
\label{tab:rvs1} %
\centering     %
\tiny          %
\setlength{\tabcolsep}{8pt}
\begin{tabular}{lrrrrrl}
\hline\hline %
~~~~~~BJD      & RV\,$^{\mathrm{a}}$~ & $\sigma_{\rm RV}$\,$^{\mathrm{b}}$ & BS ~~ & $\sigma_{\rm BS}$ ~~ & Phase & Instrument \\ [1pt]%
($2,456,000+$) & (\ms)                & (\ms)                              & (\ms) & (\ms)                &       &            \\
\hline \\         %
%\multicolumn{7}{c}{\bf HATS-13} \\   [2pt]                 %
%\hline \\%
%
\input{\hatcurhtr{13}_rvtable.tex}
\hline %
\end{tabular}
\tablefoot{
    \tiny{Note that for the iodine-free template exposures we do not
        measure the RV but do measure the BS.  Such template exposures
        can be distinguished by the missing RV value. The Subaru/HDS
        observation of \hatcur{13} without a BS measurement has too
        low S/N in the I$_{2}$-free blue spectral region to pass our
        quality threshold for calculating accurate BS values. We also
        exclude from the table one I$_{2}$-free Subaru/HDS observation
        of \hatcur{13} which had too low S/N to provide an accurate BS
        measurement. \\ [2pt]
    $^{\mathrm{a}}$
    The zero-point of these velocities is arbitrary. An overall offset
    $\gamma_{\rm rel}$ fitted independently to the velocities from
    each instrument has been subtracted. \\ [2pt]
    $^{\mathrm{b}}$
    Internal errors excluding the component of astrophysical jitter
    considered in \refsecl{sec:globmod}.
}}
\end{table*}

\begin{table*}
\caption{Relative radial velocities and bisector spans for \hatcur{14}} %
\label{tab:rvs2} %
\centering     %
\tiny          %
\setlength{\tabcolsep}{8pt}
\begin{tabular}{lrrrrrl}
\hline\hline %
~~~~~~BJD      & RV\,$^{\mathrm{a}}$~ & $\sigma_{\rm RV}$\,$^{\mathrm{b}}$ & BS ~~ & $\sigma_{\rm BS}$ ~~ & Phase & Instrument \\ [1pt]%
($2,456,000+$) & (\ms)                & (\ms)                              & (\ms) & (\ms)                &       &            \\
\hline \\         %
%\multicolumn{7}{c}{\bf HATS-14} \\   [2pt]                 %
%\hline \\%
%
\input{\hatcurhtr{14}_rvtable.tex}
\hline %
\end{tabular}
\tablefoot{
    \tiny{As in Table\,\ref{tab:rvs1}
}}
\end{table*}

\end{appendix}

\end{document}